\documentclass[journal]{IEEEtran}
\IEEEoverridecommandlockouts
\usepackage{array}
\usepackage{makecell}
\usepackage{diagbox}
\usepackage{cite}
\usepackage{amsmath,amssymb,amsfonts}
\usepackage{amsthm,bm}
\usepackage{graphicx}
\usepackage{setspace}
\usepackage{textcomp}
\usepackage{float}
\usepackage{stfloats}
\usepackage{hyperref}
\hypersetup{colorlinks, citecolor=purple, filecolor=purple, linkcolor=purple, urlcolor=purple }
\usepackage{url}
\usepackage{mathtools}
\def\BibTeX{{\rm B\kern-.05em{\sc i\kern-.025em b}\kern-.08em
 T\kern-.1667em\lower.7ex\hbox{E}\kern-.125emX}}

\usepackage{stackengine}
\usepackage{algorithm}
\usepackage{enumerate}
\usepackage{booktabs}
\usepackage{subfig}
\usepackage{booktabs}
\newcommand{\tabitem}{~~\llap{\textbullet}~~}
\usepackage{multicol}
\usepackage[noend]{algpseudocode}
\usepackage[table,xcdraw]{xcolor}
\usepackage{pgfplots}
\makeatletter
\def\BState{\State\hskip-\ALG@thistlm}
\makeatother
\usepackage{multirow}

\definecolor{Blues}{RGB}{0,0,0}
\definecolor{mygray}{gray}{.8}
\definecolor{mygray2}{gray}{.7}
\definecolor{mygray3}{gray}{.6}

\algnewcommand\algorithmicforeach{\textbf{for each}}
\algdef{S}[FOR]{ForEach}[1]{\algorithmicforeach\ #1\ \algorithmicdo}
\usepackage{array}
\begin{document}

\title{Semantic Communications for Future Internet: Fundamentals, Applications, and Challenges}
\author{Wanting Yang, Hongyang Du, Zi Qin Liew, Wei Yang Bryan Lim, Zehui Xiong, Dusit Niyato, Xuefen Chi, Xuemin Sherman Shen, Chunyan Miao
\thanks{This research is supported in part by the National Research Foundation (NRF), Singapore and Infocomm Media Development Authority under the Future Communications Research Development Programme (FCP), and DSO National Laboratories under the AI Singapore Programme (AISG Award No: AISG2-RP-2020-019), under Energy Research Test-Bed and Industry Partnership Funding Initiative, part of the Energy Grid (EG) 2.0 programme, under DesCartes and the Campus for Research Excellence and Technological Enterprise (CREATE) programme, Alibaba Group through Alibaba Innovative Research (AIR) Program and Alibaba-NTU Singapore Joint Research Institute (JRI), and in part by the supported by Wallenberg-NTU Presidential Postdoctoral Fellowship. This work is also supported by the SUTD SRG-ISTD-2021-165, the SUTD-ZJU IDEA Grant (SUTD-ZJU (VP) 202102), the SUTD-ZJU IDEA Seed Grant (SUTD-ZJU (SD) 202101), and the Ministry of Education, Singapore, under its SUTD Kickstarter Initiative (SKI 20210204). This work is also supported in part by National Natural Science Foundation of China under Grant 62271228, 
the Jilin Scientific and Technological Development Program under Grant 20220101103JC, 
Natural Science Foundation of Sichuan Province under Grant 2022NSFSC0897, and the China Scholarship Council CSC NO. 202106170088.

W.~Yang is with the Department of Communications Engineering, Jilin University, Changchun, China, and also with Information Systems Technology and Design Pillar, Singapore University of Technology and Design, Singapore. Email: yangwt18@mails.jlu.edu.cn. 

H. Du, WYB.~Lim, and D.~Niyato are with the School of Computer Science and Engineering, the Energy Research Institute @ NTU, Interdisciplinary Graduate Program, Nanyang Technological University, Singapore -mail: {hongyang001,limw0201}@e.ntu.edu.sg;dniyato@ntu.edu.sg.

ZQ.~Liew is with with Alibaba Group and the Alibaba-NTU Joint Research Institute, Nanyang Technological University, Singapore. Email: ziqin001@e.ntu.edu.sg.

Z.~Xiong is with Information Systems Technology and Design Pillar, Singapore University of Technology and Design Singapore. Email: zehui\_xiong@sutd.edu.sg. D.~Niyato is with School of Computer Science and Engineering, NTU, Singapore. Emails: dniyato@ntu.edu.sg.

X.~Chi is with the Department of Communications Engineering, Jilin University, Changchun 130012, China. Email:  chixf@jlu.edu.cn.

Xuemin Sherman Shen is with the Department of Electrical and Computer
Engineering, University of Waterloo, Waterloo, ON, Canada, N2L 3G1 Email: sshen@uwaterloo.ca.

Chunyan Miao is with the School of Computer Science and Engineering,
Nanyang Technological University, Singapore, also with the Alibaba–NTU
Joint Research Institute, Nanyang Technological University, Singapore, and
also with the Joint NTU–UBC Research Centre of Excellence in Active
Living for the Elderly, Nanyang Technological University, Singapore E-mail:
ascymiao@ntu.edu.sg.
}}
\maketitle
\IEEEpeerreviewmaketitle


	\begin{abstract}
With the increasing demand for intelligent services, the sixth-generation (6G) wireless networks will shift from a traditional architecture that focuses solely on a high transmission rate to a new architecture that is based on the intelligent connection of everything.
Semantic communication (SemCom), a revolutionary architecture that integrates user as well as application requirements and the meaning of information into data processing and transmission, is predicted to become a new core paradigm in 6G. While SemCom is expected to progress beyond the classical Shannon paradigm, several obstacles need to be overcome on the way to a SemCom-enabled smart wireless Internet. In this paper, we first highlight the motivations and compelling reasons of SemCom in 6G. Then,  we provide an overview of SemCom-related theory development.  After that, we introduce three types of SemCom, i.e., semantic-oriented communication, goal-oriented communication, and semantic-aware communication. Following that, we organize the design of the
communication system into three dimensions, i.e., semantic information (SI) extraction, SI transmission, and SI metrics. For each dimension, we review existing techniques and discuss their benefits and limitation, as well as the remaining challenges. Then, we introduce the potential applications of SemCom in 6G and portray the vision of future SemCom-empowered network architecture.
Finally, we outline future research opportunities. In a nutshell, this paper provides a holistic review of the fundamentals of SemCom, its applications in 6G networks, and the existing challenges and open issues with insights for further in-depth investigations.


	\end{abstract}
\section{Introduction}\label{introduction}
\subsection{Motivation}

As we revisit the development path from the first-generation (1G) to 5G communications, it is evident that the conventional focus has been to optimize \textit{data-oriented} performance metrics, such as communication data rate and bit error probability, while ignoring service-, goal-, or semantic-related metrics. For example, 3G focuses on mobile broadband development. It promises a thousand times the data rate of 2G, whereas 4G unlocks high-speed Internet streaming, delivering a thousand times the data rate of 3G.
The motivation for this convention is traced back to the time when Shannon first demonstrated that reliable communication is possible in noisy channels in the classical information theory (CIT) literature~\cite{shannon1948mathematical}. Shannon believed that ``the \textit{semantic} aspects of communication should be regarded as \textit{irrelevant} to the engineering problem". The reason is that the meaning of a message can be related to ``certain physical and conceptual entities" and that involving the meaning in a mathematical model may affect the generality of the theory \cite{shannon1948mathematical}. 

 With continuous technological progress following CIT, the advent of the 5G has brought about a breakthrough in communication network design~\cite{li2017intelligent}, enabling a variety of services from digital twins, edge computing, the Internet of Things (IoT), and more, through the supporting technologies  such as ultra-reliable and low-latency communications (URLLC), massive machine type communications (mMTC), and enhanced mobile broadband (eMBB) communications.
However, content-centric data-driven communication architecture is increasingly seen as a barrier to providing end-users with services that demand high quality of experience (QoE). This is especially so given that the emerging applications of the 6G will be human-centric, data-, and resource-intensive. One such application is the Metaverse \cite{xu2022full}, which has been envisioned to be the future Internet. Just as we navigate the web pages of today's Internet, we will soon explore the virtual worlds of Metaverse through a head-mounted display (HMD) or navigate the augmented physical world through Augmented Reality (AR) glasses. The Metaverse is formed via the synchronization of the virtual and physical worlds, and the result is that one's actions in the virtual and physical domains will be inextricably linked. Driven by Artificial Intelligence (AI), edge intelligence, virtual and augmented reality, as well as blockchain technology \cite{lim2022realizing,du2022attention,jeon2022blockchain},  the user-centric QoE metrics required for the successful implementation of Metaverse calls for a rethink of the classical information theory (CIT) driven communication networks, because the massive data from new applications increases significantly the processing latency of conventional communication networks~\cite{du2022rethinking}. Specifically, the following difficulties in 6G networks should be addressed:
\begin{enumerate}
    \item[{\textbf{D1)}}] The emergence of new services, e.g., Metaverse, requires
6G network to support the wireless transmission of massive volumes of data.
    \item[{\textbf{D2)}}] 6G applications with a massive number of nodes, e.g., collaborative robots and hyper-intelligent IoT, require fast system responses and reliable, efficient information interaction.
    \item[{\textbf{D3)}}] {More network resources are consumed for real-time updating of information and analysis of user data to ensure a better service experience.}
\end{enumerate}

In response, a novel paradigm known as semantic communication (SemCom) is inspired as a brand new technology in 6G to breakout the ``Shannon's trap", which identifies and utilizes the \textit{meaning} of messages during Internet communication. In contrast to conventional data-oriented communication networks, the capacity of which is improved at the cost of system complexity, SemCom enables all communication participants to lighten the network burden via transmitting the most relevant information for the receivers or the goal of communication task after the pre-processing of the data based on the advanced AI technology~\cite{lan2021semantic,shi2021semantic,kountouris2021semantics}. The development of SemCom and the advancement of 6G are mutually reinforcing, so as to bring solutions to the three difficulties mentioned above. On the one hand, the availability as well as connectivity of distributed computation and ubiquitous AI networks in 6G will allow SemCom to be feasibly deployed at scale~\cite{strinati20216g}. On the other hand, SemCom overcomes traditional communication constraints and will enable unprecedented improvements in network performance. Thus, with the successful training and development of SemCom, the visions of 6G, e.g., lower latency than 5G and enhanced reliability, can be fully realized. Specifically, the SemCom has the following abilities to address {\textbf{D1)}}, {\textbf{D3)}}, and {\textbf{D3)}}, respectively.
\begin{enumerate}
    \item[{\textbf{A1)}}] Reduce the wireless data transmission burden of 6G network.
    \item[{\textbf{A2)}}] Enhance efficiency of 6G network control and management.
    \item[{\textbf{A3)}}] {Use the semantics of information to design effective network resource allocation schemes.}
\end{enumerate}

However, while the mutually reinforcing convergence properties in 6G and SemCom have attracted the attention of the academic community, there is not yet a comprehensive survey paper that provides a complete overview of the developments, challenges, and future trends for the SemCom-enabled 6G and Beyond networks. As SemCom is a relatively nascent topic, our survey aims to serve as a useful and insightful guide for future studies to researchers and practitioners alike that look to incorporate SemCom concepts into future communication architectures.
\subsection{Comparisons and key contributions}
 \begin{table*}
\small
 \centering
 \caption{{A comparison of contribution between relevant surveys and our survey.}}
 \begin{tabular}{|c|m{0.08cm}|m{7.5cm}|m{7.5cm}|}
  \hline
  \multicolumn{2}{|c|}{} & \bfseries Key contributions  & \bfseries Main limitations \\
  \hline
\cite{strinati20216g} & \raisebox{-10\normalbaselineskip}[0pt][0pt]{\rotatebox{90}{{{Comprehensive survey}}}}  & {{\quad}} \newline \vspace{ -0.6cm} \newline \tabitem Highlight 6G use cases, services and related KPIs \newline \tabitem Motivate the need of SemCom and suggest an efficient cross-layer design architecture \newline
 \tabitem Motivate a paradigm shift  towards semantic and goal-oriented communications \newline \tabitem Highlight the importance of learning-based approach in SemCom & { {\quad}} \newline \vspace{ -0.6cm} \newline \tabitem Fail to provide a roadmap for the  system design in a cross-layer architecture \newline \tabitem Only provide a highly abstract theoretical analytical model for the encoding and decoding in the semantic and goal-oriented communication system, without any practical approaches for specific applications\\ \cline{1-1} \cline{3-4}
 \cite{lan2021semantic} & & {{\quad}} \newline \vspace{ -0.6cm} \newline \tabitem Define semantic and effectiveness encoding according to the destination type \newline \tabitem Introduce two SemCom architectures: layer-coupling approach and SplitNet approach \newline \tabitem Discuss the potential role of KG technique in SemCom & {{\quad}} \newline \vspace{ -0.6cm} \newline \tabitem  Only a few encoding methods for natural language and model/gradient compression are elaborated closely \newline \tabitem The communication part of SemCom, such as transmission and decoding processes, is rarely covered\\ \cline{1-1} \cline{3-4}
 \cite{qin2021semantic}& & {{\quad}} \newline \vspace{ -0.6cm} \newline \tabitem Compare the conventional and semantic communication systems and theories \newline \tabitem Presents SemCom system components, frameworks, and performance metrics \newline \tabitem Discuss recent advancements on DL-enabled SemCom  systems for transmitting multimodal data & {{\quad}} \newline \vspace{ -0.6cm} \newline \tabitem  Goal-oriented communication studies are not included \newline \tabitem Only DL-based SemCom methods are reviewed, and there is no technical discussion about the selection and design of DL models \newline \tabitem Simply introduce error-based metrics, without discussing their usage, as well as their advantages and disadvantages\\ \hline \hline
 \multicolumn{2}{|c|}{} &  \multicolumn{2}{m{15cm}|}{\bfseries Key contributions} \\
  \hline
 \cite{shi2021semantic}  & \raisebox{-5.3\normalbaselineskip}[0pt][0pt]{\rotatebox{90}{{{Short brief}}}}  & \multicolumn{2}{m{15cm}|}{{\quad} \newline \vspace{ -0.6cm} \newline \tabitem Review classical SemCom frameworks \newline \tabitem Propose an architecture based on federated edge intelligence for supporting semantic-aware networking } \\ \cline{1-1} \cline{3-4}
\cite{kountouris2021semantics} &   &\multicolumn{2}{m{15cm}|} {{\quad} \newline \vspace{ -0.6cm} \newline \tabitem Apply SemCom to a communication scenario where the destination is tasked with real-time source reconstruction for the purpose of remote actuation} \\ \cline{1-1} \cline{3-4}
 \cite{luo2022semantic}&  & \multicolumn{2}{m{15cm}|} {{\quad} \newline \vspace{ -0.6cm} \newline \tabitem An overview of the latest deep DL-based end-to-end SemCom is given and the open issues that need to be tackled are discussed explicitly} \\ \cline{1-1} \cline{3-4}
\cite{zhang2022toward} &  &  \multicolumn{2}{m{15cm}|} {{\quad} \newline \vspace{ -0.6cm} \newline \tabitem Conceive an intelligent semantic communication-empowered ubiquitous-X 6G framework \newline \tabitem Present three promising application scenarios for SemCom}\\ \hline \hline
\raisebox{-5.3\normalbaselineskip}[0pt][0pt]{\rotatebox{90}{{{Our paper}}}} & \raisebox{-8.1\normalbaselineskip}[0pt][0pt]{\rotatebox{90}{{{Comprehensive survey}}}}& \bfseries Overlapping Contributions & \bfseries Distinct Contributions\\ \cline{3-4}
& &{\quad} \newline \vspace{ -0.6cm} \newline \tabitem  Classify SemCom into \textit{three} categories, and present their system models and application scenarios \newline \tabitem Provide an overview of SemCom related theory development 
\newline \tabitem Discuss the challenges from three design dimensions of SemCom system
\newline \tabitem Based on the ubiquitous-X 6G framework proposed in~\cite{zhang2022toward}, we envision the 6G Internet with examples \newline \tabitem Identify and outline a series of directions for future research of SemCom& {\quad} \newline \vspace{ -0.6cm}  \newline \tabitem Review \textit{four} types of SE method in SemCom, and discuss their pros and cons, as well as suitable scenarios \newline \tabitem Summary the existing technical DL-based SE models, and discuss their benefits and limitations \newline \tabitem Discuss the communication-related techniques and challenges for SemCom \newline \tabitem Classify semantic metrics into three basic types, and discuss their limitations and usages \newline \tabitem Discuss the potential links between the SemCom and promising 6G applications  \\
 \hline
 \end{tabular}
 \label{Comparison}
\end{table*}
Due to the recent attention in SemCom, some review papers have emerged to address this topic.  
In \cite{strinati20216g}, the authors state the need to integrate the semantic and effectiveness levels in traditional communications, and suggest an efficient cross-layer design architecture.  Moreover, according to the level of communication achieved, they classify the communications beyond Shannon into SemCom and goal-oriented communication. However, for the encoding and decoding in semantic and goal-oriented communications, only highly abstract  analytical models are provided. Although they emphasize the importance of machine
learning in SemCom, the related technical details, such as the neural networks (NNs)  suitable for semantic information (SI) extraction of different data types, are missing.
In~\cite{lan2021semantic}, the authors introduce two SemCom architectures. One is the layer-coupling
approach, which is similar to the cross-layer design architecture proposed in~\cite{strinati20216g}. The other is the SplitNet approach, which is used in most existing DL-based SemCom studies. However, similar to~\cite{strinati20216g}, most of their work focuses on conceiving the
concepts and roadmaps, with relatively less focus on technical details. Moreover, different from~\cite{strinati20216g}, they classify SemCom into two types according to the receiver. The encoding process in the two types is called semantic encoding and effectiveness encoding. A few encoding methods for natural language and
model/gradient compression are reviewed, while the transmission and decoding processes are not discussed in detail.  In \cite{qin2021semantic}, the authors in~\cite{qin2021semantic} compare the conventional and semantic communication systems and theories, and present SemCom system components, frameworks, and
performance metrics. Then, they review recent advancements in DL-enabled SemCom systems for transmitting multimodal data. In their work, they mainly focus on semantic-oriented communication, and the studies about goal-oriented communications are not covered. Moreover, the usage, as well as the benefits and limitations of semantic metrics and the techniques of DL-based SemCom, are not discussed in detail. 

In addition, there are several short         briefs~\cite{shi2021semantic,kountouris2021semantics,luo2022semantic} providing insights from different perspectives with regards to the design of SemCom systems, e.g., FL-enabled SemCom networks~\cite{shi2021semantic}, goal-oriented SemCom systems~\cite{kountouris2021semantics}, and DL-enabled E2E semantic networks~\cite{luo2022semantic}.  However, these studies review the works from a certain perspective only and do not provide a comprehensive review of the challenges and techniques.

To this end, we aim to provide a comprehensive survey for the implementation of SemCom in 6G, by thoroughly reviewing the existing studies and discussing the 6G applications in potential SemCom-empowered network architecture. In our paper, we classify all communications that take into account the semantic or effectiveness layer into SemCom. Meanwhile, we divide SemCom into three categories: semantic-oriented communication, goal-oriented communication, and semantic-awareness communication. The first two categories belong to traditional connection-oriented communication, which follows the definitions of two class communication in~\cite{strinati20216g}. The third category of SemCom defined in our paper belongs to task-oriented communication\footnote{Task-oriented communication here refers to an emerging type of communication in 6G wherein there are multiple explicit or implicit connections between different terminals and network nodes in a proactive or reactive manner~\cite{wu2021toward}. It can be regarded as a counterpart to traditional connection-oriented communication wherein it is easy to tell an explicit pair of source and destination terminals according to the content they intend to communicate~\cite{wu2021toward}.}. 
Meanwhile, to provide a clear roadmap for SemCom implementation, we   organize the design of the communication system into three dimensions, i.e., \textit{SI extraction}, \textit{SI transmission}, and \textit{SI metrics}.  For each dimension, based on the review of the state-of-the-art methods for the traditional data type, such as text, image, and audio, we summarize the lessons learned about their applicable SemCom categories and scenarios as well as their benefits and limitations. Additionally, we  discuss the remaining challenges in each dimension, respectively. Then, we highlight the potential of SemCom in 6G applications and networks. A series of future research directions are identified. Our discussion aims to shed light on the road ahead for SemCom research.

{A more detailed comparison of our paper and existing surveys is listed in Table~\ref{Comparison}. The thorough scope of our survey is presented in  Section~\ref{scope}.}

\subsection{Scope of the survey}
\label{scope}
\begin{figure*}[t]
 \centering
 \includegraphics[scale = 0.44]{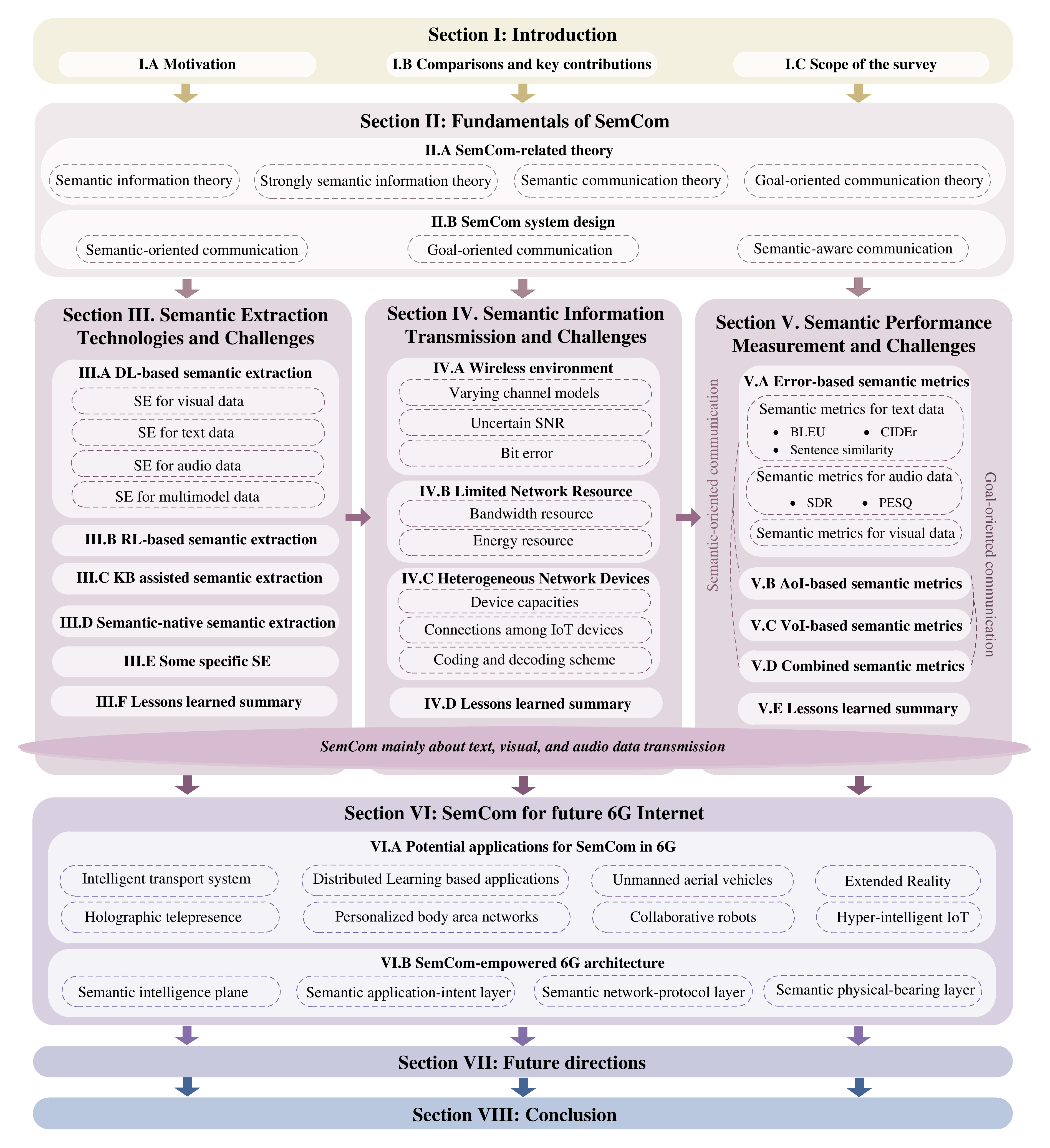}\\
 \caption{ Structure of the survey.
 }
 \label{structure}
\end{figure*}

{The scope of the survey is shown in Fig.~\ref{structure}. In Section~\ref{sec:2}, we first provide a holistic
overview of the SemCom-related theory development from semantic information theory to goal-oriented communication theory. Meanwhile, we identify the three categories of SemCom and the corresponding system design. Then, in Sections~\ref{sec:4}-\ref{semantic metric}, we discuss the state-of-the-art techniques and remaining challenges in SI extraction, SI transmission, and SI metrics, respectively. In Section~\ref{sec:4}, we first review four general semantic extraction (SE) methods in Section~\ref{DL-based SE}-Section\ref{semantic-native SE}. Among them,  DL-based SE and  RL-based SE mainly apply to semantic-oriented communications, and KB-assisted SE and semantic-native SE can be employed in goal-oriented communications. Meanwhile, in Section~\ref{sec:spse}, we also take two typical examples to illustrate the role of SE in semantic-aware communications. In Section~\ref{transmission challenges}, we focus on the transmission process. We review the techniques and challenges in terms of the wireless environment, limited network resources, and heterogeneous network devices in Section~\ref{sec:4_1}-Section~\ref{4_3}, respectively. In Section~\ref{semantic metric}, we  first discuss three basic types of semantic metrics: error-based semantic metrics, age-of-information (AoI) based semantic metrics, and value-of-information (VoI) based semantic metrics in Section~\ref{sec:error}-Section~\ref{sec:VoI}, respectively. Then, in Section~\ref{sec:Combined}, we review existing combined semantic metrics based on the three basic types. In Section~\ref{sec:application}, and Section~\ref{Sec:6G Internet}, we aim to highlight the potentials of SemCom in 6G. In Section~\ref{sec:application}, we introduce the potential applications for SemCom in 6G, and discuss the possible roles of SemCom in each application. Furthermore, in section~\ref{Sec:6G Internet},  we discuss the implementation of SemCom in 6G Internet with some specific applications based on the ubiquitous-X 6G framework suggested in~\cite{zhang2022toward}. At last, we identify and outline a series of directions for future research of SemCom in addition to the challenges ahead in Section~\ref{Section7}. Section~\ref{Section8} concludes the survey.}

\begin{table*}
\small
 \centering
 \caption{{List of common abbreviations.}}
 \begin{tabular}{|c|c||c|c||c|c|}
  \hline
 Abbr. & Description & Abbr. & Description & Abbr. & Description \\
  \hline
  \hline
SI  & Semantic information  & SE  &  Semantic extraction & ML & Machine Llearning\\
DL &  Deep learning & RL & Reinforcement learning & KB & knowledge base \\
KG & Knowledge graph& CR & Compression ratio &
 CE & Cross entropy \\
 MSE & Mean square error &
GAN & Generative adversarial net &
DNN & Deep neural network \\
CNN & Convolutional neural network &
CV & Computer vision &
NLP & Natural language processing\\

\hline
 \end{tabular}
 \label{tbl:mae}
\end{table*}

\section{Fundamentals of SemCom}\label{Section3}
\label{sec:2}
\subsection{{SemCom-related theory}}
\label{theory}
The concept of \textit{semantics} is initially introduced in the studies on semiotics~\cite{zhong2017theory}. In~\cite{morris1938foundations}, the authors define semiotics as a triple combination of \textit{syntactics}, \textit{semantics}, and \textit{pragmatics}. Syntactics focuses on the interrelation of the formal features for signs (visual and linguistic) without considering the meaning. Semantics specializes in the meaning of the signs at different levels. Pragmatics concentrates on the relationship between the utility of the signs with respect to the user in the sign system~\cite{zhong2017theory,ch1971writings}. Comparable to the triple-definition for the signs, Weaver~\cite{weaver1953recent} identifies three levels of communication as below to further characterize the syntactic, semantic, and pragmatic features of communications~\cite{{zhang2021toward}}.

\begin{figure*}[t]
 \centering
 \includegraphics[scale = 0.46]{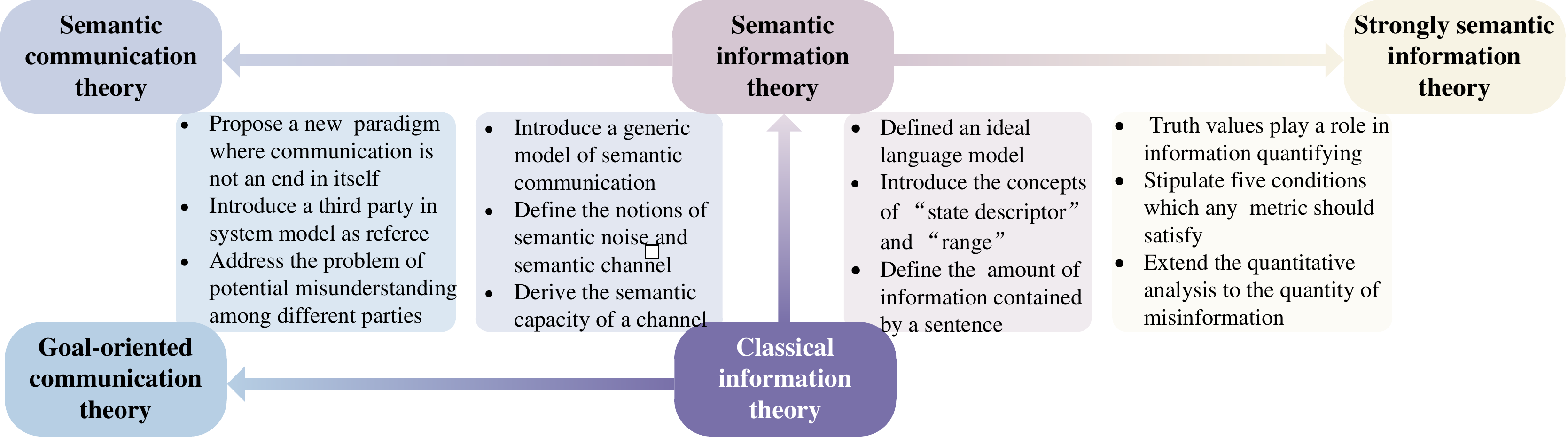}\\
 \caption{Development of classical SemCom.
 }
 \label{classicSemCom}
\end{figure*}

\begin{enumerate}[Level A]
\item How accurately can the symbols of communication be transmitted? (The technical level.)
\item How precisely do the transmitted symbols convey the desired meaning? (The semantic level.)
\item How effectively does the received meaning affect conduct in the desired way? (The effectiveness level.)
\end{enumerate}
Shannon's CIT achieves a big success in deriving a rigorous mathematical theory of communication based on probabilistic models, wherein the concept of \textit{information} is defined as \textit{what can be used to remove uncertainty} and the analysis is based on mutual information in the \textit{entropy} domain.
However, CIT focuses only on the technical level. Therefore, some researchers follow Shannon's work and make an attempt to extend it to the semantic level and effectiveness level. The development of classical SemCom has been highlighted in Fig.~\ref{classicSemCom}.
\subsubsection{{Semantic information theory}}
The authors in~\cite{bar1953semantic,elias1954outline}  make the first effort in contributing to the \textit{Theory of semantic information} (TSI) in 1950s. They propose an ideal language model which consists of $n$ nouns and $k$ adjectives. An arbitrary noun $A$ and an arbitrary adjective $a$ can be conjugated via a verb. For instance, $Aa$ means ``$A$ is $a$" or ``$A$ has property $a$". Besides, there exists five connections in the proposed model: $\sim$ (Not), $\vee$ (Or), $\wedge$ (And), $\to$ (If ... then), and $\equiv$ (If and only if). In this sense, many sentences can be generated based on the above conjunctions. Following the definition of \textit{information} in CIT, the amount of SI for a word can be defined as a function of the number of sentences that the word can imply in the considered language model, i.e., the more sentences a word can imply in the language model, the more SI the word contains~\cite{zhong2017theory}.
Moreover, to quantify the amount of information of a sentence $x$, they further propose a concept named ``state descriptor $Z$", which is defined as the conjunction of one noun and one adjective (positive or negative)~\cite{zhong2017theory}. The $range$ of the valid sentences for sentence $x$ is denoted by $R(x)$. By introducing a measurement function for a description denoted by $m(Z)$, where $0 \leqslant m\left( {{Z}} \right) \leqslant 1,\sum\nolimits {m\left( {{Z}} \right)}=1 $~\cite{zhong2017theory}, the measurement for a sentences $m(x)$ equals the sum of all $m(Z)$ within $R(x)$ and, similar to Shannon theory of
syntactic information, the amount of SI for a sentence can be calculated as its entropy, i.e., $I\left( i \right) = - {\log _2}m\left( x \right)$.

By this point, one significant limitation can be found that it completely ignores the motivation and purpose of the communication at hand. In fact, there is no radical difference between TSI and CIT~\cite{zhong2017theory}. In this sense, a false sentence that happens to say much may also be highly informative, as the SI, in such a measurement for SI amount, is not meant to imply truth~\cite{bar1953semantic}. 
\subsubsection{{Strongly semantic information theory}}
To address the above issue, the study of~\cite{floridi2004outline}  develops a \textit{Theory of Strongly Semantic Information} (TSSI). Compared to the ``weakly" TSI, the truth values play a role in TSSI. Define $f(s)$ as the degree of discrepancy of statement $s$ from the actual situation. The author in~\cite{floridi2004outline}  stipulates five conditions that any feasible and satisfactory metric should satisfy as below.
\begin{enumerate}[C 1]
\item For a true $s$ that conforms most precisely and accurately to the actual situation , $f(s) = 0$.
\item For an $s$ that is made true by every situation, i.e., a tautology, $f(s) = 1$.
\item For an $s$ that is made true in no situation, i.e., a contradiction, $f(s) = -1$.
\item For a contingently false $s$, $-1 < f(s) < 0$.
\item For a contingently true $s$ that is also made true by situations other than the actual one, $0 < f(s) < 1$.
\end{enumerate}
The details about calculations for the degree of inaccuracy and vacuity in C4 and C5 can be referred to~\cite{floridi2004outline}. Based on the degree of discrepancy, the degrees of informativeness for statement $s$ is calculated as $g\left( s \right) = 1 - f{\left( s \right)^2}$. Clearly, the more a statement deviates from 0, the more informative it is, which is more in line with the instincts of humans. However, it can only perform the quantitative analysis for the complete class of propositions in logical space and fails to provide rigorous metrics. The authors in~\cite{d2011quantifying} further improved this work based on the available works on truthlikeness, which measures the degree of being similar to the truth~\cite{niiniluoto2012truthlikeness},  via extending the quantitative analysis to the semantic concepts of the quantity of misinformation, wherein \textit{SI} is defined as true semantic content and \textit{semantic misinformation} is defined as false semantic content. The related concept of information is rooted in the SI framework using information flow~\cite{barwise1997information}.
By employing prior works on truthlikeness for classical systems~\cite{oddie1986likeness,niiniluoto1987truthlikeness}, the proposed SI quantifying method can support a broader range of use cases. However, dealing with non-classical systems is still an open issue. Nevertheless, the transformation of the measurement from uncertainty into message content in~\cite{floridi2004outline,d2011quantifying}  has made a milestone step in the development of semantic information theory.

\subsubsection{{Semantic communication theory}}
The authors in~\cite{bao2011towards} initially put forward a \textit{Theory of Semantic Communication} based on SI quantifying method~\cite{bar1953semantic}, aiming to achieve semantic-level communications. They propose a SemCom model for a basic type source that can just make factual statements in propositional logic. %
In their model, the source and destination are modeled as a 4-tuple of $<$ world model $W$, background knowledge $K$, inference $I$, and message interpreter $>$. Moreover, the Shannon entropy $H(W)$ is employed to quantify the information amount of the source, i.e., semantic entropy. Furthermore, they consider a finite set of allowed messages $X$, which can be seen as the set of available semantic codes. In this regard, semantic coding is the process of mapping from the observed values of the world model to a specific message. The strategy is a conditional probabilistic distribution $P(X|W)$, and deterministic coding is encoding the observed value $w$ into $x$ with the highest $P(x|w)$. Furthermore, the relationship between the semantic entropy and message entropy is $H\left( X \right) = H\left( W \right) + H\left( {X|W} \right) - H(W|X)$, where $H\left( {X|W} \right)$ measures semantic redundancy of coding, and $H(W|X)$ measures semantic ambiguity of the coding~\cite{bao2011towards}. The major difference from CIT is that the SI measure is based on the \textit{logical probabilities} which are determined by the background knowledge and inference, instead of \textit{statistical probabilities}. Secondly, the \textit{side information}, i.e., the destination’s prior knowledge about the source, can also be considered in the coding process to reduce the code length. More detail about encoding method based semantic entropy can be found in~\cite{basu2014preserving}.
Moreover, by denoting the received message by $y$, the semantic channel can be characterized by the distribution of $p\left( {y|x} \right)$. Furthermore, different from the CIT, the \textit{semantic channel capacity} for the discrete memoryless channel is dependent on three elements. The first one is the mutual information $I(X;Y)$ between $X$ and $Y$, which is also the channel capacity for CIT. The second one is the degree of semantic ambiguity introduced in semantic encoding with $K_s$ and $I_s$, i.e., ${H_{{K_s},{I_s}}}(W|X)$. The last one is the \textit{average} logical information of the received messages, which is determined by $K_d$ and $I_d$, i.e., $\overline{{H_{{K_d},{I_d}}}(Y)}$. If $K_s$ ($I_s$) and $K_d$ ($I_d$) do not match, excessive \textit{semantic noise} is generated. For deriving the limit of semantic channel capacity, the authors in~\cite{bao2011towards} simplify the model by assuming $K_s$ = $K_d$ and $I_s$ = $I_d$, and the upper bound is given as $C = \mathop {\sup }\limits_{P(X|W)} \left\{ {I(X;Y) - H(W|X) + \overline {H(Y)} } \right\}$.

These works can be seen as an initial but pioneering exploration of SemCom. However, it is only a model-theoretical framework, which could be unrealistic for practical communication scenarios. More relevantly, in the above work, the information amount is merely quantified based on classical Shannon entropy, which has no concerns with the essence of SI, the meaning factor, and thus is inconsistent with the original vision of SemCom.
\subsubsection{{Goal-oriented communication theory}}
In contrast to the study of~\cite{bar1953semantic}, which focuses on the extension of Shannon's CIT, the study of~\cite{juba2008universal1} focuses on the development of the classical communication system model. In both Shannon's classical system model and the SemCom model~\cite{bao2011towards}, all the communication parties need to have a common language or background. Faced with the increasing interaction among the diversified computers at that time, the study of~\cite{juba2008universal1} tried to make an attempt to make progress on the \textit{universal SemCom}, wherein the communication parties are expected to obtain a common understanding via learning each other's behavior without any prior common language. In~\cite{juba2008universal1}, the authors focus on a particular communication model between Alice and Bob, where Bob is a probabilistic polynomial time bounded interactive machine with the goal of solving a hard computational problem, and Alice has unbounded computational power and is willing to help Bob. Meanwhile, they speak different languages and expect to discuss via some binary channel. To solve this problem, the authors introduce a ``trusted third party", which knows both languages of Alice and Bob and can give finite encoding rules to translate for this discussion. The results of the theoretical analysis show that Alice can help Bob if and only if the problem that Bob wants to solve is in PSPACE~\cite{lund1992algebraic}, (i.e., the solutions to the problem are verifiable for Bob). Although the above assertions are in a restricted setting, it first highlights that communication is not an end in itself,
but rather a means to achieve some general goals among the communicating parties.

\begin{figure}[t]
 \centering
 \includegraphics[scale = 0.46]{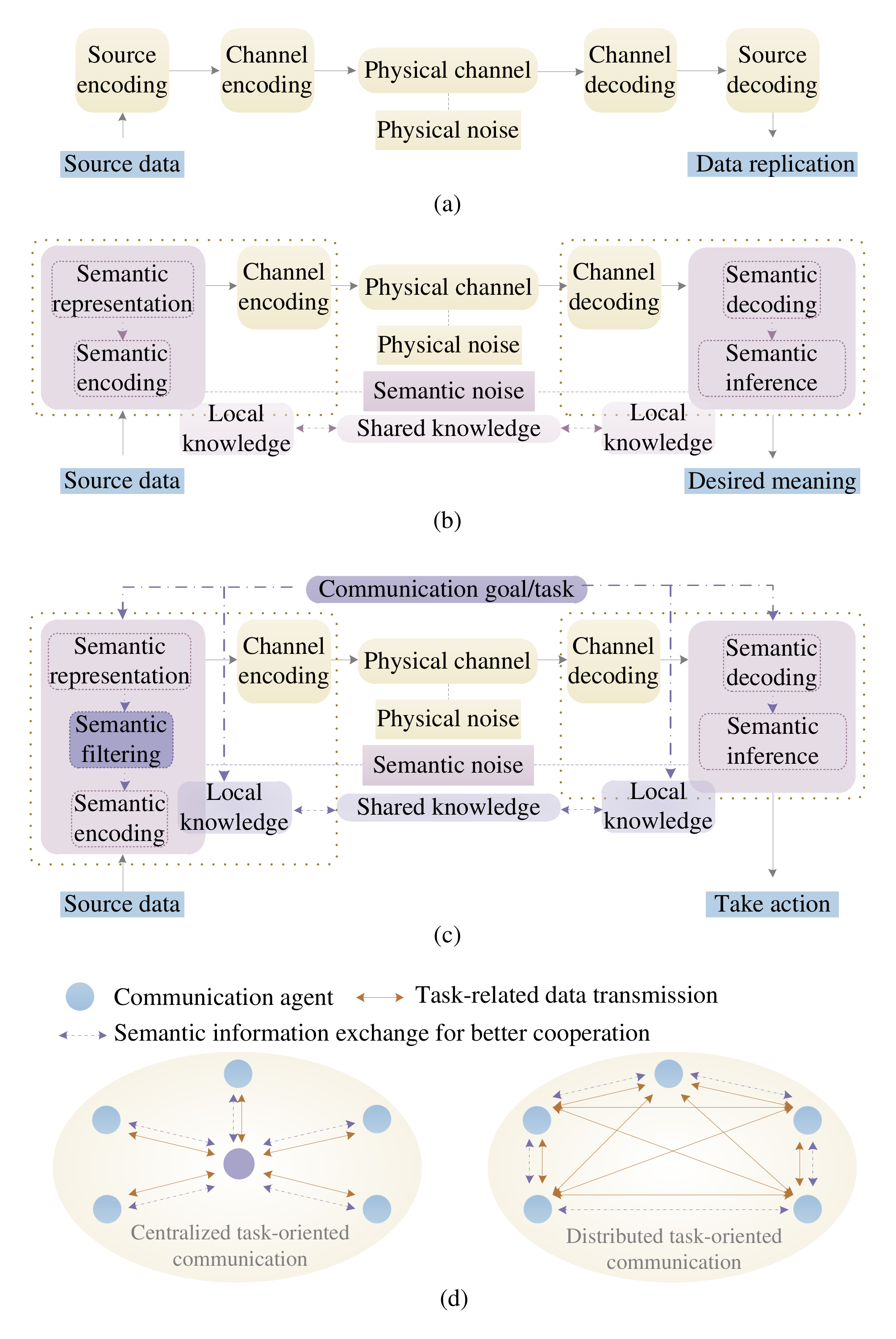}\\
 \caption{{General communication system models. (a) system model for classical communications; (b) system model for semantic-oriented communications; (c) system model for goal-oriented communications; (d) scenario model for semantic-aware communications. }
 }
 \label{systemmodel}
\end{figure}

Based on the formulated communication model in~\cite{juba2008universal1}, the authors make an extension to study the general goals of communication and first propose the conception of ``\textit{goal-oriented communication}" in~\cite{juba2008universal}. In this work, they clarify two definitions related to the \textit{goals} in communication. One is \textit{meta-goal}, which captures the intents of communicating agents, and the other is \textit{syntactic goal}, which captures effects that can be observed by an agent. The results show that the meta-goals with different syntactic versions are also achievable, i.e., two communicators do not (necessarily) share a common language under some technical conditions. Based on this, a novel architecture could be enabled for the communication among multiple agents with different protocols, wherein the trusted party called ``interpreter" played an essential role. It should be noted that in the above communication model, while the communication parties did not share a common language, they are assumed to be \textit{``sufficiently helpful"}. In~\cite{goldreich2012theory}, the authors further generalize the above work. At this level of generality,  misunderstandings might occur between the communication parties. In this work, the third party is renamed as \textit{referee}, which hypothetically
monitors the conversation between communication parties and assesses whether or not the goal has been achieved. Moreover, they identify and highlight a new concept called \textit{sensing}, which captures the communication parties' ability to simulate the referee's assessment. Based on the concept, they propose a design principle for communication systems, which could achieve polynomial overhead in the description length of the desired strategy. In~\cite{juba2011semantic}, the authors claim that the construction of universal users from such sensing functions is equivalent to the design of an \textit{on-line learning} algorithm. However, the above works mostly rely on the ``\textit{try and check}" paradigm. They can only provide guidance on the design of the protocol or strategy for the simplistic communication system, such as the conversation between a server and a printer. Although the series of works only focus on a mathematical theory of goal-oriented communication for traditional computer communication models, the system model proposed in their work has laid the foundation for modern goal-oriented communication.

Even though the studies about SemCom theory are still facing simple logic language models or the server-printer communication scenario, the exploration of SemCom systems is not constrained by this. Thanks to 
the advanced AI technologies, there has been a surge in the research on improving the design of SemCom systems to meet more practical scenarios. In this section, we first focus on three typical generic SemCom system models and the concerns in the performance evaluation that are different compared to traditional communications. Then, in the following three sections, we organize the design of the communication system into three dimensions, i.e., \textit{SI extraction}, \textit{SI transmission}, and \textit{SI metrics}. The available technologies and the remaining challenges are reviewed and discussed in Section~\ref{sec:4}, Section~\ref{transmission challenges}, and Section~\ref{semantic metric}, respectively.


\subsection{{SemCom system design}}
As mentioned at the beginning of Section~\ref{theory}, the traditional communication systems only focus on the first level (i.e., technical level) in the three levels of communications identified by Weaver and Shannon. 
 SemCom is proposed to integrate the remaining two higher levels into the design of communication systems. In our survey, we classify the existing works for SemCom into three categories according to the level and the role of the SemCom, i.e., semantic-oriented communication, goal-oriented communication, and semantic-aware communication.
 The comparison of the communication models can be shown in~Fig.~\ref{systemmodel}. Next, we describe the details of the general system models for SemCom.

\subsubsection{Semantic-oriented communications}
Different from the content-blind classical communication systems, what matters in semantic-oriented communication design is the accuracy of the semantic content of source data, instead of the average information associated with the possibilities of source data that can be emitted by a source~\cite{strinati20216g}. As such, as shown in Fig.~\ref{systemmodel}(a) and Fig.~\ref{systemmodel}(b), the main changes in the semantic-oriented communication system lie in the data processing phase before sending and after receiving. The traditional source encoding is designed to find a method to convert source data into shortcodes. Meanwhile, since the transmitted message is blind to the underlying meaning, a good source encoding method means that it can cope with more possibilities of source data, which is in line with the information quantification in CIT. However, in SemCom, the definition of ``information" needs to be modified. 

\begin{figure}[t]
 \centering
 \includegraphics[scale = 0.46]{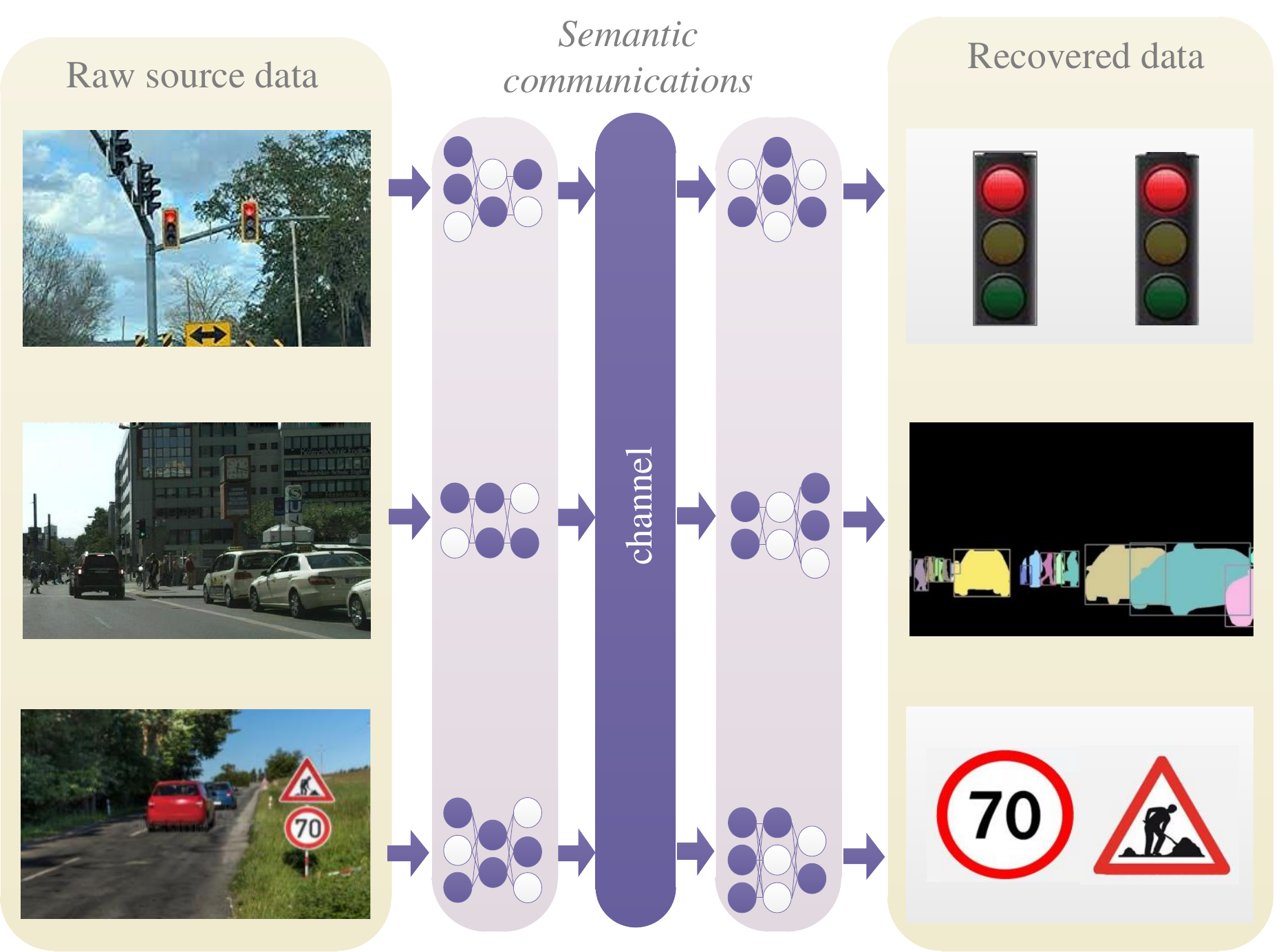}\\
 \caption{Example of semantic-oriented communication.
 }
 \label{sematic-orientedexample}
\end{figure}

As stated in~\cite{dretske1981knowledge}, information is the commodity capable of yielding knowledge, and the information that a signal carries is what we can learn from it. 
In this sense, a module of \textit{semantic representation} is introduced before encoding in SemCom, which is responsible for capturing core information embedded in source data and filtering out the unnecessary redundancy information. In many studies, the function of \textit{semantic representation} and \textit{semantic encoding} are integrated into one module called semantic encoding, which jointly plays a similar role to source the shortcodes in traditional communications. Similarly, the combined role of \textit{semantic inference} and \textit{semantic decoding} is equivalent to that of source decoding. In general SemCom scenarios, decoding is the inverse process of encoding, which is based on the AI technologies, such as Transformer and auto-encoder which are powerful with prior knowledge. Since the objective of SemCom is to enable the receiver to successfully infer SI, we regard the joint semantic encoding and decoding process as SE. {Take an example of image transmission for the transportation system in Fig.~\ref{sematic-orientedexample}. In traditional communications with the goal of image replica,  the compressed image based on a content-blind approach is expected to preserve all the details of the original image. In contrast, in semantic-oriented communication, the SE process can filter out irrelevant image details for different tasks before transmission by performing the appropriate image processing techniques, thereby relieving the network burden without compromising the system's performance.}

Moreover, as with human conversation, effective conversation requires common knowledge of each other's language and culture. In SemCom, the \textit{local knowledge} of the communication parties needs to be shared in real-time to ensure that the processes of understanding and inference can be well matched for all the source data. If the local knowledge fails to match, \textit{semantic noise} generates, which leads to semantic ambiguity, even in the absence of syntactic errors during the transmission in physical transmission.

\subsubsection{Goal-oriented communication}
\label{sec:goal}
\begin{figure}[t]
 \centering
 \includegraphics[scale = 0.46]{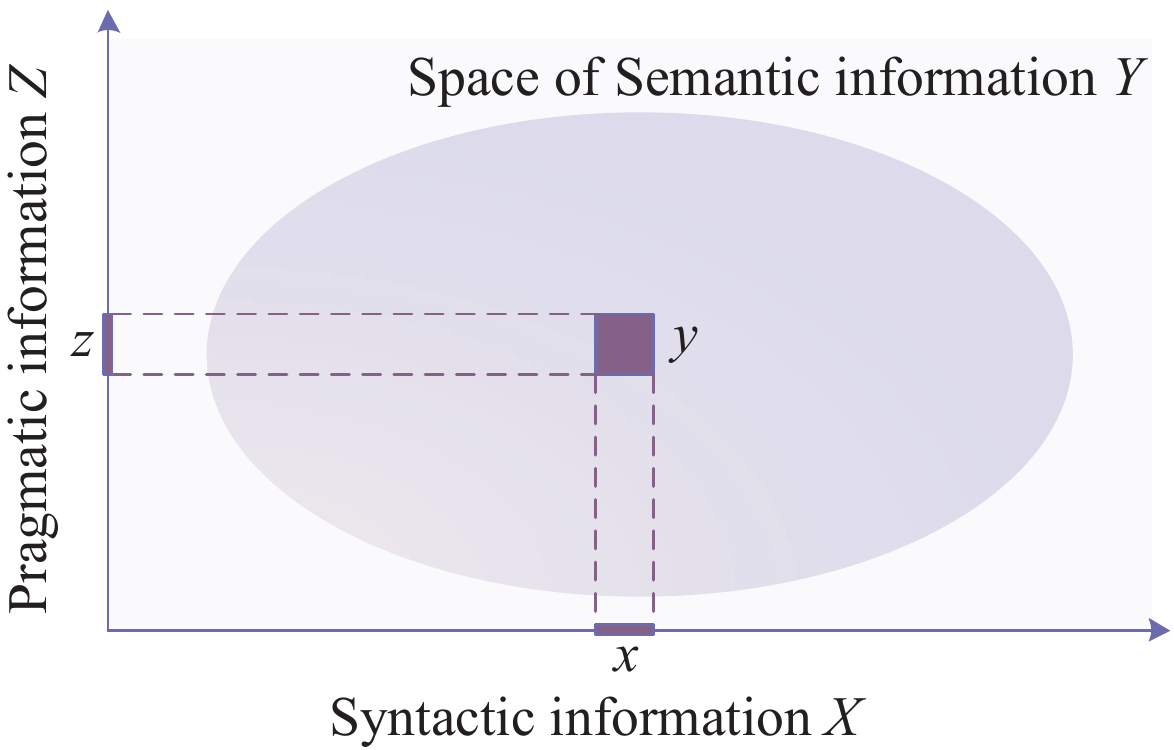}\\
 \caption{ Relationship among syntactic information, semantic information,and pragmatic information~\cite{zhong2017theory}.
 }
 \label{Relationship}
\end{figure}

Recall the triple definition for signs, i.e., syntactics, semantics, and pragmatics. In the above semantic-oriented communications, SE mainly focuses on  SI, whereas in goal-oriented communication, it is necessary to capture \textit{pragmatic information}. In~\cite{zhong2017theory}, the authors illustrate the mutual relationship among the three information types. As shown in~Fig.~\ref{Relationship}, the pragmatic information can be treated as part of all the SI that can be conveyed by syntactic information that can be treated as raw data generated by the sources in communications.  It is only relevant to a certain goal of communication. Therefore, we also refer to pragmatic information as SI for the sake of a concise presentation. 

Therefore, as shown in Fig.~\ref{systemmodel}(c), the main difference between SE in goal-oriented communication and semantic-oriented communication lies in that the goal of the communication task needs to play an important role in SE as well. Meanwhile, the communication goals also have to be involved in the local knowledge of the communication parties,  which helps further filter out the irrelevant SI in each transmission, when the communication goal changes frequently. {Take the image transmission as an example. The features (i.e., SI) of the images required for different tasks, such as classification based on different attributes, detection of different targets, or simply replication, are different. Thus,  in a transmission system with multiple tasks, perhaps only a local feature of an image needs to be transmitted each time for a certain task in goal-oriented communication. In contrast, in semantic-oriented communication, due to the non-goal-specific SE, the extracted SI should include the features for all the possible tasks, which inevitably results in information redundancy and waste of resources during transmission.}

{By comparing  Fig.~\ref{systemmodel}(b) and Fig.~\ref{systemmodel}(c), another difference is the output of the SemCom system. For semantic-oriented communication, the output of the system is the recovered meaning of the transmitted message. Then, the receiver takes the next step according to the meaning of the received message, but this process is not considered in the design of the communication system. In contrast, the output of the goal-oriented communication system is a direct action to be performed.
Recall the semantic-oriented communication example of image transmission in the transportation scenario as shown in Fig.~\ref{sematic-orientedexample}, the results inferred by the receiver may be a combination of feature maps similar to the ones on the right of Fig.~\ref{sematic-orientedexample}. In contrast, in goal-oriented communication, the output of the inference module is the action execution instruction, such as acceleration, braking, the angle for the steering wheel, and flashing headlights, to respond to pedestrians, roadblocks, and traffic signal status changes. In summary, goal-oriented communication focuses on the effective level and aims to accomplish the task in the desired way given limited network resources,  rather than the SI accuracy focused on the semantic level in semantic-oriented communication.}

Moreover, similar to semantic-oriented communication, the local knowledge and communication goal of all the communication parties need to be maintained to be consistent, otherwise, the resulting semantic noise 
may cause the task to fail.

\subsubsection{{Semantic-aware communication}}
\label{semantic-aware}
{As shown in Fig.~\ref{systemmodel}(b) and Fig.~\ref{systemmodel}(c), both semantic-oriented and goal-oriented communications establish a connection between two specific agents. They belong to the traditional connection-oriented communication, wherein it is easy to tell an explicit pair of source and destination agents according to the content that they intend to communicate~\cite{wu2021toward}. In contrast, semantic-aware communication in this survey refers to the SemCom that plays a role in task-oriented communications, such as automatic driving and unmanned aerial vehicle swarm. }

{In task-oriented communication, multiple agents cooperate to accomplish a task in a centralized or distributed way as shown in Fig.~\ref{systemmodel}(d). The semantic-aware communication in the task establishes multiple explicit or implicit connections among different terminals in a proactive or reactive manner to enhance the knowledge among agents. In other words, semantic-aware communications can be treated as a kind of ``overhead" for the task for better collaboration to facilitate the completion of tasks. In semantic-aware communication, the SI here is obtained by analyzing the agent behavior and the current environment in performing the task, instead of extracting it from a  data source. For instance, in autonomous driving, the SI can represent  the risk of a collision between two vehicles, which is determined jointly by the vehicle location and kinematic information, traffic density, road conditions, traffic light deployment, etc. Moreover, the SI can also be a description of the views captured by a series of successive cars when they pass the exit of a walled subdivision. After aggregating the SI, a semantic representation of a continuous view over the subdivision exit can be obtained, which can facilitate exit monitoring and activity tracking along the exit lane within the subdivision~\cite{bista2018semantic}.
In semantic-aware communication, there may be no explicit transceivers or a complete pairwise semantic encoding and decoding processes. Therefore, there is not yet a general system model for semantic-aware communication.}

\section{Semantic extraction technologies and challenges}
\label{sec:4}
As discussed in Section~\ref{introduction}, the achieved transmission rates in  conventional communication systems are approaching the Shannon limit and the remaining available spectrum resources are becoming increasingly scarce~\cite{xie2021deep}~\cite{sana2021learning}.
 The key to SemCom being pushed forward to address the bandwidth bottlenecks lies in that it converts the \textit{transmission-before-understanding} communication paradigm to the \textit{understanding-before-transmission} communication paradigm. In this way, SE can be integrated into the communication model to achieve SemCom~\cite{zhang2021toward,lan2021semantic}, which allows only the information of 
 interest to the receiver for transmission, rather than raw data, thereby alleviating bandwidth pressure and enhancing privacy preservation by reducing and hiding the redundant data to be exchanged. 
 
In fact, SE is not a brand-new topic, but it has been evolving~\cite{merono2015semantic},~\cite{rachana2021literature}. Some comparable works have been explored in other research fields, such as \textit{semantic segmentation} in computer vision, which is used to cluster parts of images together which belong to the same
object class~\cite{thoma2016survey}, \textit{semantic computing}, which addresses the derivation and matching of the semantics of computational content and that of user intentions to retrieve, use, manipulate, or even create the content~\cite{sheu2010semantic}, and \textit{semantic web}, which can be considered as a knowledge graph formed by combining the linked data with intelligent content and is widely used in recommendation systems to facilitate intelligent and integrated user experience~\cite{hitzler2021review},~\cite{chen2021recommendation}. 
Compared to~\cite{thoma2016survey,sheu2010semantic,hitzler2021review,chen2021recommendation}, \textit{SemCom} is another key field for SE. In this field, 
all the communication parties have to be highly aligned in semantic representation and interpretation, which imposes challenges for SE.
In addition, information in the 6G communication system features a strong time-sensitive nature and is highly demanding in terms of accuracy~\cite{she2020deep}, which is also a stringent requirement not found in other fields.

{Hence, in the following, we merely concentrate on SE methods in SemCom. In Section~\ref{DL-based SE} and Section~\ref{RL-based SE}, we introduce two general SE  methods for semantic-oriented communications. Then, in Section~\ref{sec:KB-assistant} and Section~\ref{semantic-native SE}, two general SE methods for goal-oriented communications are presented, where the communication goals are integrated into the SE. At last, we take two examples to illustrate the SE in semantic-aware communication in Section~\ref{sec:spse}.}

\subsection{DL-based SE}
\label{DL-based SE}
Following the success of DL in the individual block optimization in the physical layer~\cite{qin2019deep, o2017introduction, ye2017power, chun2019deep, guo2021canet}, DL-based end-to-end communication systems have emerged as another potential direction to outperform the conventional communication structure in block error rate (BLER) and BER performance~\cite{park2020end,ye2021deep,dorner2017deep}. Inspired by this, some researchers further introduce the DL-enabled method in fields of computer vision (CV)~\cite{mnih2014recurrent,szegedy2015going,wang2017residual}, (natural language processing) NLP~\cite{bahdanau2014neural,luong2015effective,vaswani2017attention} and speech processing~\cite{purwins2019deep,ogunfunmi2019primer,haeb2019speech} into end-to-end communication system as SE approaches, which pioneered the \textit{modern} SemCom study~\cite{xie2020deep}.

\subsubsection{SE for visual data }
Due to the large volume of image data, the authors in~\cite{lee2019deep} first focus on an image transmission scenario, where an IoT device transmits images to the server to perform recognition. The IoT device maintains a direct,
point-to-point wireless link to the server. Different from the conventional communication models where multiple modules are cascaded, they propose a DL-constructed \textit{joint transmission-recognition scheme} (JTRS) with the design metric of \textit{recognition accuracy}. In the designed scheme, the ResNet architecture~\cite{he2016deep} is employed due to its favorable performance and few parameters. In order to complete feature extraction before transmission, the deep neural network (DNN) of ResNet is split into two parts. The first few layers function as a feature extractor (i.e., semantic extractor) at the transmitter, and the rest of the layers serve as a recognizer at the receiver.
Furthermore, to achieve the adaptive semantic extraction in noisy channels, the joint semantic-channel coding (JSCC) is implemented by using the DNN as channel encoders and decoders, which is discussed in detail in Section~IV. 

\begin{table}
\small
 \centering
 \caption{{Performance Comparison for JTRS~\cite{lee2019deep}(CR = 0.04).}}
 \begin{tabular}{|c|c|c| c | c |c|}
  \hline
  & JTRS & JPEG & CS-DR & CS-R  & \\
  \hline
  \hline
 runtime &  7e-5~s & 7e-2~s & 1e-3~s & 4e-1~s &\\  \cline{1-5}
 accuracy & \multicolumn{1}{>{\columncolor{mygray3}}c}{0.9} & 0.5 & \multicolumn{1}{>{\columncolor{mygray}}c}{0.1} & \multicolumn{1}{>{\columncolor{mygray}}c}{0.14}&\raisebox{-1\normalbaselineskip}[0pt][0pt]{\rotatebox{90}{{{analog}}}}\\\cline{1-5}
 ${\text{SNR}}_{\min}$&  -4~dB & 5~dB & 15~dB & 15~dB &\\
\hline
\hline
 runtime &  1e-2~s & 9e-3~s & 1e-3~s & 4e-1~s &\\  \cline{1-5}
 accuracy & \multicolumn{1}{>{\columncolor{mygray2}}c}{0.9} & 0.7 & 0.47 & \multicolumn{1}{>{\columncolor{mygray}}c}{0.14} & \raisebox{-1\normalbaselineskip}[0pt][0pt]{\rotatebox{90}{{{digital}}}} \\\cline{1-5}
 ${\left( {{{{E_b}} \mathord{\left/
 {\vphantom {{{E_b}} {{N_0}}}} \right.
 \kern-\nulldelimiterspace} {{N_0}}}} \right)_{\min }}$&  0~dB & 5~dB & 5~dB & 4~dB &\\
\hline
 \end{tabular}
 \label{tbl:performance}
\end{table}

To demonstrate the effectiveness of the DNN-constructed JTRS, the authors in~\cite{lee2019deep} compare the scheme with three other cascaded compression-and-recognition schemes given the similar compression ratio (CR) of 0.04. 
The three baseline schemes are JPEG-compressed scheme (JPEG), compressed sensing with direct recognition (CS-DR), and compressed sensing with reconstruction (CS-R). {Table~\ref{tbl:performance} shows the complexity (in terms of runtime), the highest recognition accuracy, and the corresponding channel condition thresholds of the four schemes for digital and analog transmissions, respectively\footnote{Here the analog transmission means that the data values are directly used to modulate the signal without going through the steps of quantization. The digital transmission means the data values need to be quantized and converted to bits before modulation and transmission.}. 
From Table~\ref{tbl:performance}, we can see that, due to the excessively low CR, the schemes of CS-DR and CS-R have almost lost the capability to do recognition even under favorable channel conditions. Among the three baselines, only the scheme of JPEG can achieve the accuracy of more than 50\% in digital transmission with LDPC codes. In contrast, the proposed JTRS can achieve the accuracy of up to 0.9 under poor channel conditions in both analog and digital transmission. Surprisingly, JTRS performs better in analog transmission than in digital transmission. More encouragingly, due to the lack of quantification and bit conversion process before modulation and transmission, the runtime of JTRS in the analog transmission is far lower than those in other methods, which means that the DL-based SemCom has an inherent advantage in low--latency communications. }

However, this scheme is only designed to operate under a specific SNR level. When channel conditions change, the SE model needs to be retrained or refined, which introduces considerable additional overhead. In the traditional communication system, the general source encoders and channel encoders can achieve an adaptive CR and the channel coding rate according to the SNR to achieve optimal performance given the limited bandwidth. To fill this gap between  SemCom and traditional communication, the authors in~\cite{xu2021wireless} consider a point-to-point image transmission system with SNR feedback. They
integrated the \textit{attention mechanism}~\cite{mnih2014recurrent, wang2017residual} that is widely used in CV into SE. The attention mechanism adopts an additional neural network to rigidly select certain features or assign different weights to different features in the original neural network. 
In their proposed design, the joint semantic-channel encoding is performed by a single network, which consists of two modules: a feature extraction (FE) module and an attention feature (AF) module. The FE module is used to learn features from the input images. The AF module then takes the output of the FE module and SNR as its input and produces a sequence of scaling parameters. The product of the outputs of the feature learning module and the attention feature module can be seen as a filtered version of the feature learning module output. The decoder is similarly designed. 
In the simulation, the authors compare the performance of the attention-based DL JSCC scheme trained under the uniform distribution of SNR from 0~dB and 20~dB and five basic DL based JSCC schemes trained at the SNR of 1~dB, 4~dB, 7~dB, 13~dB, and 19~dB, respectively. From the results, the peak signal-to-noise ratio (PSNR) curve achieved by the proposed scheme can be seen as the upper envelope of the other PSNR curves of the baseline scheme trained at different SNRs, which demonstrate the higher robustness, versatility, and adaptability to the wide range of SNR of the attention-based approach. 

The above two works focus on image recognition and image recovery, respectively. The authors in~\cite{hu2022robust} focus on the applications of image classification against semantic noise. Exploiting the heavy spatial redundancy of image data, they propose a resource-efficient SE model with an asymmetric encoder-decoder architecture. The encoder employs a masked autoencoder (MAE) with vision Transformer (ViT) architecture~\cite{he2021masked}. The MAE can reconstruct an image from partial observations. Specifically, in the proposed architecture, a portion of the original image is masked and disregarded first. Then, the unmasked portion is embedded with the information about their position in the original image, which then goes to Transformer blocks to extract the image features~\cite{he2021masked}. As the encoder only needs to process the portion of unmasked patches, which significantly reduces the memory consumption. 
On the contrary, the input of the decoder is the full set of tokens consisting of encoded features of unmasked patches and the masked tokens, which is a shared and learned vector suggesting the presence of the patches that are to be predicted~\cite{hu2022robust}. Moreover, different from the above end-to-end SE model, the decoder can be designed independently of the encoder, as the decoder is only used to perform the image reconstruction task, which allows for greater flexibility in the system design. 

Meanwhile, MAE can also defend against malicious attackers, i.e., by adding semantic noise to images. Since MAE randomly masks partial patches of the image during the encoding process, the impact of semantic noise added in the patches of the image can be eliminated to some extent~\cite{hu2022robust}.
In addition, to further strengthen the resistance to malicious attacks, the authors in~\cite{hu2022robust} propose a codebook for encoded feature representation, which consists of multiple discrete basis vectors trained together with the encoder and decoder parameters. Based on the well-trained codebook, the continuous encoded features output by the encoding neural networks are mapped into the discrete indices of basis vectors by a nearest neighbour search~\cite{hu2022robust}. Hence, the distortion caused by semantic noise can be corrected with a high probability during discrete representation at the transmitter, which greatly enhances the robustness of the communication. In the training process,  adversarial learning is employed where the semantic noise is generated by fast gradient sign method. 

\begin{table}
\small
 \centering
 \caption{{Comparison of the number of symbols for an encoded image~\cite{hu2022robust}.}}
 \begin{tabular}{|c|c|c|c|}
  \hline
  & JPEG+LDPC & MAE & Ratio  \\
  \hline
  \hline
 total symbols &  20432 & 196 & 0.95\%  \\  
\hline
 \end{tabular}
 \label{tbl:mae}
\end{table}

{Since MAE effectively reduces the spatial redundancy of images, the number of symbols of an image to be transmitted is 0.95\% of that of an image encoded by the traditional scheme (JPEG+LDPC), as shown in Table~\ref{tbl:mae}. Due to the effective SE, the MAE scheme can achieve a classification accuracy of 0.6 even with an SNR of -6~dB. In contrast,  the classification accuracy of the traditional scheme (JPEG + LDPC) is close to zero with SNR ranging from -6~dB to 6~dB due to the limited bandwidth. Only when the SNR reaches 14~dB does the traditional scheme achieve a classification accuracy of 0.6. 
However, the size of the JEPG images used for training and testing is relatively small (5108 bytes). Considering the complexity of the training process, the feasibility and the effectiveness of
this scheme have to be further verified. 
Nevertheless, this result fully demonstrates the importance of SemCom in improving communication performance by reducing the data transmission burden with effective SE.}

\subsubsection{SE for text data} 
Inspired by the success of DL in NLP such as machine translation, the authors in~\cite{farsad2018deep} pioneer the implementation of SemCom for text transmission. They consider a simple system model, where a transmitter sends sentences to a receiver using the limited number of bits over an erasure channel. In the proposed scheme, the words are first represented by an embedding vector using GloVe~\cite{pennington2014glove}, which is the pre-trained lookup table available for extracting SI. Then, motivated by the success of the sequence-to-sequence learning framework in machine translation~\cite{bahdanau2014neural,wu2016google}, the long short-term memory (LSTM)-based encoder and decoder are employed, wherein the embedding vector of the previously estimated word is taken as the input for the next step and the beam search algorithm is used to find the most likely sequences of words~\cite{wu2016google,graves2012sequence}. In this sense, the SI can be embedded into the sentence recovery. Compared with Gzip and Huffman, LSTM-based SE achieves the lowest word error rate for a given coding length with a high bit-drop rate. Meanwhile, under a certain bit-drop rate, due to the effectiveness of extracted information, the superiority of LSTM-based SE  becomes more remarkable as the length of the sentence increases.
However, the word representation models like Glove or Word2Vec~\cite{mikolov2013efficient} only capture the relationship among words and fail to describe syntax information~\cite{xie2021deep}. Therefore, the proposed method can only describe the probability of a certain word coming after another in a sentence, which makes it hard to deal with complex sentences. 

In the face of the above challenge, a newly proposed architecture called Transformer has attracted a great deal of attention, as it can extract both the SI and syntax from the whole sentences effectively~\cite{xie2021deep}. 
The Transformer network is combined with \textit{multi-head attention} mechanisms, which allows it to extract multiple characteristics of input sentences in parallel~\cite{sana2021learning}. Therefore, compared with the recurrent neural network (RNN)-based architectures, such as LSTM, the Transformer network achieves lower computational complexity and more parallelizable computations while learning long-range dependencies in input sentences~\cite{sana2021learning}~\cite{vaswani2017attention}. Hence, in the recent works~\cite{xie2020lite,xie2021deep}, the Transformer networks replace the RNN networks, and the channel models are extended to additive white Gaussian noise (AWGN) channels and fading channels. In their work, more expert semantic metrics, such as BLEU and sentence similarity, (which are introduced in Section~\ref{semantic metric}) are employed to measure the SemCom performance.  The superiority of the scheme in terms of semantic metrics under the low SNR region demonstrates the effectiveness of Transformer in SE for text data. 

However, the standard Transformer has a fixed attention structure, {which makes it treat all inputs indifferently and limits its adaptability in the learning process.} In fact, in a sentence processing system, some words or phrases are more likely to cause semantic ambiguity due to polysemy or noise interference. With this in mind, the authors in~\cite{zhou2021semantic} propose a
flexible SE approach based on Universal Transformer (UT)~\cite{dehghani2018universal}, by introducing an adaptive circulation mechanism in the Transformer to break the original fixed structure. Compared to the standard Transformer, UT is integrated with the Adaptive Computation Time (ACT) model~\cite{graves2016adaptive}. The ACT model dynamically adjusts the number of
computational steps required to process each input symbol in the standard RNN, according to a  halting probability predicted 
at each step. Such a dynamic halting mechanism allows UT-based SE to give loop play to its own circulation mechanism for each input symbol (i.e., per-symbol self-attentive RNN) and flexibly respond to different SI and varying physical channels through different cycles. 

In~\cite{zhou2021semantic}, the authors compare the performance in terms of BLEU of both the SemCom schemes with UT-based SE approach and classical Transformer-based SE approach, with the traditional source coding and channel coding cascaded schemes with fixed-length coding (5-bit) for source coding and Turbo coding or Reed-Solomon coding for channel coding. For both traditional schemes, the BLEU score keeps staying pretty low over a wide range of SNR, and is only significantly improved when the SNR is increased above 15~dB. In contrast, both SemCom schemes achieve remarkably higher BLEU scores under a variety of changing channel conditions. Specifically, since the adaptive circulation mechanism facilitates a more accurate capture of SI,
the UT-based algorithm consistently scores higher than the Transformer-based algorithm over the full SNR region.

\subsubsection{SE for audio data}
With the success of E2E SemCom focusing on images and text, the authors in~\cite{tong2021federated} further investigate the SemCom for the audio signal. In~\cite{tong2021federated}, the authors design an audio SE based on a DL-based NLP model named Wav2Vec~\cite{schneider2019wav2vec}. The semantic encoder consists of two cascaded convolutional neural networks (CNNs), called FE and feature aggregator (FA), respectively. The FE is responsible for extracting the rough audio features from the raw audio vector, and the FA is responsible for combining the rough audio features into a higher-level latent variable that contains semantic relations among contextual audio features~\cite{schneider2019wav2vec}. Accordingly, the semantic decoder is also based on Wav2Vec architecture, which consists of two symmetrical CNNs to the encoder called feature decomposer (FD) audio generator (AG), respectively. { This scheme can reduce the MSE to below 2e-4,  when the SNR is above 0~dB. However, due to the simplicity of SE model, the extracted SI is somewhat limited. As the SNR  increases, there is no obvious downward trend in MSE.}  Moreover, similar to the LSTM model employed in image SE, the SE model is trained under AWGN channels with a fixed channel coefficient, which makes it challenging to guarantee decent performance under more complicated channel conditions. 

At the same time, similar to the evolution of text semantic encoder, the authors in~\cite{weng2021semantic,weng2021semantic2} further integrate the $attention$ $mechanism$ named SE-ResNet into SE, and the encoder and decoder are constructed by one or multiple sequentially connected SE-ResNet modules. The term ``SE" in ``SE-ResNet" represents a squeeze-and-excitation network, which is treated as an independent unit and employed to assign high values to the weights corresponding to the essential information during the training phase. In particular, the squeeze operation is to aggregates the 2D spatial dimension of each input feature, and the excitation operation is to learn and output the attention factor of each feature by capturing the inter-dependencies. Meanwhile, the residual network is adopted to alleviate the gradient vanishing issue due to the network depth. With the simulation, it can be shown that the proposed SE approach shows better performance under various fading channels and SNRs compared to the CNN-based methods. However, similar to the CNN-based SE model, the SE-ResNet-based one still fails to implement a dynamic SE that adapts to the channel condition varies.

Later, the authors in~\cite{weng2021Recognition} further focus on speech recognition tasks for the English language. In~\cite{weng2021Recognition}, the original speech sample sequence is converted into a spectrum before feeding into the transmitter. Moreover, they introduce a transcription of a single speech sample sequence, where each token represents a character in the alphabet or a word boundary. Based on the spectrum and transcription, they design the encoder and decoder. The semantic encoder is constructed by the CNN and the gated recurrent unit-based bidirectional RNN (BiRNN)~\cite{schuster1997bidirectional} modules. The CNN is utilized for data compression and the BiRNN is utilized to extract the text-related semantic features before transmission. The channel encoding and decoding are performed by the dense layer, and the semantic decoding is responsible for decoding the recovered text-related semantic features into the text transcriptions. The text-related semantic features are referred to as a probability matrix with the probability that each token corresponds to each letter.
Considering the limited number of letters in the English alphabet, the semantic decoder is designed as a greedy decoder, wherein the maximum probability in all the steps is indexed and the corresponding token is employed to construct the final transcription.  {With the simulation, the SemCom-based speech recognition achieves a much lower character-error-rate and word-error-rate under a low SNR region, compared to the traditional
communication systems. In traditional communication systems, the speech signals are transmitted directly and then transcribed into text at the receiver with automatic speech recognition (ASR) module~\cite{amodei2016deep} or the speech signals are first transcribed into text at the transmitter by ASR module and then transmitted. However, as SNR increases, the superiority of the algorithm becomes diminishing due to an unavoidable error floor generated by DL~\cite{jiang2019deepturbo}.}
\begin{figure}[t]
 \centering
 \includegraphics[scale = 0.46]{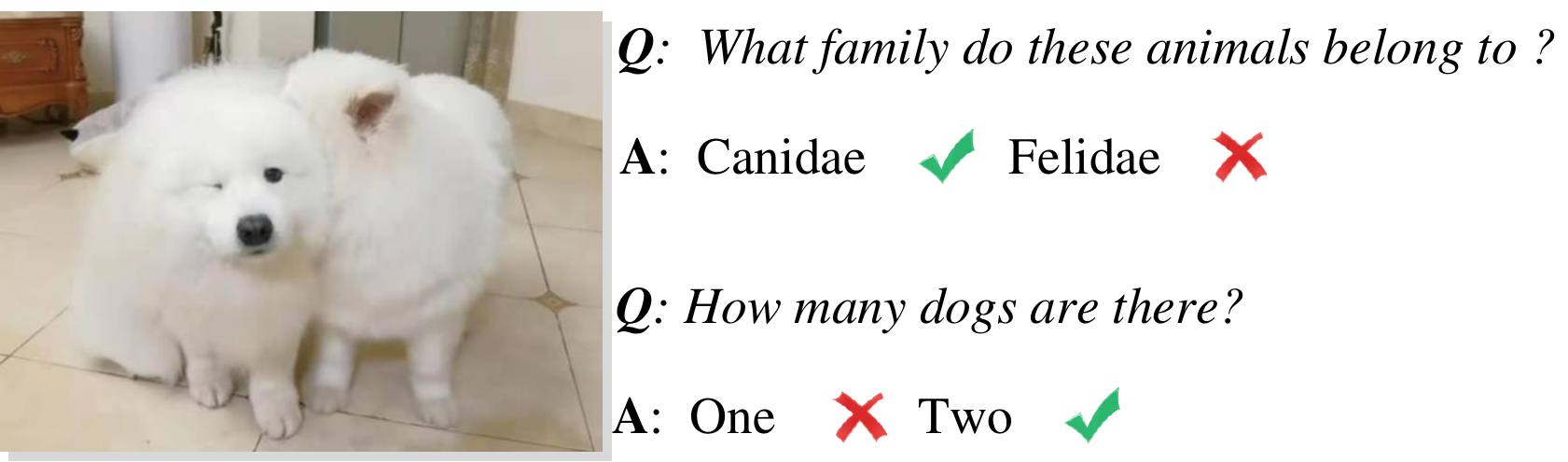}
 \caption{An example of a simple VQA task.}
 \label{VQA}
\end{figure}
\begin{table*}
\small
 \centering
 \caption{{Summary of  DL-based SE for typical data types.}}
 \begin{tabular}{|m{0.08cm}|m{0.08cm}| m{5cm} | m{5cm} |m{5cm}|}
  \hline
  \multicolumn{2}{|c|}{} & \qquad \quad \bfseries NN Architecture  &\qquad \qquad \qquad \bfseries Benefits & \qquad \qquad\quad \bfseries Limitations  \\
  \hline
  \hline
&  \raisebox{-2.5\normalbaselineskip}[0pt][0pt]{\rotatebox{90}{{{ recognition}}}} & \makecell { {\it ResNet-CIFAR 10}~\cite{lee2019deep} \\ \\$(\underbrace {{\text{feature extractor}}}_{{\text{encoder}}}{\text{+}}\underbrace {{\text{recognizer}}}_{{\text{decoder}}})$ }&  {The scheme greatly reduces the compression rate while ensuring recognition accuracy and dramatically reduces system complexity and processing latency. } & This scheme is applicable only to a specific SNR range and requires retraining when the channel changes, thus introducing additional overhead.\\\cline{2-5}

\raisebox{-1\normalbaselineskip}[0pt][0pt]{\rotatebox{90}{{{image}}}}& \raisebox{-2.2\normalbaselineskip}[0pt][0pt]{\rotatebox{90}{{{transmission}}}} & \makecell {{\it Attention-integrated DNN}~\cite{xu2021wireless} \\ 
$\underbrace {\underbrace {{\footnotesize \text{CNN}}}_{{\text{FE module}}} \to \underbrace {{\text{\footnotesize Attention - based DNN}}}_{{\text{AF module}}}\mathop {\hbox{\ \hbox{$\mid$}\kern -1em
\lower .5em \hbox{$\leftarrow$}}} \limits^{{\text{SNR}}}  \ldots }_{{\text{encoder(decoder)}}}$} &
By integrating SNR  into SE, the scheme can operate successfully over a wide range of SNRs with lower computational/storage complexity than that of the basic DNN-based structures.& The effectiveness of the scheme is just demonstrated in AWGN channel. The robustness and adaptability of the algorithm are still to be verified and studied under more general models.\\ \cline{2-5}
&  \raisebox{-2.2\normalbaselineskip}[0pt][0pt]{\rotatebox{90}{{{classification}}}} & \makecell{{\it MAE with ViT and codebook}~\cite{hu2022robust} \\ $(\underbrace {{\text{MAE}}  {\footnotesize \to } {\text{Vi}}{{\text{T}}^{\mathop {\footnotesize {\hbox{\ \hbox{$\mid$}\kern -1em
\lower .5em \hbox{$\leftarrow$}}}}  \limits^{{\text{\footnotesize codebook}}} }}}_{{\text{encoder}}}{\text{+}}\underbrace {{\text{Vi}}{{\text{T}}^{\mathop {\footnotesize {\hbox{\ \hbox{$\mid$}\kern -1em
\lower .5em \hbox{$\leftarrow$}}}} \limits^{{\text{\footnotesize codebook}}} }}}_{{\text{decoder}}})$ }& Based on MAE, the scheme achieves high SE efficiency by reducing the image spatial redundancy and resists largely the interference of semantic noise on classification with codebook. & The complexity of the system is high, which poses difficulties in training the SE model. For large image transmission, the feasibility and effectiveness of this scheme are yet to be verified. \\
  \hline
  \hline
\multicolumn{2}{|c|}{} & \makecell{{\it LSTM}~\cite{farsad2018deep} \\ \\ $(\underbrace {{\text{BiLSTM}}}_{{\text{encoder}}} + \underbrace {{\text{LSTM}}}_{{\text{decoder}}})$}& Compared with Gzip and Huffuman, the scheme can achieve remarkable low word error rate for large-size sentences and high bit-drop rates. &  The scheme can only capture the relationship among words and fails to describe
syntax information. This makes it hard to deal with complex sentences.\\
\cline{3-5}
\multicolumn{2}{|c|}{\raisebox{-3.8\normalbaselineskip}[0pt][0pt]{\rotatebox{90}{{{text transmission}}}}} & \makecell{{\it  Transformer}~\cite{xie2021deep}\\ $(\underbrace {\mathop {{\text{Transformer}}}\limits^{\mathop {{\text{\footnotesize multi-head attention}}}\limits_  \Downarrow   } }_{{\text{encoder}}\left( {{\text{decoder}}} \right)})$ } &  The multi-head attention  module in Transformer can capture long-range dependencies in  sentences in parallel with low complexity, thus extracting accurate SI. & The attention structure in Transformer is fixed, which makes it hard to deal with 
 noise interference or polysemy, such as  ``mouse" has a different meaning in computing and biology.\\
\cline{3-5}
\multicolumn{2}{|c|}{} & \makecell{{\it Universal Transformer}~\cite{zhou2021semantic} \\ $(\underbrace {\mathop {{\text{Transformer}}}\limits^{\mathop  \Downarrow \limits^{{\text{\footnotesize multi - head attention}}} } {\text{  }}\mathop {\hbox{\ \hbox{$\mid$}\kern -1em
\lower .5em \hbox{$\leftarrow$}}} \limits^{{\text{ACT}}} }_{{\text{encoder(decoder)}}})$} & The scheme can be considered as a per-symbol
self-attentive RNN, which can capture more precise SI and flexibly respond to varying channel conditions. & Due to the loop play introduced by the adaptive circulation mechanism, computational complexity increases, which causes extra processing latency and computing resource demand. \\
\hline
\hline
& & \makecell{{\it CNN}~\cite{tong2021federated}  \\ $(\underbrace {\underbrace {{\text{CNN}}}_{{\text{FE}}} \to \underbrace {{\text{CNN}}}_{{\text{FA}}}}_{{\text{encoder}}} + \underbrace {\underbrace {{\text{CNN}}}_{{\text{FD}}} \to \underbrace {{\text{CNN}}}_{{\text{AG}}}}_{{\text{decoder}}})$} & The SE model is simple and easy to train. The scheme is remarkably competitive at low SNRs. & The basic SE model fails to extract semantically enriched information and adapt to changing channel conditions.\\
\cline{3-5}
&\raisebox{-0.5\normalbaselineskip}[0pt][0pt]{\rotatebox{90}{{{transmission}}}} &  \makecell{{\it Squeeze-and-excitation network}~\cite{weng2021semantic}
\\$(\underbrace {\mathop {{\text{SE-ResNet}}}\limits^{\mathop {{\text{\footnotesize 
 attention}}}\limits_ \Downarrow  }  \cdots }_{{\text{encoder}}} + \underbrace {\mathop {{\text{SE-ResNet}}}\limits^{\mathop {{ \text{\footnotesize attention}}}\limits_ \Downarrow  }  \cdots {\text{CNN}}}_{{\text{decoder}}})$} & Due to the introduction of the attention mechanism, SE-ResNet-based SE model can achieve higher performance in terms of PESQ and SDR at any given SNR than CNN-based one. & This scheme can only perform the training of SE model under a fixed SNR. A dynamic and flexible SE model that adapts to channel changes remains to be studied.\\
\cline{2-5}
\raisebox{2.5\normalbaselineskip}[0pt][0pt]{\rotatebox{90}{{{speech}}}}& \raisebox{-2.3\normalbaselineskip}[0pt][0pt]{\rotatebox{90}{{{recognition}}}}& \makecell{\textit{CNN \& RNN}~\cite{weng2021Recognition} \\ \\ $(\underbrace {{\text{CNN}} \to {\text{BiRNN}}}_{{\text{encoder}}} + \underbrace {{\text{greedy}}}_{{\text{decoder}}})$}& The scheme achieves a much lower character-error-rate
and word-error-rate compared to the traditional communication systems under low SNRs. & The scheme becomes sub-optimal as SNR increases, since the DL-based methods always generate an avoidable error floor. \\
\hline
\hline
\raisebox{-8.5\normalbaselineskip}[0pt][0pt]{\rotatebox{90}{{{multi-model data}}}} & \raisebox{-7.3\normalbaselineskip}[0pt][0pt]{\rotatebox{90}{{{VQA}}}}& \makecell{{\footnotesize \textit{ResNet \& LSTM \& MAC network}~\cite{xie2021task}} \\$(\begin{array}{*{20}{c}}
  {\underbrace {\operatorname{Re} {\text{sNet - 101}}}_{{\text{image encoder}}}} \\ 
  {\underbrace {{\text{Bi - LSTM}}}_{{\text{text encoder}}}} 
\end{array} + \underbrace {{\text{MAC network}}}_{{\text{decoder}}})$} & Compared to the traditional method, where the recovered image and text are input to MAC, the end-to-end scheme achieves significantly higher answer accuracy. & Since the scheme assumes perfect channel state information, it is not robust to channel changes in a real-world environment \\ \cline{3-5}
& &
\makecell{ \textit{Transformer}~\cite{xie2021task2}
\\ \\ $\begin{array}{l}
\mathop {\underbrace {{\rm{Transformer}}}_{{\rm{image encoder}}}{\rm{  }}\underbrace {{\rm{Transformer}}}_{{\rm{text encoder}}}}\limits_ +  \\
\underbrace {\begin{array}{*{20}{c}}
{\begin{array}{*{20}{c}}
{{\rm{Transformer}} \to }\\
{\mathop {{\rm{Transformer}} \to }\limits^ \uparrow  }
\end{array}}&{\mathop {{\rm{FC}}}\limits_{\scriptstyle{\rm{Information}}\hfill\atop
\scriptstyle{\rm{Fusion}}\hfill} }
\end{array}}_{{\rm{decoder}}}
\end{array}$ }  & The scheme achieves comparable answer accuracy in both perfect and imperfect channel state information and is considerably higher than traditional methods. & The complex model design introduces extra computing latency and computational resource consumption for training, especially for text encoding. The size of the images used for training and testing is small. The superiority of this model over traditional methods for VQA with large image sizes is yet to be verified.\\ \hline 
\end{tabular}
 \label{tbl:DLSE}
\end{table*}
\subsubsection{SE for multimodel data}
In addition to the three representative data above, the authors in~\cite{xie2021task} take the visual question answering (VQA) task as an example and investigate a SemCom system for multimodal data transmission. 
In a VQA task, some users transmit images while the others transmit texts to inquire about the information of the images. The answer is obtained at the receiver.
In~\cite{xie2021task}, they consider a simple communication scenario with an image transmitter, a text transmitter, and a receiver. Similar to the above works for image and text, the proposed image transmitter employs the ResNet-101 network~\cite{he2016deep} pre-trained on ImageNet~\cite{russakovsky2015imagenet} and the proposed text transmitter employs the Bi-LSTM network. Nevertheless, the design of the decoder is not well studied. Since the SI from both users is correlated, the decoder needs to merge the text and image SI as well as  answer the vision questions. To address the issue, the authors adopt the memory, attention, and composition (MAC) neural network~\cite{hudson2018compositional} to deal with the correlated data. Specifically, each MAC cell consists of three units.
The \textit{control unit} first generates a query based on the received text SI by an attention module, then the \textit{read unit} receives the query and searches the corresponding key from image SI by another attention module~\cite{xie2021task}. Finally, the \textit{write unit} integrates the information and outputs the predicted answers to the questions~\cite{xie2021task}. {Compared to the traditional method, where the recovered image and text are input to MAC, the end-to-end scheme achieves significantly higher answer accuracy.  However, since the scheme assumes perfect channel state information due to the lack of attention mechanism, it is not robust to channel changes in a real-world environment.} Furthermore, in~\cite{xie2021task2}, the authors unify the semantic encoding structure for both image transmitter and text transmitter based on Transformer. Meanwhile, they propose a new semantic decoder network that consists of two modules: the query module and the information fusion module. The query module adopts \textit{layer-wise Transformer}, which consists of Transformer encoder layer and Transformer decoder layer. Different from classical Transformer, layer-wise Transformer in ~\cite{xie2021task2} takes the
output tokens of each encoder layer as the input of each decoder layer, which can exploit more   keywords in the
text information and the corresponding regions in the image information. The fusion module then fuses both information to get the answer. {Compared to \cite{xie2021task}, the scheme achieves comparable answer accuracy in both perfect and imperfect channel state information and is considerably higher than traditional methods. However, the complex model design introduces extra time consumption and computational resource consumption for training, especially for text encoding. Meanwhile, the size of the images used for training and testing is small. The superiority of this model over traditional methods for VQA with large image sizes is yet to be verified.} An instance of a simple VQA task can be found in Fig.~\ref{VQA}.

\vspace{0.2cm}

\subsection{RL-based SE}
\label{RL-based SE}
Intuitively, the learning process guided by sophisticated semantic metrics can facilitate more accurate SE. However, many existing semantic metrics in other fields are \textit{non-differential}. To overcome the stringent requirements of DL for the loss function to be differentiable, RL is treated as a promising alternative. 

RL is regarded as a promising paradigm to address the issues with user-defined, task-specific, and non-differentiable task metrics in some other fields~\cite{ranzato2015sequence,rennie2017self,sutton2018reinforcement}.
Considering the success of RL in sequence-generation tasks~\cite{bahdanau2016actor,ren2017deep,yu2017seqgan}, the authors in~\cite{lu2021reinforcement,lu2021rethinking} make the first attempt to integrate RL into the end-to-end SemCom system for text transmission, where the encoder-decoder scheme can be viewed as the agent that interacts with an external ``environment", (i.e., sentences).
In the general RL framework, the tasks required to be converted into a Markov decision process (MDP), which consists of five elements: state, action, policy, reward, and long-term return~\cite{sutton2018reinforcement}. In their proposed encoding-decoding scheme, the LSTM is employed to provide the policy.
Similar to the MDPs for other sequence-generation tasks, the state is defined as the recurrent state of the decoder and the previously generated words. In this sense, the transition between two adjacent states is determined by the next generated token. Meanwhile, the action of the RL agent is to generate a new token, and thus the action space is the dictionary dimension. Moreover, the semantic metrics of the whole recovered sentence can be intuitively treated as the long-term return

However, the determination of the immediate reward function form is particularly tricky. Unlike most RL-based strategies with well-defined rewards at each time step, the rewards during decoding cannot be directly measured until the end of a sentence. To overcome this challenge, several methods have been proposed. The first is using the Monte Carlo search to obtain the reward in each time step~\cite{silver2016mastering, yu2017seqgan}. The second  is training another neural network to estimate the reward or for an incomplete sequence~\cite{konda2000actor}. However, the above methods are more time-consuming and resource-consuming and introduce the risk of divergence in a huge action and state space~\cite{lu2021reinforcement}. Moreover, quantifying the reward value in each time step may be inconsistent with ensuring the semantic meaning of the whole sentence. Thereby, in~\cite{lu2021reinforcement}, the authors adopt a newly emerging approach named self-critical sequence training (SCST)~\cite{carpi2019reinforcement}. The idea of SCST is to utilize the output of its own test-time inference algorithm to normalize the long-term rewards it experiences, rather than to focus on estimating the reward, or how the reward function should be normalized~\cite{carpi2019reinforcement,luo2020better}. In~\cite{lu2021reinforcement}, the mean long-term return (i.e., the semantic metric for the whole sentence) from a group of selected samples is used to normalize the rewards and treated as the baseline term in the objection function, which enables stable and self-supervised training at the cost of nearly no extra computations. Meanwhile, it is necessary to note that the policy network is not updated until the end of a complete transmission of a sentence.

In the simulation, the proposed RL-based scheme is trained with the semantic similarity metric of CIDEr, and the performance is evaluated by BLEU scores from 1-gram to 4-gram. The size of gram means the length of the phrase considered in calculating the similarity between the reference and candidate sentences, which is detailed in Section~\ref{semantic metric}. By comparing the proposed scheme with the DL-based SemCom trained with cross-entropy loss, one can be found that, with the increasing size of the gram, the superiority of RL-based algorithms over DL-based scheme becomes more significant. This demonstrates the capability of the proposed scheme to catch the underlying semantics, since the longer phrases generally carry more abundant semantic meanings.
\begin{figure*}[t]
 \centering
 \includegraphics[scale = 0.38]{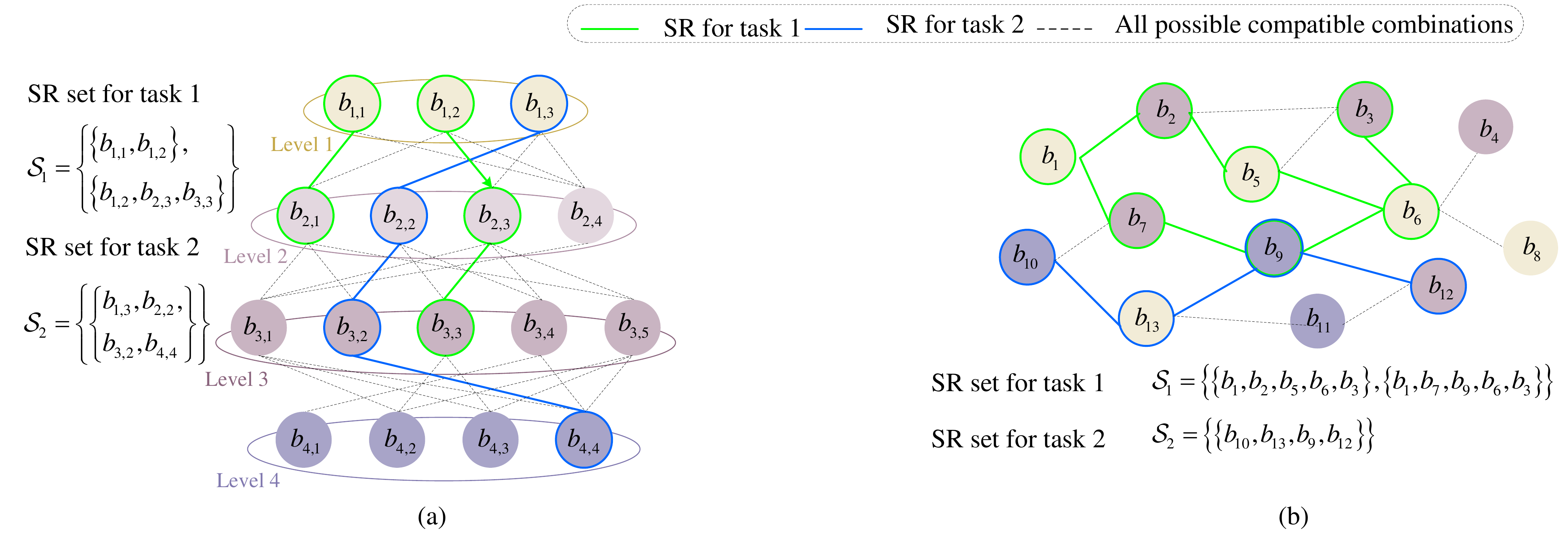}\\
 \caption{{ Two types of KB models~\cite{farshbafan2022curriculum,wang2021performance,zhou2022cognitive }.}
 }
 \label{kbmodel}
\end{figure*}

\subsection{KB-assisted SE}\label{sec:KB-assistant} {Intuitively, given raw data, the SI may be distinct for different communication goals~\cite{lan2021semantic}. ples of image transmission depicted in Fig.~\ref{example}, the receiver may have to perform different tasks, such as classification based on different attributes, detection of different targets, or simply replication. In this sense, the SI applied to  different tasks is different, but is still highly correlative. Therefore, if we employ a general DL-based SE model for multi-tasks, the extracted SI may be redundant for specific tasks. Otherwise, repeatedly
performing SE on the same raw data based on multiple
DL-based SE models that are discussed in Section~\ref{DL-based SE}
may cause much system redundancy. To address this issue, a suitable approach is to extract all the SI units conveyed by the raw data and correspond individual communication goals to the different combinations of SI units. To this end, a shared KB is required to be established at the transmitter and receiver in advance before the task request is sent. Meanwhile, the process of SE can be treated as refining the importance of each SI unit to individual communication goals.}

KB is a technology that has been widely used in automated AI systems to store the data with formal representation allowing for inference~\cite{rosa2018knowledge,lin2020kbpearl,zheng2021knowledge}. In general, a typical KB consists of a computational ontology, facts, rules, and constraints~\cite{strinati20216g}.  Particularly for the SemCom system, the KB should be composed of SI, goals of the communication tasks, and the possible ways of reasoning that can be understood, recognized, and learned by all the communication participants~\cite{yang2021semantic, shi2021semantic}. Specifically, the KB can be employed to record the relationship between each SI unit and each task, as well as quantify the level of importance of SI for different tasks, thus instructing the SE under different channel conditions, when the communication task changes.

Following this, in~\cite{yang2021semantic}, the authors first manage to establish a simple KB based on CNN for an image classification task and accomplish the KB-assisted SE. In their work, CNN  is treated as a SI generator, wherein the feature maps for each layer output indicate different aspects of SI of the source images, such as the color, the texture. 
Since the parameters of a well-trained CNN model can identify the optimal form of feature maps that represent the original image (i.e. SI), the gradients of the CNN's output can be treated as the importance weights of the feature map to different classes~\cite{yang2021semantic}. Thereby, the KB is established by storing the importance weights of all feature maps for each class. Next, based on the KB,  semantic encoding can be accomplished by refining the SI that is relevantly related to the specific task.  In addition, since encoding and decoding are mutually reversible processes, the scheme is also implemented in an end-to-end manner. In this sense, the KBs in both the transmitter and receiver should be synchronized by a shared KB at an authoritative third party or a virtual KB. If the two KBs on the two sides mismatch, the \textit{semantic noise} may be generated during the SI inference~\cite{basu2014preserving}. Since KB-assisted SE focuses only on the goal-related SI, the KB-assisted SemCom with CR\footnote{The value of CR means the percentage of feature maps that are ignored.} of $98\%$ can still achieve more than 40$\%$ classification accuracy gains compared with the conventional communications at 10~dB. However, it still has room for enhancement, such as the optimization of neural network structure and loss function~\cite{selvaraju2017grad,fong2018net2vec,johnson2016perceptual}.

{In addition to the study of the KB establishment of the semantic KB, in works~\cite{farshbafan2022curriculum,wang2021performance,zhou2022cognitive } on resource allocation in SemCom scenarios, some ideas are proposed for the KB storage model of the KB.  From now, there are two available kinds of KB models as shown in Fig.~\ref{kbmodel}.
In~\cite{farshbafan2022curriculum},  a
hierarchical structure is proposed for the semantic KB of a task set, wherein the indivisible units of SI are called beliefs. 
The higher the level that a belief belongs to, the more SI it contains.
For a task in the considered task set, there may be multiple feasible semantic representations (SRs) and each SR only includes one belief from each level of the hierarchy~\cite{farshbafan2022curriculum}. However, such a hierarchical structure is hard to incorporate the relationship between multiple descriptions and a task. Moreover, in the hierarchical structure, a belief in a higher level is completely dependent on a belief in its previous level. Therefore, it is not flexible enough to represent the combination of several beliefs belonging to discontinuous levels. 
In~\cite{shi2021semantic},  the authors point out that a graphical structure is one potential solution adopted to model semantic knowledge, where any two SI units can be linked by an edge if necessary. In~\cite{wang2021performance,zhou2022cognitive}, the authors firstly employ the graphical structure to model the semantic knowledge for text transmission according to the grammatical structure of sentences. In their works,  the tokens encoded with fixed bit length are treated as vertexes, and the relationship between two tokens is reflected by the edge. However, the issue of modeling a generic semantic  KB  is still open. }

\vspace{0.2cm}

\begin{figure}[t]
 \centering
 \includegraphics[scale = 0.46]{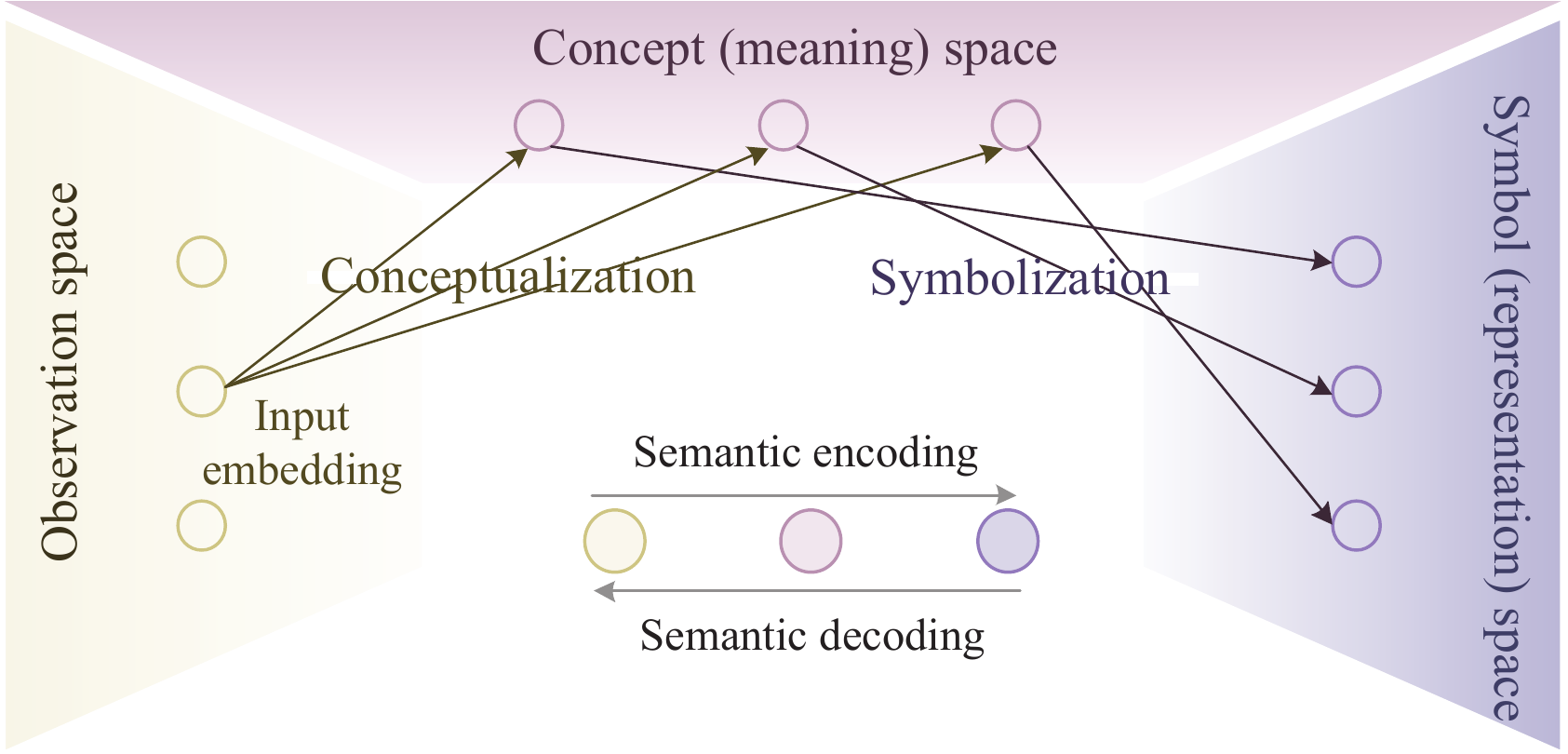}\\
 \caption{ An illustration of triangle of meanings model~\cite{seo2021semantics,ogden1923meaning}.
 }
 \label{toymodel}
\end{figure}

\subsection{Semantic-native SE}
\label{semantic-native SE}
All the above three SE methods rely on well-trained neural networks based on a large amount of labeled data, which makes their works only feasible for communication systems with unvarying SI. Hence, they are powerless for scenarios where semantics vary over time or communication context, and such scenarios are more common in real life~\cite{seo2021semantics}. Specifically, In this sense, transforming ``passive learning" into ``active learning" is particularly imperative for SE in communication with varying semantics and context.

Indeed, there have been some primary research studies fitting the above idea called \textit{emergent communication}~\cite{lazaridou2020emergent}, wherein the semantics and goal-oriented representations are not predefined and are required to be learned during the iterative communication between multiple intelligent agents~\cite{seo2021semantics,lazaridou2020emergent}. However, most of the works merely focus on some simple and specific AI tasks such as image-related referential games~\cite{lazaridou2016multi}, where the accomplishment of SE may be spurious owing to the inscrutable patterns of the single transmitted objects~\cite{mu2021emergent}.

In~\cite{seo2021semantics}, the authors open up the black box of SE with the focus on a point-to-point communication scenario between two agents who can communicate in both directions. In analyzing the reliability (which is measured by the recognition accuracy in the considered scenario),
they introduce the \textit{triangle of meanings}~\cite{ogden1923meaning} for human communication architectures in linguistics. As shown in Fig.~\ref{toymodel},
the vertices of the semantic triangle connect the three spaces of the observation of the input, concept
(or meaning), and symbol (or representation)~\cite{seo2021semantics}. The edge from an input embedding to its concept is termed conceptualization, and
the edge from the concept to its symbol is termed symbolization, while their opposite directions represent deconceptualization and desymbolization~\cite{seo2021semantics}. Based on this model, they propose two SemCom systems (indexed by System 1 and System 2, respectively). System 1 can be summarized by a multi-triangular model with a shared input embedding. The conceptualization process can be interpreted as a stochastic soft decision or the likelihood of a decision in ML, which plays a similar role in unconscious pattern recognition to the ML in the aforementioned methods. In addition, the symbolization process is assumed to be predetermined among the agents. Due to the fact that rational speakers are self-aware of what they say, in System 2, the authors infuse \textit{contextual reasoning}~\cite{bell1992pragmatic} process for each agent. In linguistics, contextual reasoning is often computationally described using the rational speech act model~\cite{goodman2013knowledge,kao2014formalizing,goodman2016pragmatic}, which is rooted in the Gricean view of language use\cite{grice1975logic}. In the proposed system, contextual reasoning is equivalent to communicating with a virtual agent that mimics and simulates its listener, which allows the agents to communicate effectively and efficiently based on reasoning. 
To demonstrate the significance of contextual reasoning, the authors abstract both the systems into stochastic models, and derive the bit-length of semantic representation in the two systems with Shannon coding. With the experimental results, it can be seen that the bit-length of semantic representation is significantly reduced with high reliability.

\begin{table*}
\small
 \centering
 \caption{Summary of generic semantic extraction methods.}
 \begin{tabular}{|m{4.5cm} m{3.5cm} | m{4.5cm} m{3.5cm} |}
  \hline
  \multicolumn{4}{|c|}{\it Semantic-oriented communication}  \\[1ex]
  \hline
\multicolumn{2}{|c|}{\textbf{DL-based SE}} & \multicolumn{2}{c|}{\textbf{RL-based SE}} \\
\hline
 
\vspace{0.2cm}
  
  \begin{minipage}[b]{0.5\columnwidth}
		\centering
		\raisebox{0.03\height}{\includegraphics[width=\linewidth]{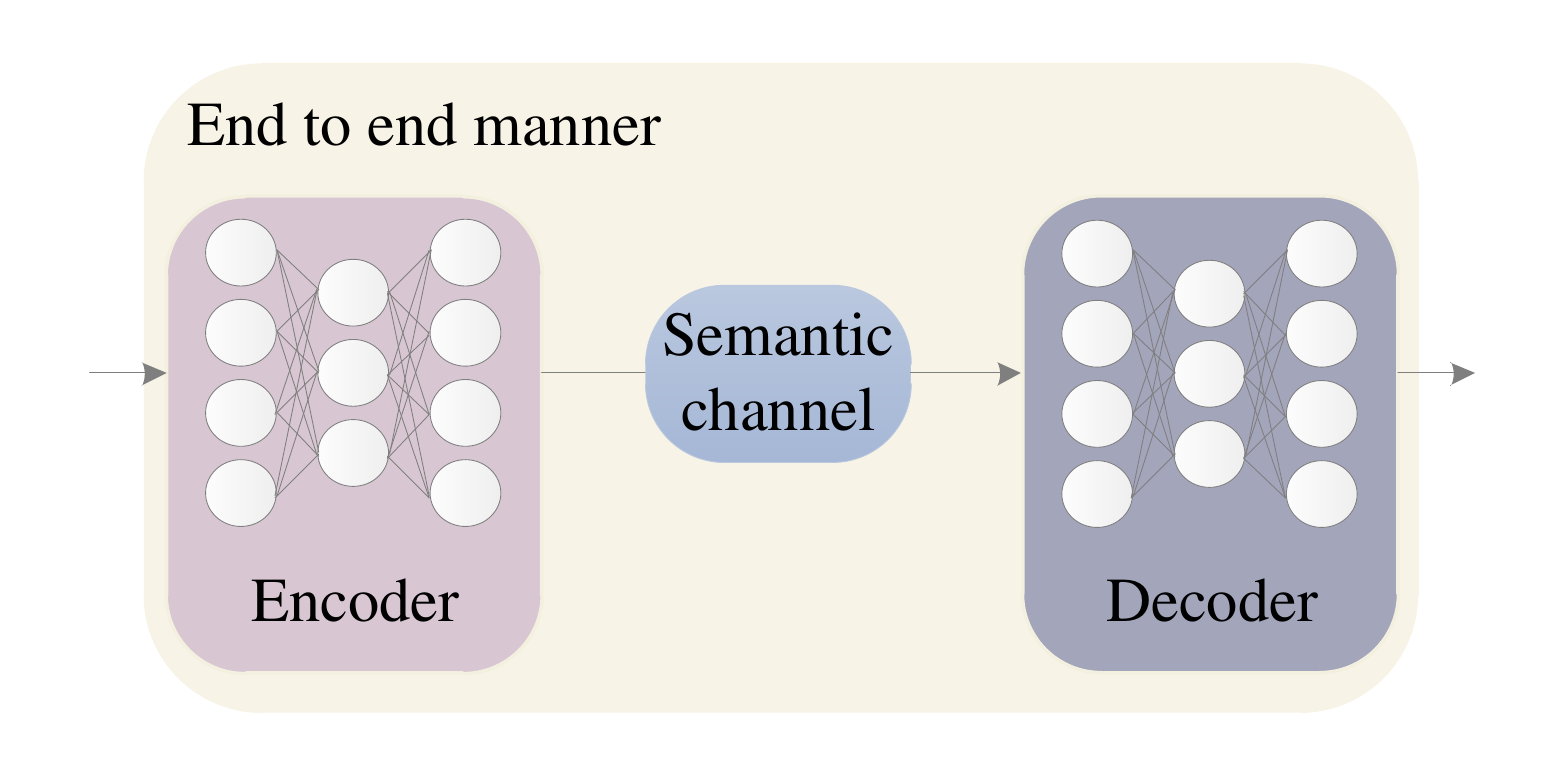}}
  
	\end{minipage} & \textbf{Description}: \newline The encoder
and decoder are usually modeled as two separate learnable
NNs, and linked through a random channel, which are
trained jointly. The dataset used for training can be seen as their shared background knowledge.& 
   
  \vspace{0.2cm}
  
 \begin{minipage}[b]{0.5\columnwidth}
		\centering
		\raisebox{0.03\height}{\includegraphics[width=\linewidth]{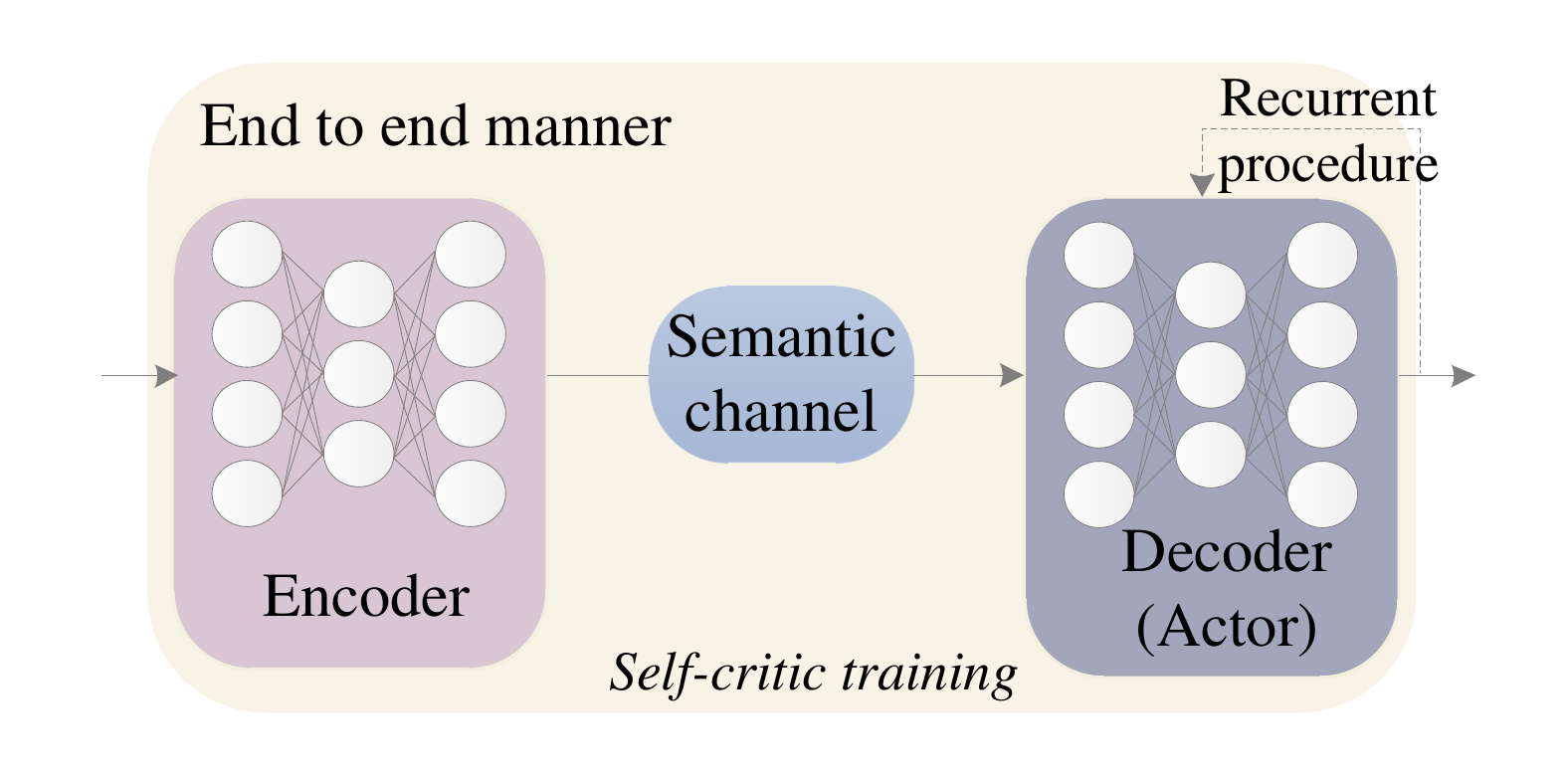}}
  
	\end{minipage}  & \textbf{Description}: \newline It is developed on the basis of DL-based SE. The decoding process is converted into a recurrent procedure. By employing self-critic training, the non-differentiable metrics, such as BLEU, can guide the learning process directly. \\
  \multicolumn{2}{|c|}{\parbox{8cm}{{\textbf{Pros}: \newline  \tabitem Achieve lower CR while preserving the relevant information \newline \tabitem Significant superiority in the low SNR region \newline \tabitem Reduce processing latency in analog transmission without compromising communication performance \newline \textbf{Cons}: \newline  \tabitem Become sub-optimal in ideal channel conditions due to the error floor of DL \newline  \tabitem The loss Function for guided learning in training can only be used for differentiable MSE and CE}}} & \multicolumn{2}{c|}{\parbox{8cm}{{\textbf{Pros}: \newline  \tabitem Achieve more precise SE guided by the specialized semantic metrics \newline \tabitem Time-related metrics, such as AoI can also be integrated into the reward to guide SE due to the online paradigm of RL \newline \tabitem Also features the  pros of DL-based SE  \newline \textbf{Cons}: \newline  \tabitem Frequent interactions with the environment of RL greatly increases the training complexity \newline  \tabitem applicable only to sequence-generation tasks, such as sentence recovery}}}  \\
  \hline
  \hline
  \multicolumn{4}{|c|}{\it Goal-oriented communication}\\[1ex]
  \hline

\multicolumn{2}{|c|}{\textbf{KB-assisted SE}} & \multicolumn{2}{c|}{\textbf{Semantic-native SE}} \\
\hline
 
 \vspace{0.2cm}
  
  \begin{minipage}[b]{0.5\columnwidth}
		\centering
		\raisebox{0.03\height}{\includegraphics[width=\linewidth]{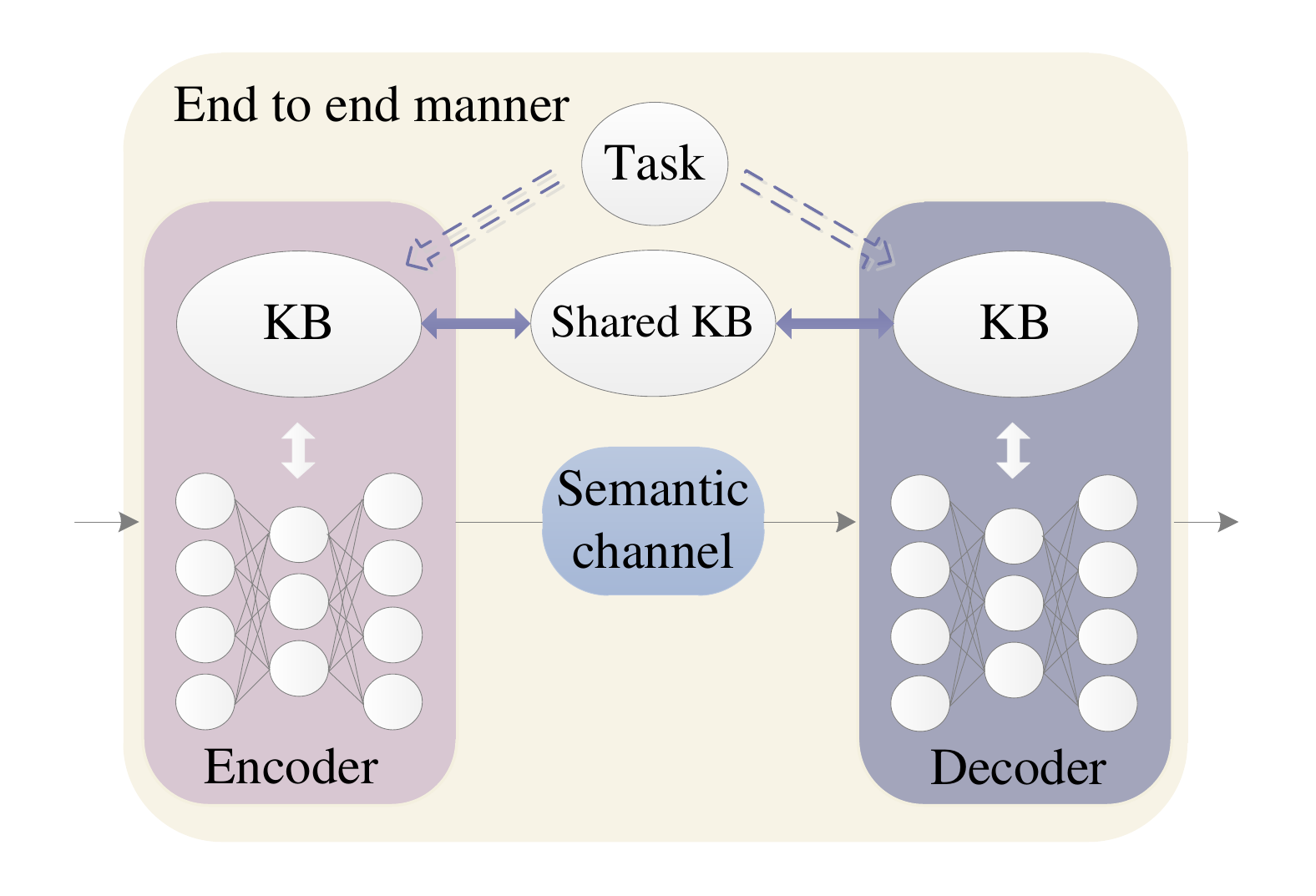}}
  
	\end{minipage}   & \textbf{Description}: \newline The KB stores all the SI units conveyed by the raw data and the importance of each SI unit to different tasks, which is well-constructed before communication link establishment. In each transmission, only the task-related SI is transmitted according to the KB and channel states. &  
  \vspace{0.2cm}
  
  \begin{minipage}[b]{0.5\columnwidth}
		\centering
		\raisebox{0.03\height}{\includegraphics[width=\linewidth]{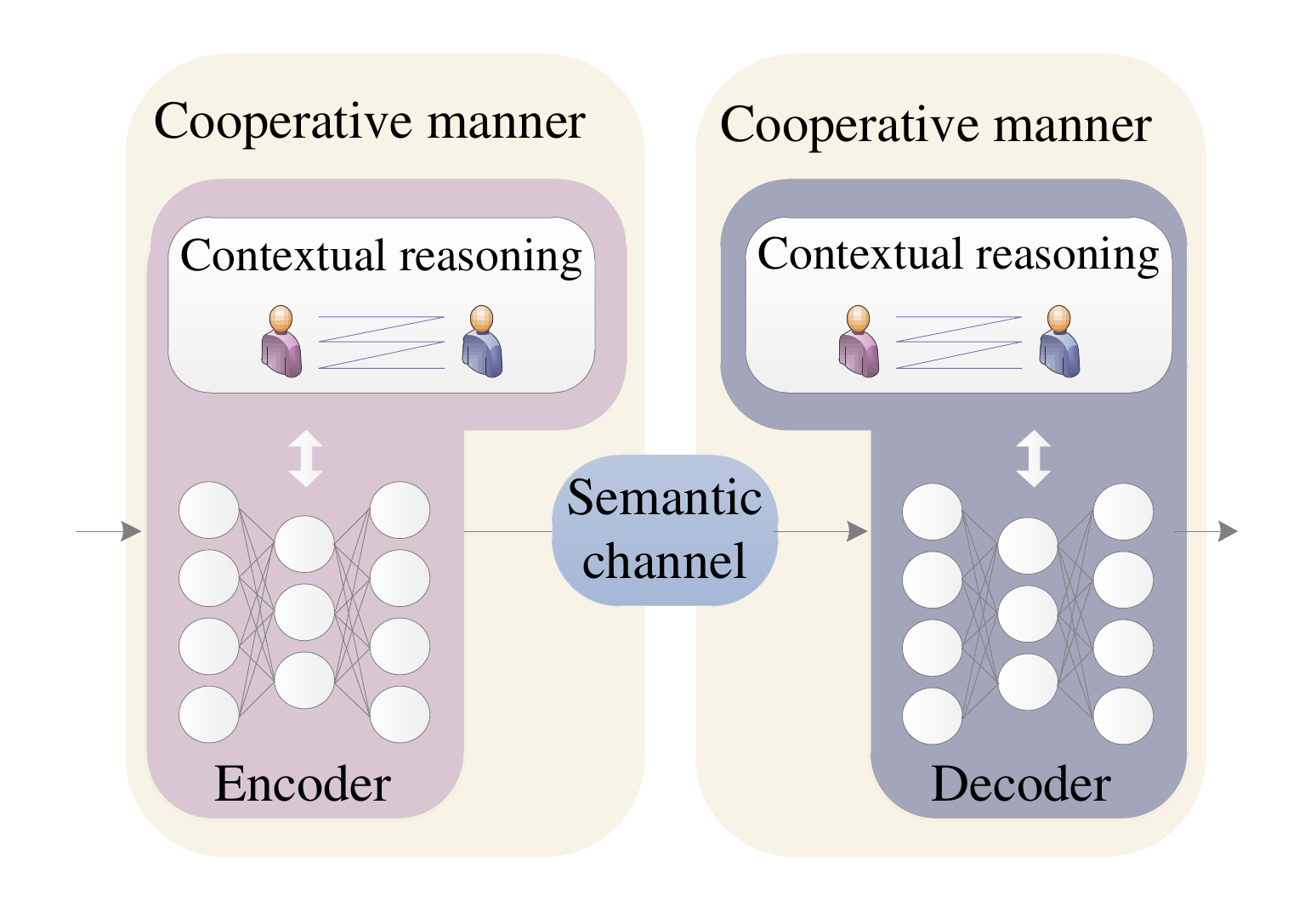}}
  
	\end{minipage} & \textbf{Description}: \newline It is developed based on emergent communication. It converts ``passive learning" to ``active learning". SI and background knowledge are learned through interaction and feedback between the communicating parties, which does not depend on an existing database.\\
\multicolumn{2}{|c|}{\parbox{8cm}{{\textbf{Pros}: \newline  \tabitem Allow for flexible and more precise task-specific SE \newline \tabitem Applicable to complex communication scenarios with multiple goals \newline \tabitem Lay the foundation for SemCom-aware resource allocation due to the quantified data size and importance of SI units \newline \textbf{Cons}: \newline  \tabitem Applicable only to the non-real-time on-demand services \newline  \tabitem The construction of KB is computation-intensive }}}  &  \multicolumn{2}{c|}{\parbox{8cm}{{\textbf{Pros}: \newline  \tabitem Adaptive to changes in the communication context and goal, reducing human intervention   \newline \tabitem Background knowledge does not need to be shared in real time\newline \tabitem Some other features such as channel states and QoS requirements can be considered in the learning process \newline \textbf{Cons}: \newline \tabitem The training process is time-consuming and computing resource intensive \newline  \tabitem  Convergence of training is hard to be ensured}}} \\ 
   
  \hline
 \end{tabular}
 \label{tbl:4methods}
\end{table*}

\subsection{Some specific SE}
\label{sec:spse}
\begin{figure*}[t]
 \centering
 \includegraphics[scale = 0.5]{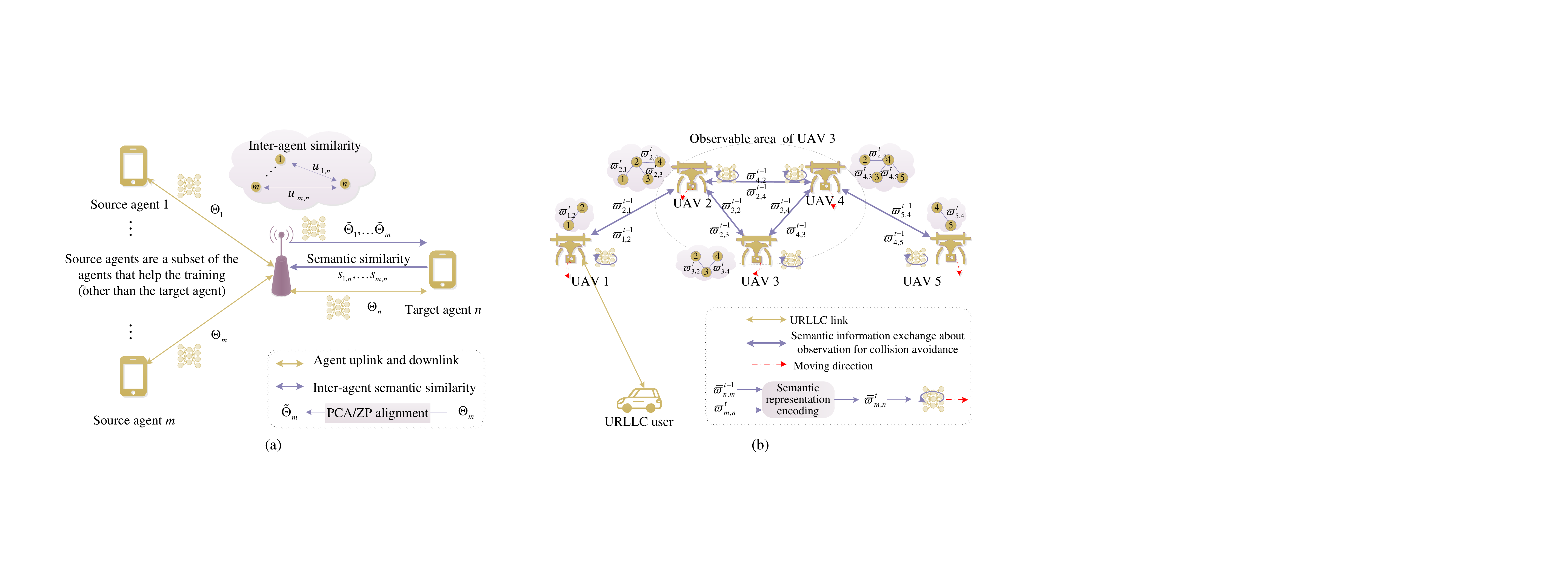}\\
 \caption{Examples for SE in semantic-aware communication~\cite{lotfi2021semantic,yun2021attention}.
 }
 \label{example}
\end{figure*}

The above four SE methods shown in Table~\ref{tbl:4methods} can be generalized to semantic-oriented and goal-oriented communication systems in different scenarios. However, there is no  general approach to SE in semantic-aware communication. {In this subsection, we first give two typical examples as shown in Fig.~\ref{example} to illustrate the motivation of introducing semantic awareness to communication, as well as the role of SE in semantic-aware communication.}

In~\cite{lotfi2021semantic}, the authors focus on a federated DRL task as shown in Fig.~\ref{example}(a), where multiple heterogeneous agents participate in model training in a cooperative manner under the coordination of a central controller. The agent requiring training is called the target agent, and the agents that help the target agent with its training are called source agents. {In the proposed scheme, SemCom plays a role in the construction of a  knowledge graph that records similarities among all the agents.  Based on the KG, a subset of source agents with high similarity are selected strategically to contribute to the training of the target agent. }
In their work,  they employ semantic relatedness to measure the similarity of underlying learning tasks of the agents~\cite{tao2021repaint}. In the knowledge transfer domain, it can quantify the extent to which the transferred knowledge from a source agent can potentially help the target agent to find its optimal policy. {In this sense, semantic relatedness can be regarded as a kind of SI exchanged between the central controller and agents. It is defined as the average return value received by the source agent from a target environment in a limited number of training episodes. Meanwhile, the training process can be treated as its corresponding SE method.} However, since the average return value is obtained from limited training steps, the metric may be inaccurate, especially for a complex target environment with a large state space. Therefore, the structural similarity is jointly considered during the similarity KG construction. The structural similarity can be measured by the central controller based on the cosine similarity of the received network parameters of the policy from  agents~\cite{visus2021taxonomy}.
Besides,  since the DNN parameter dimensions of heterogeneous agents should be aligned before comparing the similarity, the principal
component analysis (PCA) method and the zero-padding (ZP) method are employed for compression and  expansion of the source agents' DNN parameters~\cite{goodfellow2016deep}, respectively. With the simulation, the promising performance of the semantic-aware CDRL schemes~\cite{lotfi2021semantic} over bandwidth-constrained wireless networks has been demonstrated by comparing with the uniform or random resource block allocation.

{In another example in~\cite{yun2021attention}, the authors focus on a non-terrestrial ultra-reliable and low-latency communication (URLLC) system (as shown in Fig.~\ref{example}(b)), where an unmanned aerial vehicle (UAV) swarm employs the centralized training and decentralized execution (CTDE) multi-agent deep reinforcement learning (MADRL) to serve a moving ground user while avoiding inter-UAV collisions. In their work, SemCom is not employed to enhance the URLLC performance directly, but to be integrated to the differentiable inter-agent learning (DIAL)~\cite{foerster2016learning} between UAVs to avoid inter-UAV collisions. DIAL is widely used in CTDE MADRL framework. In the traditional DIAL,  the agents exchange their respective observable states, which are treated as an input of their actor models. Then during the training process, the raw state data are progressively translated into meaningful information for better inter-agent collaboration.
However, training from scratch could be inefficient. 
To that end, the authors in~\cite{yun2021attention} perform SE
on the observable state before exchanging
information. Specifically, each UAV $n$ constructs a local star-topological graph, where the leaf nodes are all its observable UAVs. The SI about UAV $m$ at step $t$ is recorded as the weight of the corresponding edge (i.e.,  ${\bar \varpi _{n,m}^t}$ in Fig.~\ref{example}(b)), which reflects the level of attention to  UAV $m$ , when UAV $n$ takes its action. 
Considering the fact that UAVs should pay equal attention to each other for collision avoidance, the attention paid to UAV $m$ by UAV $n$ is derived based on an RNN with the two inputs. One is the attention feature extracted by UAV $n$ from its observable state about UAV $m$ based on the self-attention mechanism~\cite{vaswani2017attention}. The other is the SI about UAV $n$ at last step  that  is sent by UAV $m$  to UAV $n$. From the simulation, it can be seen that using SI as the input of actor model instead of raw state data can significantly improve the training efficiency.}


\subsection{Lessons learned summary}
\subsubsection{Lessons learned for DL-based SE}
{Comparing the above DL-based SE methods as summarized in Table~\ref{tbl:DLSE}, we can observe that the attention mechanism plays an essential role in SE performance enhancement due to its excellent performance in capturing long-range dependence of the input. 
The advantage of DL-based SE is that it can extract important information from the whole raw data, and then re-aggregate and re-extract it at different layers, thus effectively removing redundant information. In this sense, compared with the content-blind traditional encoding and decoding, DL-based SE can achieve much lower CR without losing relevant information. Therefore,  under the low SNR region, the superior performance of DL-based SemCom is especially remarkable.  Moreover, with the well trained end-to-end  SE model, the DL-based SemCom achieves good performance in analog communication. Due to the absence of quantization and complex modulation processes, such as 16-QAM and QPSK, the data processing delay before transmission can be significantly reduced, which shows its potential for low-latency communications.}

{However, the DL technique has an inherent weakness, i.e., the unavoidable error floor~\cite{jiang2019deepturbo}. Therefore, under ideal channel conditions, the DL-based SemCom is often suboptimal compared with traditional communication. Hence, how to overcome the performance bottleneck under the high SNR region is worth future investigation.} Moreover, during the end-to-end training, the back-propagation through transceivers requires the loss function in the DL paradigm to be \textit{differentiable}~\cite{lu2021rethinking}. In this sense, all the above studies still apply the commonly used loss functions in DL (i.e., cross entropy (CE) and mean square error (MSE)) to train neural networks, which leaves the existing works far from the desired SemCom. In other words, all the above-mentioned architectures merely achieve semantic coding for a reliable and efficient transmission~\cite{lu2021rethinking}. With such end-to-end architectures, as the semantic and channel encoders and decoders need to be trained jointly, the SE and recovering are treated as a black box~\cite{seo2021semantics}. Due to the lack of \textit{explainability and interpretability} for the available DL-based SE, the informativeness of the extracted SI is hard to measure and it is also unclear how to make relevant improvements.
\subsubsection{Lessons learned for RL-based SE}
{Comparing the DL-based SE and RL-based SE, the main difference of RL-based SE lies in that the decoding of the whole sentence is converted into a recurrent procedure. That is, the output of the RL-based decoder is a single word, and the decoded word is the input for the next word to be decoded. Such a recurrent procedure can reinforce the learning of correlations among the words within a sentence during the training process, thus allowing the decoding policy to learn relevant features of the non-differentiable semantic metric function.} Moreover, as RL can be treated as an online paradigm, in addition to error-based metrics, some other metrics such as AoI-based metrics and transmission delay can also be integrated into the reward, which is another promising advantage for RL-based SE methods~\cite{kang2021task}. 

Such a method is naturally applicable to sequence-generation tasks, such as sentence recovery.
In~\cite{lu2021reinforcement},\cite{lu2021rethinking}, the authors also discuss the generalization capability for non-sequence tasks with an example of image transmission. They propose a pixel-level recurrent decoding scheme, where the state of MDP is defined as the intermediate decoded image obtained by increasing or decreasing the pixel values with a small number. {However, the effectiveness and necessity of such a conversion for image decoding for the performance enhancement of SemCom are unclear. Moreover, even if such a conversion from the decoding process to a recurrent procedure is feasible to implement in practice, it increases the decoding time.}

Moreover, learning the optimal policy through interactions with the environment inevitably increases the training complexity. It is still a critical challenge to train such a complex model from scratch for high-dimensional tasks~\cite{lu2021rethinking}. In the above works, the initial parameters are utilized the pre-trained model with stochastic gradient descent on the deterministic loss function. {Simulation results show that the accuracy is improved by 3\% in the middle SNR region with the RL-based SemCom based on the non-differential semantic metric optimization by comparing that with the DL-based counterpart.} However, for the more complicated language models like Transformer, whether the RL-based SE is still feasible needs to be further explored and studied.

\subsubsection{Lessons learned for KB-assisted SE}
{In fact, KB-assisted SE also heavily relies on the DL models. 
Different from the well-trained DL-based model that performs real-time SE for the new raw data, KB-assisted SE requires the KB for raw data to be well-constructed and synchronized among the parties before communication link establishment. This makes it merely suitable for non-real-time on-demand services. Moreover, compared to the single model training in DL-based SE, the construction of a sophisticated KB is a much more computation-intensive task.  This means that the KB cannot be updated frequently, so this method is more suitable for scenarios where the data source is stable.}

{In addition, as discussed earlier, by introducing communication goals into the SE, KB-assisted SE can improve multi-task system efficiency in two aspects. One is to improve communication efficiency by extracting flexibly the SI related to a certain task in each transmission. The other is to improve computational efficiency by avoiding repetitive SE of raw data. Beyond these two advantages, the construction of KB also lays the foundation for resource allocation for complex SemCom scenarios with multiple tasks. Since the size and importance of each SI unit can be recorded in the KB model, the SemCom-aware resource allocation for each SI unit can be done with different quality of service (QoS) requirements, such as delay and reliability. This differs from the resource allocation in traditional communication where all the data packets are treated equally, which is discussed in detail in Section~\ref{transmission challenges}.} Furthermore, it is worth noting that the semantic KB construction relies heavily  on the explainability of SE models. However, most of the available SE models have a black-box nature. To this end, enhancing the  explainability  of SE is the key to breaking the bottleneck of existing SemCom research.
\subsubsection{Lessons learned for semantic-native SE}
{ Comparing the above four SE methods, semantic-native SE is the most similar to the way of human conversational communication. The first three SE methods require that the background knowledge of the  
 transmitters and receivers are fully synchronized before SE model training and determination, e.g., the semantic encoder and decoder models are trained based on the same dataset. In contrast, semantic-native SE relaxes this constraint. Contextual reasoning for the agent is more like the process of capturing and inferring the way of thinking of a human during a conversation with a stranger. As communication parties become   ``familiar" with each other, the background knowledge of the agents converges.}
Undoubtedly, the semantic-native SemCom system is with a high degree of flexibility and adaptivity, which is more toward the vision of an intelligent and autonomous 6G network~\cite{calvanese20196g}. However, the above analysis is again only based on a representative model. It remains a huge challenge to put it into practice. 
\subsubsection{Lessons learned for specific SE}
{From the above two examples of a multi-agent collaborative task discussed in Section~\ref{sec:spse}, we can  see that SemCom in the above tasks only serves to facilitate cooperation and indirectly improve task performance.  In contrast to semantic-oriented and goal-oriented communications, SE in semantic-aware communication is not performed on the raw data generated directly by the source agent. The relevant SI for cooperation is obtained through analyzing the properties of the task itself and the behavior of the agents. This makes the process of SE have to be more tailored and it is difficult to find a universal and unified approach. However, it is clear that semantic-aware communication will play an essential role in task-oriented communication, as exchanging SI can enhance the knowledge between agents for better collaboration. Moreover, an efficient SE  can greatly reduce the communication overhead  caused by exchanging information between agents.}
\section{Semantic Information Transmission and Challenges}\label{transmission challenges}
\begin{figure}[t]
 \centering
 \includegraphics[width = 0.5\textwidth]{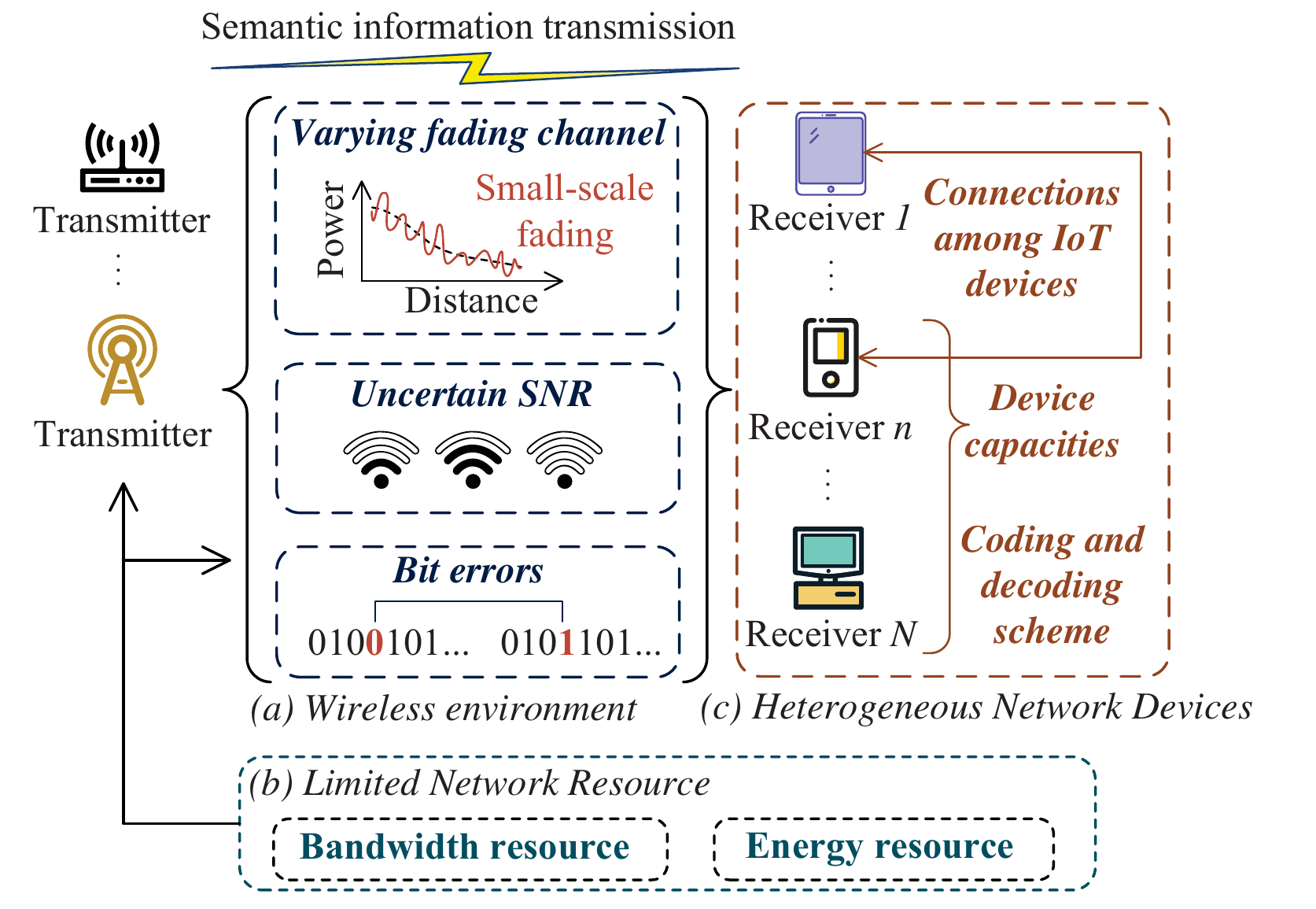}\\
 \caption{Semantic information transmission and challenges related to wireless environment, limited network resource, and heterogeneous networks.}
 \label{hy1}
\end{figure}

The focus of discussion in this section shifts from semantic-related to communication-related challenges and techniques. While conventional and SemCom systems use different methods to encode and decode information, they both face the same communication constraints, such as unpredictable channel conditions, limited transmission, and processing resources. However, unlike prior works in conventional communication systems, the solutions in SemCom are required to address new challenges in modern communication systems. {In the following, as shown in Fig.~\ref{hy1}, we discuss the challenges and techniques related to the wireless environment, limited network resources, and heterogeneous networks in SemCom.}

\subsection{Wireless environment}
\label{sec:4_1}
{The fading effect of the wireless channels has a great negative impact on the stability of data transmission, whether in conventional communication or in SemCom. To mitigate the negative effects of fading channels, in conventional communication systems, source and channel coding schemes are carefully designed. Specifically, source coding encodes the data into a sequence of symbols with optimized length, and channel coding adds redundant symbols to the sequence to detect and recover data corruption during wireless transmission. In the SemCom systems, the source and channel coding can be connected more tightly with the help of AI. Jointly designing and training the source and channel coding are shown to benefit data transmission in the DL-based communication systems \cite{farsad2018deep}.} However, the AI-based methods cannot currently be explained by explicit mathematical expressions. To overcome this obstacle, SemCom system designers have to consider how to make connections between the complex and changing wireless environment and sophisticated SemCom mechanisms, so as to obtain insights to guide the system design. {Thus, we discuss the impact of the varying fading channel, uncertain SNR, and bit error on the SemCom performance.}

\subsubsection{Varying fading channel}
In conventional wireless communications, several classical channel models are commonly used in performance analysis, e.g., Rician, Rayleigh, and Gaussian channel models. Moreover, to unify system performance with various channel environments, generalized fading distributions are proposed, e.g., $\alpha-\mu$~\cite{yacoub2007alpha}, Fisher-Snedecor $\mathcal{F}$~\cite{du2020sum}, FTR fading models~\cite{zhang2017new}. These channel models have various parameters to represent the different conditions of the wireless environment, such as the strength of the shadow effect and the multi-path effect. However, in SemCom systems that have an end-to-end structure, modeling the channel layer is a challenging task. Most of the existing works model the channel layer in two ways: fixed channel layers with the fading models used in conventional wireless communications, and generative channel layers with the neural network.

\textbf{Fixed channel layer modeling scheme:} In the fixed channel layer modeling scheme, the channel layer is modeled as a fixed fading model that is used throughout the training process. For example, the erasure channel is used in \cite{farsad2018deep} to model the dropping of data packets. The input of the erasure channel is a binarized bit vector from the encoder. Every element in the bit vector can be $-1$ or $1$, and the dropped element will become 0 at the output of the erasure channel. A drop probability is determined before the training. Eventually, the elements of the output vector after the erasure channel are in $\{-1,0,1\}$. This process is similar to the dropout technique used to prevent the over-fitting problem in deep neural networks. Hence a dropout layer can be used to represent the erasure channel of the communication systems.
For input that is not quantized or binarized, communication channels such as AWGN, Rayleigh channels, and Rician channels are considered. Recent research works in SemCom for text \cite{xie2021deep}, speech signals \cite{weng2021semantic} and multimodal data \cite{xie2021task} consider the channel layer as AWGN, Rayleigh channels, and Rician channels, for the training process. However, the performance evaluation is done with the same channel conditions in the training, without considering the fact that the changes in the wireless environment can lead to the change of the suitable channel model. Another drawback of using a fixed channel layer modeling scheme is that, if the SemCom model is trained under a certain fading channel, it is impractical to retrain the model for each possible channel condition and load all these models to the transmitter and the receiver. Although the trained model has shown some robustness, e.g., the authors in \cite{weng2021semantic} tested the model trained with Rician channels under AWGN and Rayleigh channels while achieving MSE loss less than $1 \times 10^{-4}$, we cannot explain the upper-bound of the robustness, and are not sure whether the model will fail when the environment changes. Instead of training with fixed channel layers, the generative channel layer modeling scheme is used to capture the dynamic behavior of the channel states.

\textbf{Generative channel layer modeling scheme:} The typical generative network adopted in existing works is generative adversarial net (GAN)~\cite{mirza2014conditional}. There are two main components in GAN, i.e., a generator and a discriminator. The generator aims to generate data samples that are as similar to real data samples as possible. The discriminator will be given real and generated data randomly and it will output a label to indicate whether the given data is real or generated. During the training process, the objective loss function helps the generator to generate more realistic data and the discriminator to output more accurate labels. To generate category-specific data, the conditional GAN \cite{mirza2014conditional} is proposed where the extra context information is provided to the GAN to obtain samples of the given context. In \cite{ye2020deep}, a conditional GAN is used to model the channel conditions. The conditional GAN is provided with both the pilot information and encoded signal from the transmitter and asked to generate an output signal that is similar to the real data. To evaluate the performance of the proposed conditional GAN in real channel conditions, the trained model is tested with the WINNER II channels \cite{meinila2009winner}. It is shown that the conditional GAN model outperforms the baseline system in terms of BER and BLER, especially when the SNR is over $10$ dB. Although the Generative channel layer modeling scheme improves the adaptability of the training model to the wireless environment, more training overhead is required than the fixed channel layer modeling scheme. This motivates us to think about the tradeoff between the performance and resource overhead.

Overall, fixed channel layers and generative channel layers provide wireless channel modeling during the training of the SemCom systems. The choice of different schemes has an impact on the  performance of SemCom. One possible solution is to combine the advantages of both schemes. For example, a two-phase training strategy is proposed in \cite{dorner2017deep} to adapt to the real channels. In Phase I, the model is trained with a suitable channel model to obtain model parameters with reasonable accuracy. In Phase II, the receiver is fine-tuned over the actual channel. The fine-tuned autoencoder constantly achieves lower BLER than the autoencoder without fine-tuning. However, how to choose the optimal training solution in different wireless environments is still an open question.

\subsubsection{Uncertain SNR}
After discussing the effects of the channel environment on the SemCom performance, here we consider the impact of SNR uncertainty on the trained SemCom model. Note that the influence of the wireless environment is mainly from the channel model selection, due to the effects of shadowing or scattering. The SNR uncertainty is from the effect of noise and interference, as well as the variation in transmit power, e.g., when adaptive transmit power schemes are used. Since the fixed SNR approach is typically adopted in the training of SemCom models, e.g., for text \cite{xie2021deep}, speech signals \cite{weng2021semantic} and multi-modal data \cite{xie2021task}, we need to consider whether the change in SNR will have a negative impact on performance.

 Several works tested the robustness of the trained model to SNR. In \cite{weng2021semantic}, the model is trained with a fixed SNR of 8 dB, and is then evaluated under SNR from 0 dB to 20 dB. It is found that the model has a higher MSE loss in the lower SNR region in the test under AWGN channels, Rayleigh channels, and Rician channels. However, it is still uncertain whether a model trained at a fixed SNR value can always be applied to a wide range of SNRs. Furthermore, while the SemCom constantly achieves higher performance than the conventional communication systems, both of them suffer a poorer performance in the lower SNR region. Because low SNR environments are common in the cellular edge, shopping malls, or  suburban areas, we need to consider the performance of SemCom when the SNR is low and the accuracy of the decoded signal is reduced.

To solve the aforementioned problems by making the model robust to different SNR regions, especially to the low SNR regime, the authors in \cite{ding2021snr} proposed an SNR adaptive mechanism. In the proposed model, the SNR is estimated by a pilot signal at the receiver. The estimated SNR value is then extended to an SNR map that has the same size as the channel output feature map. Both the SNR and channel output feature maps pass through a CNN layer before they are added together. The result of the element-wise addition is used as the input of a denoising module. A few transposed convolution layers are used to reconstruct the original image. It is found that, compared to the model that is trained with fixed SNR, the model trained with the proposed SNR adaptive mechanism has a smaller gap of PSNR between the high SNR region and low SNR region. By considering the SNR information in the decoding process, the proposed model shows higher adaptability to the SNR.

Instead of adding the SNR values to the channel features, another method to enhance the robustness of the SemCom model is to scale the channel features according to different SNR values \cite{xu2021wireless}. The training method proposed in \cite{xu2021wireless} adopts channel-wise soft attention, where each channel feature is multiplied by a scaling factor. To obtain the scaling factors, the SNR value is concatenated with the context information vector extracted from the input image and fed into two fully connected layers. Each element in the output vector is a scaling factor of a feature channel. It is shown that, with the help of the soft attention mechanism, the model can achieve a higher PSNR compared to baseline models which use basic deep learning networks without the attention module. In particular, the model with the soft attention mechanism can achieve more than 35 dB PSNR when the SNR is high.

However, both of these solutions are designed to solve specific communication problems. For the generalized SemCom system, the question of how to ensure that the trained semantic model can adapt to the variable SNR is still waiting for a better answer. The boundaries of the generalization capability of semantic models need to be further investigated.

\subsubsection{Bit errors}
To accommodate the changing wireless environment and the uncertainty of SNR, many mechanisms have been designed to improve SemCom performance. Now we focus on the bit error correction mechanism~\cite{moon2020error} that can further increase the probability that SI is correctly transmitted.

{Inspired by error correction algorithms in conventional communication systems, researchers have designed several error correction schemes for SemCom to minimize the transmission errors of SI}. For example, in \cite{jiang2021deep}, hybrid automatic repeat request (HARQ) is used to reduce the transmission error of semantic text transmission. With the help of HARQ, a re-transmission is requested if the received code block has uncorrectable error. The authors in \cite{jiang2021deep} first develop models with semantic encoder from \cite{xie2021deep} and Reed Solomon (RS) channel coding~\cite{shrivastava2013error} with HARQ. Then the performance is further improved by jointly designing source-channel coding and HARQ. Specifically, the RS channel coding is replaced by a dense layer to encode the output of the semantic encoder into a bit vector. To reduce the semantic error, Sim32 encoder and decoder are developed to detect the semantic similarity between original and estimated sentences. It is shown that with the joint source-channel coding and HARQ, the model achieves lower word error rate and sentence error rate when BER is larger than 0.06. In addition, new design schemes that combine correction mechanisms in conventional communication with SemCom require further study.

\subsection{Limited Network Resource}
Several resources, such as bandwidth and transmit power, are required for data transmission. Resource allocation frameworks in conventional communication systems aim to minimize metrics such as the bit error rate, packet error rate, and outage probability. However, the SemCom values the importance of the information behind the bit flow. This motivates us to develop the new resource allocation frameworks for the novel SemCom systems. Typically, in the design of a resource allocation scheme, the QoS and the QoE should be taken into consideration to build an effective system. Specifically, the QoS aims to optimize transmission rate, delay, and throughput \cite{xuan2019multi,liang2018graph}, and the QoE focuses on user satisfaction, clarity, and fluency \cite{khan2011qoe,xu2012qoe}. In the following, we discuss the methods of resource allocation in terms of bandwidth and energy in SemCom.

\subsubsection{Bandwidth resource}
Because the bandwidth resources are precious for any communication system, an effective bandwidth allocation is necessary for achieving SemCom to improve the overall system performance. Unlike the allocation of bandwidth resources in conventional communications, the uneven distribution of SI should be taken into account in SemCom, i.e., more bandwidth should be allocated to data/agents that have more SI.

One possible solution is to jointly perform the bandwidth allocation during the training process. In \cite{lotfi2021semantic}, a CDRL algorithm is designed, where multiple agents can coordinate over a wireless network to share their policies and collaboratively learn the best policy for the respective tasks. However, due to the limited bandwidth, agents that require training (target agents) can only collaborate with limited amount agents (source agents). Therefore, the metrics used to identify the most helpful agents are important for effective resource allocation. Building on previous works that consider the structural similarity of the agent model, the authors in \cite{lotfi2021semantic} include the semantic relatedness between the agents to construct a KG to aid the task selection of agents. The inter-agent semantic relatedness is defined as the return value of the target agent after a fixed number of training steps under the source agent's policy. After jointly optimizing the training loss and wireless bandwidth allocation, a KG is obtained, in which the values of the edges between the agents capture the structural and semantic relatedness between the connected agents. The KG is then used by the base station to select the most relevant agents for collaboration during the optimization. Simulation results show that the system performance can be improved by 83\%, compared to the baseline method that does not consider the semantic relatedness between the agents. However, the issue of combining dynamic allocation of bandwidth to semantic content transmission has not been fully studied. In some SemCom systems that do not require training, the bandwidth allocation schemes need to be designed to allocate more bandwidth for more important transmitted content to ensure information quality.

\subsubsection{Energy resource}
In addition to the allocation of bandwidth resources, the allocation of energy resources is also an important issue. For the growing number of IoTs with energy harvesting capabilities, it is important to determine the importance of information with the help of semantics. Allocating more energy to transmit data containing richer SI ensures the efficient use of energy. In addition, semantic metrics can be used to determine the quality of the collected energy, which can help build an efficient network market.

Based on the sentence similarity metric proposed in \cite{xie2021deep}, a semantic based valuation function is used in \cite{liew2021economics} for energy harvesting IoT devices to derive the value of harvested energy. In the proposed system model, there are IoT devices that adopt SemCom systems and a hybrid access point (HAP) that transmits energy to nearby IoT devices. The IoT devices operate by harvesting energy from the HAP \cite{ramezani2017toward,chae2018simultaneous} to transmit text data to the HAP. However, the HAP is considered to serve only one user at a specific time. To obtain the wireless energy, the IoT devices will submit their bids and the HAP will decide the winner and payment. A truthful auction mechanism is proposed so that the IoT devices will bid according to their true valuation of the energy. Instead of using the performance metrics of conventional communications, a valuation function based on the BLEU score and the similarity score is used by IoT devices to obtain their bid values.

However, at present, the application of semantic metrics in energy resource allocation is still in the early stage. Many SemCom networks that require energy-harvesting devices to work have not been studied, e.g., a UAV-aided network working with simultaneous wireless information and power transfer protocol.

\subsection{Heterogeneous Network Devices}
For the SemCom network, the wireless communication layer has a greater impact on the system performance than that of the end-to-end conventional communications. Because many heterogeneous devices work in one SemCom network, the differences in equipment hardware and wireless environments bring challenges to the system construction.

\subsubsection{Device capacities}
\label{4_3}
To enable SemCom systems, most of the existing approaches involve installing encoders and decoders into the transmitters and receivers respectively. While the DL-based auto-encoder systems can help to extract meaningful semantic information from raw data effectively, the cost of implementation is not cheap. Particularly more computational power and communication resources are required for the training process. Research shows that scaling up deep neural networks with correct techniques almost always leads to better performance \cite{devlin2018bert,he2016deep}. However, scaling up the model increases the storage requirement to store a higher number of the model parameters. In reality, communication devices have limited computational power, communication resources, and storage capacity. Especially in SemCom networks, it is unrealistic to assume that all devices have sufficient capacity. Therefore, in SemCom networks developing effective methods to balance the performance and cost requirements for heterogeneous devices is one of the important challenges.

To make the model proposed in \cite{xie2021deep} more affordable to devices in SemCom network with limited computing capability, the authors in \cite{xie2020lite} experimented with model compression to reduce the size of the model. A joint pruning-quantization scheme \cite{xie2020lite} is proposed to compress the model effectively with the idea of model pruning \cite{liu2017learning}. In the proposed method, less significant model weights are zeroed out, and the model weights that are larger than a pruning threshold remain. To determine the pruning threshold, the model weights are first sorted in the ascending order by the weight values, and then the pruning threshold is selected such that the outcome of pruning satisfies a pre-defined sparsity ratio between 0 and 1. The sparsity ratio indicates the desired ratio of zeros of the model weights. The pruned model is then fine-tuned to recover the performance of the model. The size of the model is further reduced by network quantization that converts the model weights from 32-bit float point to $m$-bits integer, $m<32$. A calibration process is needed to prevent overflows at the activation layer. In particular, an exponential moving average (EMA) is introduced to dampen the effect of outliers in the output of activation. Similar to model pruning, the model is fine-tuned after the quantization. Remarkably, the compressed model can achieve a similar BLEU score to the uncompressed model after model pruning with a sparsity ratio of 60\% and $8$-bit integer quantization. However, the performance loss due to further compression of the semantic model needs to be systematically investigated. We need to consider the capacity of different devices in the network and performance requirements.

\subsubsection{Connections among IoT devices}
For SemCom networks containing multiple smart devices, we need to design the network according to the different wireless link environments of different devices. One solution is to consider the wireless links as intelligent agents in the training process.

The authors in~\cite{zhu2021video} propose a resource allocation algorithm for semantic video transmission in spectrum multiplexing scenarios in vehicular networks. In the proposed algorithm \cite{zhu2021video}, semantic understanding accuracy of the video transmission is optimized by a multi-agent deep Q-network. In the network, vehicle-to-infrastructure (V2I) links and vehicle-to-vehicle (V2V) links are the agents. Based on the observations about the environment states, such as channel gains and interference power under the resource blocks, the agents choose to reuse spectrum resource blocks. Then, the agents receive the reward based on the V2I average object detection accuracy and V2V average transmission rate. Simulation results show that under the same spectrum and transmit power, the proposed network constantly achieves higher accuracy of video semantic understanding than the QoS and QoE based resource allocation framework, with as high as 70\% improvement for the density of correctly detected objects.

However, when the channel model is not available, the feedback links are absent for supervised learning. To solve this problem, the authors in~\cite{park2020end} propose a meta-learning approach for the receiver to adapt to the unknown new channel condition. Meta-learning which means ``learning to learn" refers to learning the adaptation module in the receiver~\cite{park2020end}. The meta-learning method first trains the adaptation rule during the training phase. In particular, during the meta-training phase, the receiver will be meta-trained to update the decoder parameters based on the output of the physical channel. During the testing phase, the receiver will self-optimize the model parameters using the trained adaptation rule. Simulation results show that the model with meta-training can achieve a lower BLER than the model with conventional training, when more than one pilot frame is sent by the transmitter during the testing phase.

\subsubsection{Coding and decoding scheme}
The coding and decoding scheme needs to be improved based on the various channel conditions of different users in the SemCom network. 

Unlike the two-phase training strategy which fine-tunes the model with supervised data, the authors in \cite{lu2021rethinking} propose a self-supervised mechanism in consideration of the varying channel states. In particular, a message is allowed to be encoded and decoded multiple times until a stopping criterion is fulfilled. For every cycle of encoding/decoding, the encoded/decoded information will be evaluated by a confidence mechanism to determine its semantic confidence. If the encoded/decoded information reaches a pre-defined confidence threshold, the encoder/decoder will release the information for the next process. Another stopping criterion is when the cycle length of encoding/decoding reaches a pre-defined maximum cycle length. With the distillation and confidence mechanisms, the encoder and decoder can fine-tune the encoded and decoded information in a self-supervised way, regardless of the channels. 

Moreover, in a multi-user scenario, the fluctuation in resources, such as available spectrum and transmit power, can have a non-negligible impact on the SemCom performance~\cite{yang2022semantic}. To this end, the variable-length semantic encoding, which is comparable to scalable video coding~\cite{schwarz2007overview} and multiple description coding~\cite{goyal2001multiple} in conventional communication, urgently needs to be investigated to cope with the dynamic SemCom network.

\subsection{Lessons learned summary}
\subsubsection{Lessons learned for wireless environment}
{The transmission problems caused by the wireless environment in classical communications should be considered in SemCom. However, due to the particular features of the SemCom, e.g., the differential importance of different bits for the raw data, schemes designed for the conventional communications cannot be used directly. Fortunately, many schemes that are designed for conventional communication systems can inspire  the construction of SemCom systems. For example, the end-to-end semantic model training requires system designers to use neural network layers to model wireless fading channels. Compared with using a fixed channel model in the performance analysis of conventional communication systems, a generalized fading channel that includes a variety of classical channels as its special case can bring more insights. Similarly, the introduction of the generative channel layer can offer more freedom to the training of semantic models.}
\subsubsection{Lessons learned for limited network resources}
{
Different from conventional communications that consider bit transmission only, the purpose of resource allocation in SemCom is to ensure the accurate transmission of semantic information related to tasks. Therefore, the consideration of semantic information introduces a new perspective to resource allocation scheme design in 6G networks. This novel design requires an in-depth analysis of 
task requirements and joint optimization design, to enhance the system performance. As we discuss above, some literature has attempted to use semantic information to guide the resource allocation process, but this paradigm shift still leaves much room for further research.
}
\subsubsection{Lessons learned for heterogeneous network devices}
{
The heterogeneity of network devices in the network is mainly reflected in two aspects, i.e., the difference in device capacity and the difference in the communication environment of each device. If the heterogeneity in network devices is not considered when training the semantic model, the trained model cannot be directly delivered to each device. Specifically, semantic models trained for high-performance devices may be large and cannot be deployed to small-capacity devices such as mobile phones. In addition, the end-to-end semantic models are influenced by the quality of the wireless channels between the transceiver devices. Therefore, differences in communication channels also affect the deployment of semantic models. Moreover, in the design of the coding and decoding scheme, which is an important component of the semantic communication network, we also need to carefully consider the impact of heterogeneity in devices.
}
\section{Semantic Performance Measurement and Challenges}
\label{semantic metric}
\begin{figure}[t]
 \centering
 \includegraphics[scale = 0.46]{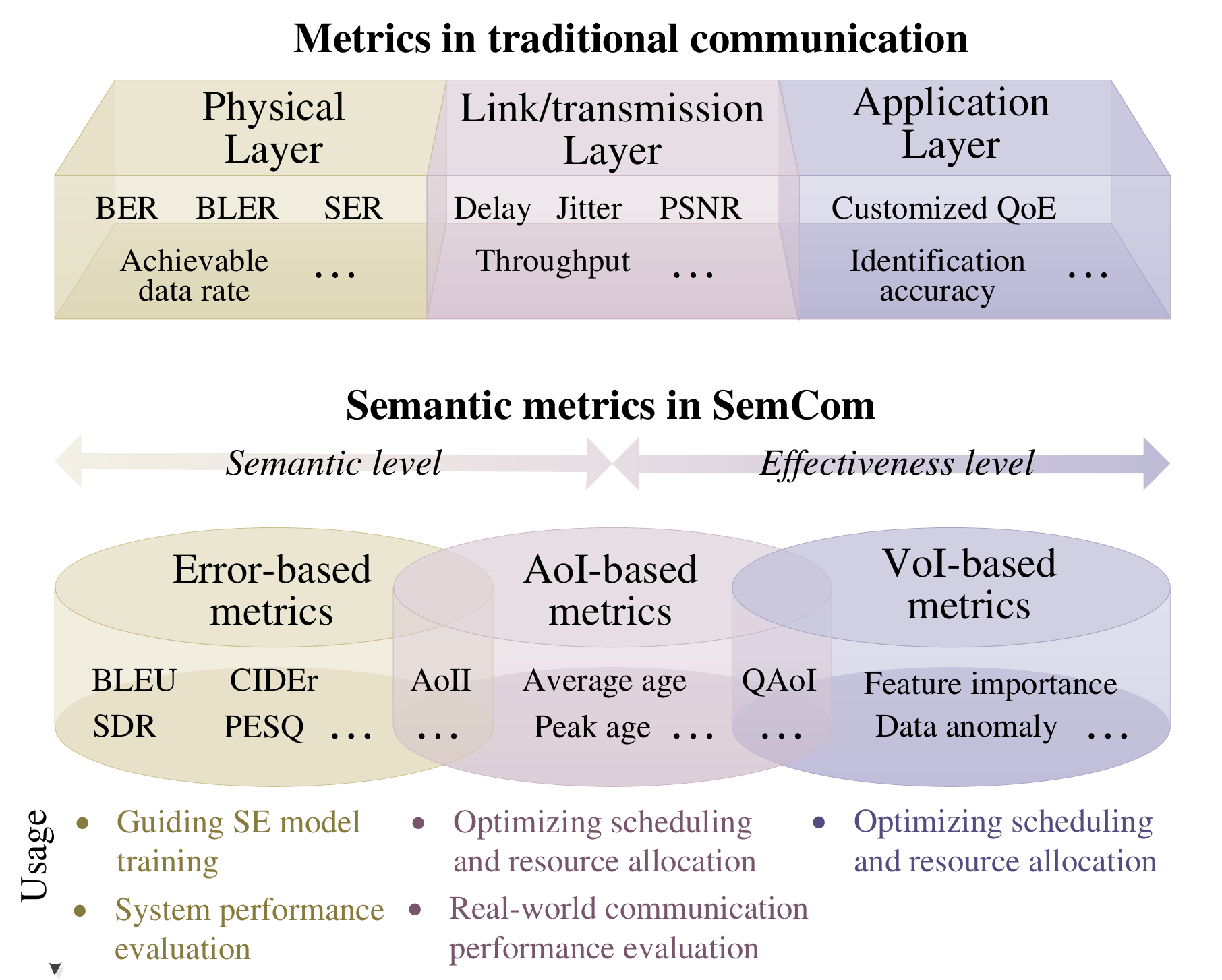}\\
 \caption{{Main types of semantic metrics.}
 }
 \label{metrictype}
\end{figure}

Network performance measure's choices have always been a nucleus concern in network design and optimization for generations. In the conventional communication system, due to the separation of transmission and data's SI and effectiveness for achieving specific goals, the communication performance tends to be evaluated from different network layers through metrics such as BER, QoS, and QoE, respectively. In constrast, in SemCom, the interlayer coupling is enhanced to a great extent~\cite{popovski2020semantic}. Hence, the new methods of evaluating communication performance in terms of semantics must be identified before the implementation of SemCom in practice. Existing SemCom evaluations mainly focus on semantic error, AoI, and VoI. Next, we go into the details about the three basic metric types and their related combined forms as shown in Fig.~\ref{metrictype}, as well as discuss the related remaining issues.

\subsection{Error-based semantic metrics}
\label{sec:error}
As aforementioned, different from the metrics of BER and SER for traditional communications, which are concerned with the accuracy of each bit and each symbol and treat all the bits and symbols as equally important, the error-based metrics in the SemCom care about whether the meaning intended by the transmitter is equivalent to the meaning understood by its destination, i.e., \textit{semantic similarity}~\cite{strinati20216g}. Moreover, the available semantic metrics are all task-specific and there is not yet a general metric for different types of embedding~\cite{qin2021semantic}. Below we discuss the error-based semantic metrics in terms of specific applications.

\subsubsection{Semantic metrics for text data}
\label{text_metric}
At the moment, text transmission has received the most attention in the study of SemCom. Semantic similarity in text transmission usually refers to the exact degree of meaning conveyed by a whole sentence. To mathematically quantify the similarity, some researchers resort to some pioneering works in NLP, and apply the following metrics in the performance evaluation of their works.
\begin{itemize}
    \item \textbf{{Bilingual evaluation understudy (BLEU)}}: Initially, BLEU is a method for automatic evaluation for machine translation~\cite{papineni2002bleu}, which is in line with what semantic measurement needs to do in the SemCom system. BLEU is used to compare $n$-grams of the candidate with the $n$-grams of the reference translation and count the number of matches, where $n$-grams represents the size of a word group. For example, for sentence ``\textit{cat is on mat}." 1-gram: ``\textit{cat}," ``\textit{is}," ``\textit{on}" and ``\textit{mat}," 2-grams: ``\textit{cat is}," ``\textit{is on}"and ``\textit{on mat}". It is first introduced into SemCom in~\cite{xie2021deep}, where the $n$-grams precision score denoted by $p_n$ depends on the difference between the minimum frequency of one element in the $n$-th grams. In this sense, the BLEU for the whole sentence is calculated as the product of the sum of the precision scores for the grams of all sizes and a brevity penalty (BP). The BP is determined by the length of the candidate (recovered) and reference (transmitted) sentences. The longer the candidate sentence is compared to the reference sentence, the lower the BLEU score is. Moreover, to make the ranking behavior more noticeable, BLEU is commonly used in its expression in the log domain.
    \item \textbf{{Consensus based Image Description Evaluation (CIDEr)}}: CIDEr is proposed as an automatic consensus metric of image description quality in~\cite{vedantam2015cider}, which is originally used to measure the similarity of a generated sentence against a set of ground truth sentences written by humans. Hence, it can also be used as a semantic metric for text transmission~\cite{lu2021reinforcement}. Similar to BLEU, the similarity between two sentences is calculated based on the set of $n$-grams presented in it. The difference is that, in CIDEr, not just one reference sentence is considered, but a set of reference sentences. When calculating sentence similarity, it takes into account the similarity between the candidate sentences and all semantically similar sentences in the reference set. 
    \item \textbf{{Sentence similarity}}: Sentence similarity is a new metric initialized in ~\cite{xie2021deep} and~\cite{jiang2021deep} for SemCom based on bidirectional encoder representations from Transformers (BERT). BERT is a state-of-the-art fine-tuned word representation model~\cite{devlin2018bert}, which employs a huge pre-trained model including billions of parameters used for extracting the SI. Fed by billions of sentences, the performance of SI extraction has been demonstrated in~\cite{devlin2018bert}. To this end, the sentence similarity is calculated directly based on the cosine similarity of the semantic features extracted by BERT.
\end{itemize}
It is to be noted that, although the BLEU and CIDEr have considered some of the linguistic laws, such as that semantically consistent words usually come together in a given corpus~\cite{lu2021reinforcement}, they remain at the level of calculating the differences of words between two sentences and have no insight into the meaning of the whole sentence~\cite{jiang2021deep}. In this sense, the metric of sentence similarity is much closer to desired SemCom paradigm, as the well-trained BERT model is sensitive to polysemy ( e.g., the word ``\textit{mouse}" with a different meaning in biology and machine), which allows it to extract information in a sentence level.

On the other hand, the non-differentiability of these metrics reduces their practicality, since they cannot apply to DL-based SE, and the computation complexity of RL-based SE is considerable. Hence, even if both BLEU and sentence similarity have been proposed in~\cite{xie2021deep}, the training pipeline in the DL-based SemCom system still adopts CE loss. Moreover, for sentence similarity, the pre-trained BERT network embedding introduces much more resource consumption in the training process and makes it hard to generalize in other tasks.

\subsubsection{Semantic metrics for audio data}
\begin{figure}[t]
 \centering
 \includegraphics[scale = 0.46]{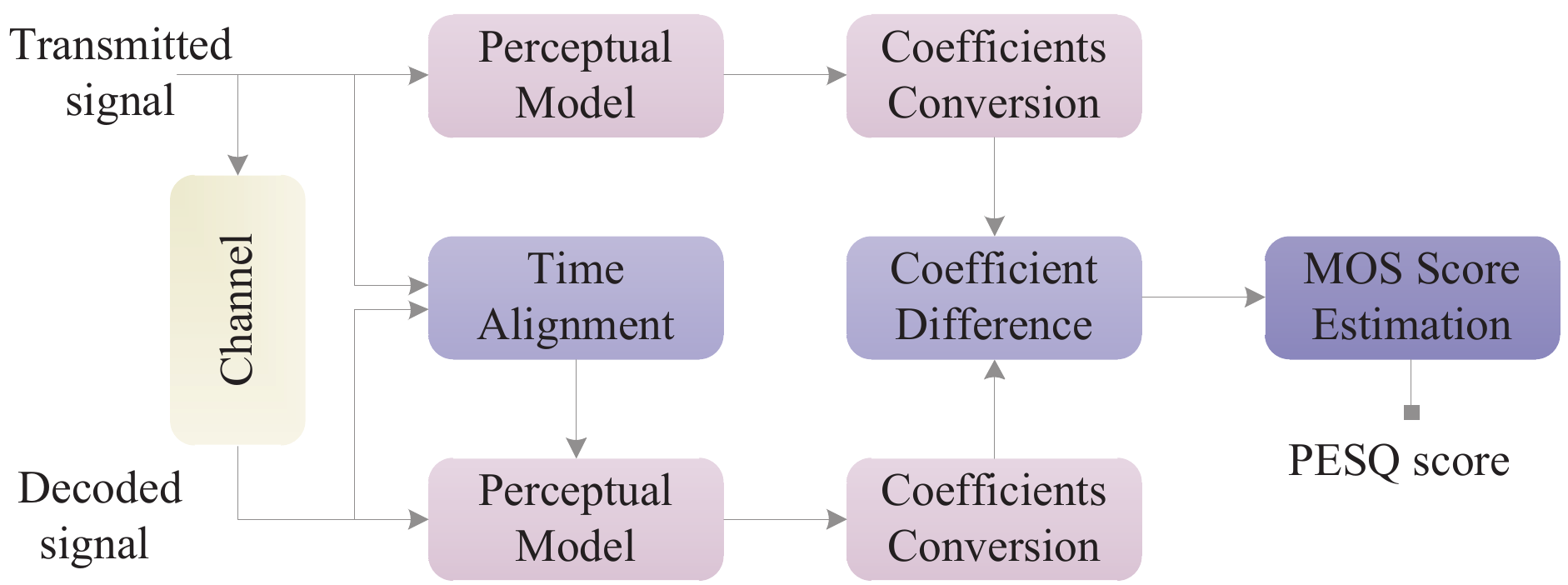}\\
 \caption{Basic diagram of PESQ.
 }
 \label{PESQ}
\end{figure}

Similar to text data, the audio data is also very close to the human natural language. Thereby, for the SemCom for audio transmission, semantic similarity can be explained by the ease with which the receiver understands the decoded audio signal, i.e.,  \textit{intelligibility}. Some similar works have been studied in the field of audio signal processing as below~\cite{vincent2006performance,rix2001perceptual,cox1997three}. 
\begin{itemize}
    \item \textbf{{Signal-to-distortion ratio (SDR)}}: SDR is originally defined based on the usual definition of the SNR with a few modifications in~\cite{vincent2006performance}. In~\cite{weng2021semantic2}, it is introduced into SemCom the recovered signal as a performance metric, which is expressed by the ${\mathcal{L}_2}$ error between the transmitted speech signal and $\mathbf{\hat{s}}$.
  Compared to MSE, the ranking behavior of the difference between $\mathbf{s}$ and $\mathbf{\hat{s}}$ in SDR is more remarkable. Besides, the numerical precision of this measurement is lower for high-performance values than for low-performance ones, which is more intuitive for the design of the measurement method. However, SDR does not go any further than MSE in terms of semantic awareness.
    \item \textbf{{Perceptual evaluation of speech quality (PESQ)}}: PESQ is a specialized quality assessment model designed for speech use across a wider range of network conditions, which has been standardized as Recommendation ITU-T P.862. It is utilized in~\cite{weng2021semantic} and~\cite{weng2021semantic2} to evaluate the performance. It combines the perceptual speech quality measure (PSQM) and perceptual analysis measurement system (PAMS)\cite{rix2001perceptual}. The basic PESQ diagram of PESQ is shown in Fig.~\ref{PESQ}. Different from the above metric which simply compares the difference between the two signals, PESQ assumes the short memory in human perception, which allows it more similar to the human behavior~\cite{weng2021semantic}. However, the method still only looks at the accuracy of the transmission instead of the semantic meaning, and thus it cannot provide effective guidance for semantic compression.
\end{itemize}
In a nutshell, none of the above approaches evaluates the performance at the level of semantic understanding. In addition, only the MSE metric is used in DL for the SE in the existing works. The SemCom has only reached the semantic encoding level so far. A semantic measurement with semantic understanding like BERT and BLEU in the text transmission is still to be studied in the field of audio SemCom.

\subsubsection{Semantic metrics for visual data}
For the communication for visual data, there are no general semantic metrics that are analogous to human perception yet. The commonly-used metrics in the SemCom for visual data are still shallow functions~\cite{zhang2018unreasonable}, such as PSNR~\cite{xu2021wireless} and structural similarity index (SSIM)~\cite{yang2021deep} employed in conventional communication systems. Moreover, compared to the text and audio data, the semantic similarity is more context-dependent, that is, it is hard to distinguish different ``senses of similarity”: is a red circle more similar to a red square or a blue circle~\cite{zhang2018unreasonable}? This imposes challenges to visual semantic metric design. Meanwhile, similar to text and audio data, the similarity judgment for visual data must also depend on a high-order structure~\cite{wang2004image}. To this end, DL-based feature capture can be considered as a potential way to assess the image semantic similarity~\cite{qin2021semantic}. In recent years, the internal activations of deep convolutional networks trained on a high-level image classification task have been considered to be often effective as a representational space for a much wider variety of tasks~\cite{zhang2018unreasonable}. For instance, features from the Visual Geometry Group (VGG) architecture~\cite{simonyan2014very} have been used in other tasks like neural style transfer~\cite{gatys2016image}, and conditional
image synthesis~\cite{dosovitskiy2016generating}. However, how to exploit this approach for SemCom performance evaluation needs to be further explored.
In addition to image transmission which is only aimed at ensuring the fidelity of the visual data, there are also many emerging visual communications for specific tasks like object recognition and attribute classification, wherein the accuracy of task execution can directly characterize the effectiveness of SemCom.
\begin{table*}
\small
\centering
\caption{Error-based Semantic metrics summary.}
\label{metrics}
\begin{tabular}{|c|m{1.5cm} m{5.5cm} m{4cm} m{4cm}|}
\hline
\multicolumn{3}{|l}{\textbf{\qquad Semantic metrics}}& Advantages & Drawbacks\\
\hline
\hline
& BLEU   & BLEU is a method for automatic evaluation for machine translation. It is used to compare word groups of different sizes of the candidate with that of the reference translation and count the number of matches. & It considers the linguistic laws, such as that semantically consistent words usually come together in a given corpus. &
It only calculates the differences of words between two sentences and has no insight into the meaning of the whole sentence.\\ \cline{2-5}

{\raisebox{-1.3\normalbaselineskip}[0pt][0pt]{\rotatebox{90}{{{text data}}}}}& CIDEr & CIDEr is proposed as an automatic consensus metric of image description quality, which is originally used to measure the similarity of a generated sentence against a set of ground truth sentences written by humans.  & Compared to BLEU, it does not evaluate semantic similarity on the basis of a reference sentence, but a set of sentences with the same meaning & Similar to BLEU, it is also based on the comparisons between word groups, and the semantic similarity can only be made at the word level. \\\cline{2-5}

& Sentence similarity         &  Sentence similarity is calculated as the cosine similarity of the semantic features extracted by bidirectional encoder representations from transformers (BERT)~\cite{{xie2021deep}} from different sentences.  & The SI considered in this metric is viewed from a sentence level owing to the sensitivity of BERT to polysemy. &  BERT  is a huge pre-trained model, which introduces much resource consumption in the training process and makes it hard to generalize in other tasks                         \\
\hline
\hline
{\raisebox{-4.3\normalbaselineskip}[0pt][0pt]{\rotatebox{90}{{{audio data}}}}} & SDR & SDR is expressed by the ${\mathcal{L}_2}$ error between the transmitted audio signal and recovered audio signal.  & The numerical precision of SDR is lower for high performance values than for low performance ones.  & SDR fails to capture the hidden SI  of the speech signal, without any further than MSE in terms of semantic awareness. \\\cline{2-5}
&PESQ & PESQ is a specialized quality assessment model designed for speech used across a wider range of network conditions, which has been standardized as Recommendation ITU-T P.862. & Instead of comparing the differences between the two signals directly, PESQ assumes the short memory in human perception. & PESQ still focuses on transmission accuracy, and thus it cannot provide effective guidance for semantic compression. \\ \hline
\end{tabular}
\end{table*}

\subsection{AoI-based semantic metrics}
\label{sec:AoI}
The difference between semantic information in communication and that of other fields such as semantic web and semantic segmentation lies in its emphasis on time sensitivity. This feature introduces new dimensions to the accuracy of semantic information, i.e., \textit{the right time}~\cite{uysal2021semantic}. Especially for some applications, such as location
tracking, control, and situational awareness, the freshness of information has a significant influence on the action execution at the receivers. In this regard, some metrics focusing on timeliness are required. 

In fact, taking AoI into account in performance evaluation can be considered as an initial attempt at SemCom~\cite{maatouk2020age}. 
 Different from the metric of delay, which primarily measures the transmission performance without concern for the content of the packets, AoI-based metrics are utilized to quantify the staleness of the information received at the destination.
The age of a packet is defined as the difference between the current time and the timestamp of the packet~\cite{kaul2012real}, which captures how unfresh the data received by the monitor is. In the traditional content-blind communication paradigm, the systems just pursue to send updates as fast as possible and ensure the minimum transmission delay. 
Undoubtedly, this requires a lot of bandwidth resources. Moreover, if the delay-QoS cannot be guaranteed, the backlog of the packets in the communication system throttles the update and leads to a monitor having unnecessarily outdated status information. Fortunately, such issues can be addressed by the scheduling scheme based on AoI minimization. This is attributable to the fact that the fresh data can be given more importance and transmitted with priority during the scheduling process. 

Moreover, owing to the stochastic features of environments, an appropriate analysis of AoI can be chosen according to the specific system, such as \textit{time-average age} and \textit{peak age}, which are presented in detail in~\cite{yates2021age}.
However, it should also be noted that there are still inherent flaws for AoI-based metrics, that is, they disregard the validity of the recovered data. For example, in some cases, the monitor is only concerned with the abnormal and abrupt states at sources~\cite{sun2019sampling}. Since AoI does not consider the value of current states for its monitor, some useless updates are transmitted to the monitor, which also results in a certain amount of resource waste.

\subsection{VoI-based semantic metrics}
\label{sec:VoI}
VoI is also a newly introduced metric in communication systems, especially for networked control systems. Before that, the concept of VoI is well-known in information analysis wherein it is defined as the price that a decision maker is willing to pay for taking the information into account~\cite{howard1966information}. For conventional communications, VoI can be defined as a measure of uncertainty reduction from the information set of the source with a successful transmission~\cite{ayan2019age}. In contrast, for communications with specific tasks, the VoI needs to be redefined. In contrast to AoI which just focuses on the timeless and ignores content, VoI is mainly utilized to measure the relevance of a piece of information to the communication tasks. In other words, VoI can be regarded as the quantified contribution of SI to effectiveness.

Take a remote temperature control system~\cite{maatouk2020age} as an example. In that setting, the central controller is not concerned with the real-time temperature variation of the sources. The goal of the system is only to make sure the controller reacts swiftly to any abnormal temperature rise. In this sense, the data for the abnormal temperature should be assigned with high VoI. Moreover, in~\cite{yang2021semantic} wherein a task of image classification is studied, the VoI here is used to measure the importance of the extracted features to the accurate classification of the images. {However, in most cases, the VoI can only be known after the task is completed. Here we revisit the two examples of semantic-aware communications in \cite{lotfi2021semantic,yun2021attention}. The VoI of SI in the federated DRL task with multiple heterogeneous agents~\cite{lotfi2021semantic} is the reduction in convergence time as well as the increase of the long-term reward after convergence. Moreover, the VoI of SI in URLLC-enabled UAV swarm collaboration~\cite{yun2021attention} is an improvement in convergence rate and reduction of inter-UAV collision probability. Therefore, for a general task, the VoI is challenging to be quantified as the AoI before communications, because the VoI is largely determined by a combination of multiple factors in the communication context. Therefore, the available VoI-based metrics are scarce, and  the VoI-based scheduling or resource allocation scheme only performs for simple tasks, such as abnormality monitoring.}

\subsection{Combined semantic metrics}
\label{sec:Combined}
As aforementioned, the above three types of semantic metrics only focus on one attribute of the information conveyed by the recovered data.
To address this limitation, burgeoning research efforts have been investigating new semantic metrics that combine multiple attributes to varying degrees~\cite{uysal2021semantic}. 

\begin{figure}[t]
 \centering
 \includegraphics[scale = 0.8]{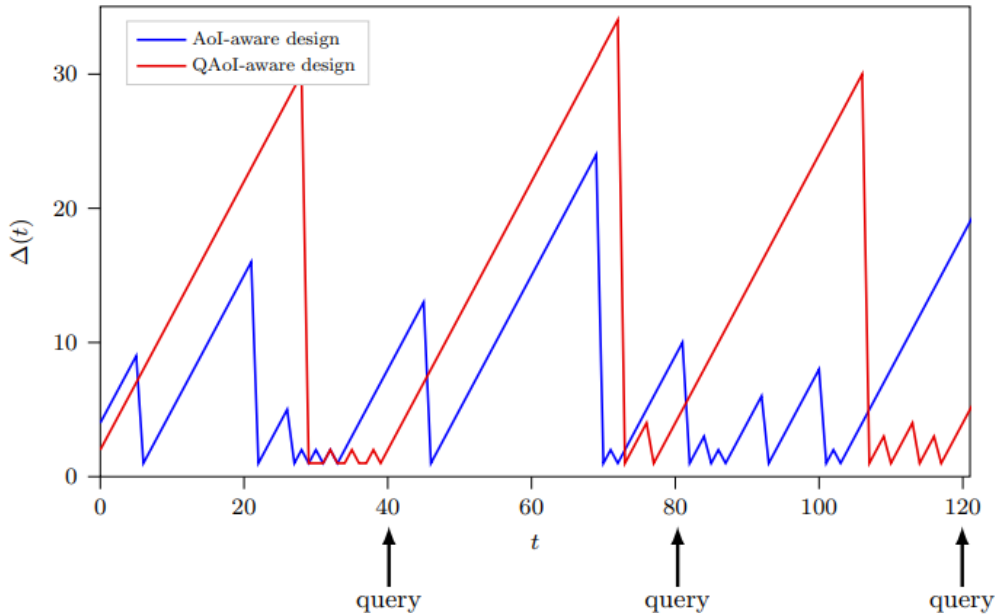}
 \caption{AoI dynamics of the PQ and QAPA policies, where PQ denotes a permanent query system, which minimizes the traditional AoI, and QAPA denotes a query arrival process aware system, which minimizes the QAoI~\cite{holm2021freshness,uysal2021semantic}}
 \label{QAoI}
\end{figure}

The authors in~\cite{maatouk2020age} integrate AoI into error-based metrics, and propose a new metric named age of incorrect information (AoII). AoII characterizes the impact of the prolongation of one inaccurate state on semantic recovering. Compared to both error-based and AoI-based metrics above, AoII is incorporated with more meaningful semantics by jointly considering the content and timeliness. Specifically, AoII considers not only the repercussions of a transient state on the overall communication goal, but also the repercussions of the states lasting for different duration. For instance, the repercussions of a long burst of error are far more severe than an instantaneous burst of error for video transmission~\cite{liang2008analysis}. The comparison of the AoI and AoII is shown in Fig.~\ref{metrics}. Meanwhile, the authors in~\cite{holm2021freshness} integrate VoI into the AoI-based metrics by considering a pull-based system and propose a new metric called age of information at query (QAoI), which reflects the freshness of the instants when the receiver actually needs the data~\cite{uysal2021semantic}. In a pull-based system, the information is effective at the receiver only at certain query instants. In this sense, the communication is expected to be query-driven, that is, the transmitter knows the query instants and optimizes the transmissions with respect to the timing of the query process. Hence, the query-driven QAoI-based on scheduling scheme is more efficient and effective for such a system than the query-blind AoI-based one to achieve sending just before the query instants. As shown in Fig.~\ref{QAoI}, the QAoI-based scheme is more likely to provide updates that are fresh when a query arrives, although its average AoI may be worse than the AoI-based scheme~\cite{holm2021freshness}.
\begin{figure}[t]
 \centering
 \includegraphics[scale = 0.6]{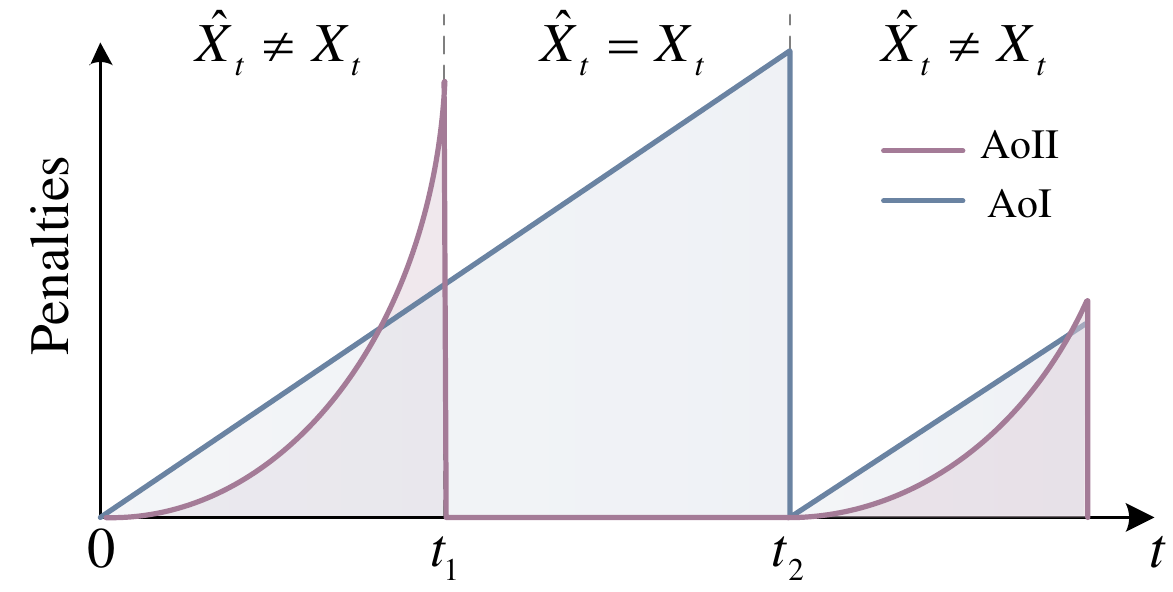}\\
 \caption{ Comparison of AoI and AoII, where $X_t$ denotes the transmitted information, and $\hat{X}_t$ denotes the estimated information inferred from the transmission~\cite{maatouk2020age}.
 }
 \label{metrics}
\end{figure}



\subsection{Lessons learned summary}
\subsubsection{Lessons learned for error-based semantic metrics}
{Overall, research on SemCom performance evaluation methods is just beginning and is lagging behind research on SE and SI transmission. Compared to the metrics, such as BER, and SER, used in traditional communications to quantify transmission accuracy, the above semantic error metrics avoid the shallow bit-by-bit, symbol-by-symbol rigid comparison. When comparing the difference between the received signal and the transmitted signal, these semantic metrics further take into account the characteristics of the data type itself, such as the syntactic structure of the text or the positive impact of the short memory in human perception on speech signal intelligibility. However, most of the above metrics are still comparing the raw signals sent and recovered, albeit from a more personalized and specialized perspective, rather than a direct comparison of the SI those imply. An ideal semantic metric might be similar to sentence similarity proposed in \cite{xie2021deep} and~\cite{jiang2021deep}, and also implies an inevitable consumption of computational resources and time for the pre-training process of the relied AI model.}

{More significantly, it is important to point out that all the existing semantic metrics are \textit{intrusive},  which involves a reference signal to accomplish the performance evaluation. Since a clean reference signal is always unavailable in real-world communications,  these metrics can only play a role in the training phase of SE models based on existing databases. Meanwhile, as discussed in~Section~\ref{RL-based SE} and Section~\ref{text_metric}, the non-differentiable, complex form of these metrics makes the training process more challenging. Hence, the practicality of these metrics is limited. Moreover, a non-intrusive semantic measurement for real-world testing, which can  blindly and intelligently estimate the objective scores, such as BLEU, and PESQ,
remains to be studied.} 
\subsubsection{Lessons learned for combined semantic metrics}
{All in all, when conducting performance evaluation based on the above three types of metrics, the data is given different importance.  Since the error-based type of semantic metrics aims at comparing the meaning conveyed by the transmitted and recovered messages, it can be treated as the metric designed for semantic-oriented communication. In contrast, VoI-based metric type goes beyond the accuracy comparison from the semantic level, and instead directly evaluates the relevance of the information to task performance from the effectiveness level. In this sense, VoI-based metrics are for goal-oriented communications. Moreover, 
the AoI-based metrics fall between the error-based metrics and VoI-based metrics, as both the meaning of the message and its impact on task execution may change over time. Combining AoI-based metrics with error-based metrics, the relevance of information can be further differentiated in terms of the time dimension.
In an ideal SemCom system, the combined metrics are expected to play a role in resource allocation to guide the filtering of irrelevant information and enhance system efficiency and performance. However, this idea has only been explored in the simplest pull-based system, and the reason is that the quantification of SI has not been achieved, and the most error-based metrics and VoI-based metrics are challenging to estimate before transmission. }

\section{SemCom for future 6G Internet}
In this section, we first discuss some potential applications for SemCom in 6G. Then, we motivate a promising SemCom-empowered network architecture in 6G.
\subsection{Potential applications for SemCom in 6G}
\label{sec:application}
\subsubsection{Intelligent transport system}
In recent years, with the development of hardware in vehicles and vehicular infrastructures, vehicles can be considered as intelligent agents with greater computing, caching and data storage capacities~\cite{li2021adaptive,ye2021joint}. This paves the way for future 6G intelligent transport systems (ITS), wherein autonomous driving and cooperative vehicle networks can be achieved without the need for human involvement~\cite{de2021survey}. {In most of the existing works, in order to enhance safety, improve assisted driving decisions,  or manage vehicles, the basic information about the vehicles and road, such as locations, braking intensity, potholes, and water puddles, is required to be broadcast periodically~\cite{bista2018semantic}. Such a non-differentiated vehicle communication for different situations can affect communication effectiveness and efficiency. To this end, SemCom has a great potential to make ITS more intelligent. }

{In ITS, the most straightforward application of SemCom is to extract the essential semantic information from the raw sensor data, such as vehicle kinematic information, road conditions, and traffic signs (as shown in Fig.~\ref{sematic-orientedexample}). For instance, there are many situations, such as a sudden appearance of a human and a sudden collision happening in front of a vehicle, that have a similar effect on vehicle driving. Therefore, by extracting the SI of a particular situation (i.e., quantifying the impact of each situation on driving),  the accuracy and timeliness of the transmission can be greatly improved because of the reduced amount of data. In addition to the compression of the data itself, as discussed in Section~\ref{sec:AoI}, the time point of data sampling is also critical in  SE. For example, when a vehicle has a sufficiently long visibility range or sufficient viewing duration along every direction, it can make the driving decision based on its own local. In this case, the driving information about other vehicles is not required to be exchanged frequently. In contrast, if a vehicle's visible duration is very short, then when the vehicle gets the vision, it may not have sufficient information or time to make a correct reaction. In such a case, information exchange needs to be more timely. More specifically, the essential information about safety, such as braking intensity, should be given a higher priority~\cite{nanda2019internet}. Moreover, as discussed in Section~\ref{semantic-aware}, to monitor the traffic flow of walled subdivision, the SI about a continuous view over the subdivision exit can be shared with the vehicles about to pass the subdivision. To this end, determining when to communicate, and what information to share in a ``quantitative" manner is exactly what SemCom is required to do.  }

\subsubsection{Distributed learning based applications}
 With increased computing power on end devices and growing privacy concerns of users, distributed learning, such as federated learning (FL) has become a dominant paradigm for privacy-preserving machine learning~\cite{samarakoon2019distributed}. It has penetrated all aspects of human's lives~\cite{zhang2021optimizing}, such as medical diagnosis, cyberattack detection, and BS association. 
As deep neural networks usually contain millions of weight parameters, the frequent exchanges of DNN models or gradients between terminals and servers incur costly communication overheads, which poses challenges for communication networks, especially for the uncertain wireless environment and limited wireless resources. 

{Fortunately, SemCom can reduce unnecessary communication overhead in two ways, thereby improving performance within limited wireless resources. Firstly, the model parameters and gradients can be compressed in a semantic-aware manner, such as gradient sparsification~\cite{amiri2020federated} and model parameter pruning~\cite{jiang2019model}, where a subset of the original model parameters is extracted considering the semantics or importance of the parameters for model accuracy and convergence speed. For example, in~\cite{amiri2020federated}, gradient sparsification is adopted to compress the model at the transmitter by setting all but $k$ elements with the highest magnitudes of entries to zero. Since only the positions of the non-zero elements are to be sent, the receiver can recover the received data in a more reliable manner with advanced noisy measurements. Secondly, as per the two examples of semantic-aware communication in Section~\ref{sec:spse}, the agents in distributed learning can exchange their own semantic features via SemCom to enhance their knowledge of each other. The semantic features can be extracted from their learning models~\cite{lotfi2021semantic}, their observable environments~\cite{yun2021attention}, their missions, and so on. Based on the semantic features of small data volume, an optimal subset of semantically-related agents for collaboration can be found. Therefore,   the exchange of models with large data volumes between irrelevant agents can be effectively reduced.}

\subsubsection{Unmanned aerial vehicles}
UAVs have attracted lots of attention especially while being served as aerial base stations (BSs) or as relays~\cite{hayat2016survey,cheng2019space,zhou2020deep}. Unlike static ground BSs or relays, UAVs can be flexibly deployed to satisfy various QoS requirements and balance load amongst users. Moreover, UAV swarms collaboratively can complete missions with greater efficiency and economy as compared to single UAV
systems. However, the energy constraints of UAVs impede their ability to facilitate long-term communication. Meanwhile, the collision problems in UAV swarm navigation have also been a key concern in the studies of UAV networks.

Fortunately, because  SemCom can reduce the amount of information that needs to be transmitted, an efficient communication framework among UAVs can be implemented. For example,  when the UAVs are being served as relays, diversity gain can be achieved with the help of cooperative communication protocols, such as decode-and-forward and amplify-and-forward~\cite{laneman2004cooperative}. {In this sense, a novel semantic process-and-forward method is motivated to be proposed to cater to SemCom. In addition to the traditional relay function, the UAV can be deployed as a semantic encoder and/or decoder. For example, when one of the sending and receiving parties cannot enable SemCom due to insufficient memory or computing power, UAV can carry out encoding or decoding instead to reduce the data volume of a certain link without compromising communication performance.  Certainly, this requires the UAV to know the background knowledge of both sides of the communication. This also poses new challenges for the joint optimization of communication, computing, and caching resources. Moreover, in the case where both parties can enable SemCom, the UAV can perform the semantic decoding for the received signal based on the background knowledge of the transmitter, and then it can re-code the signal based on the background knowledge of the receiver. This greatly reduces the overhead of synchronizing background knowledge for sending and receiving and the semantic noise caused by unsynchronized background knowledge~\cite{luo2021autoencoder}.} Moreover, as discussed in Section~\ref{sec:spse}, SemCom can also play an essential role in UAV swarm navigation. As shown in~\cite{yun2021attention}, by introducing SemCom, the UAV swarm navigation based on graph attention exchange network can achieve  $6.5$x lower latency with the target $10^{-7}$ error rate compared to the state-of-the-art CTDE based method. 

\subsubsection{{Extended reality}}
{
Advances in 6G network technologies provide technical support for next-generation Internet services. In particular, the possibility of synchronization of the physical and virtual world through extended reality (XR) has led to the birth of Metaverse, which has been dubbed as the successor to the Internet. The performance of XR is heavily dependent on the collection and processing of data that reflects or describes human movements and changes of surroundings, e.g., shifting rendered targets, displaying particular videos, and giving the corresponding tactile feedback. To guarantee the ideal immersive Metaverse service experience for users, the end-to-end latency and data rates requirements have to be strictly met~\cite{du2022exploring,liu2022slicing4meta}.

To this end, SemCom can be seen as an enabler for XR-based Metaverse access~\cite{ismail2022semantic}. In the SemCom paradigm, the data tracked by the end devices, such as head movement, arm swing, gestures, and speeches, need to be extracted semantically first. This allows the end device to transmit the information concerned by the XR server for operation after understanding and filtering out the irrelevant information to save bandwidth and reduce computing latency at the XR server. Meanwhile, the XR server can also extract SI based on the user's preference, ignoring irrelevant details in the face of bandwidth constraints, thus reducing downlink pressure.}

\subsubsection{{ Holographic telepresence (HT)}}
{As another technique to deliver next-generation services to users, holographic telepresence (HT) can project realistic, full-motion, real-time three-dimensional (3D) images of distant human beings or physical objects with a high level of realism rivaling of the physical presence~[177][191]. It can be applied not only to virtual conferencing, virtual games, and the realm of entertainment, but also to remote repair, and remote surgery~[177][192]. However, like the immersive XR application, to ensure a real-enough virtual and seamless service experience, HT also requires stringent QoS.} Moreover, almost all the human senses, such as smell and taste, are expected to be transmitted through future networks for a fully immersive experience. 

For such communication-and-computation intensive services, the traditional content-blind communication paradigm results in a waste of bandwidth resources and computing resources. To this end, the understanding-before-transmission paradigm of SemCom can be regarded as a promising method to alleviate the pressure of bandwidth and the receiver's end processing. This is attributable to the fact that SemCom allows the SI extraction that is adaptive to network conditions and enhances transmission reliability, so that changes in the network conditions cannot be perceived by users, thus ensuring high quality service experience.


\subsubsection{Personalized body area networks}
Personal data management as well as transmission of wearable devices are future trends that will affect how personal services and procedures develop. An important element is the wireless body area network (WBAN). Defined formally by the IEEE $802.15$ (Task group $6$) as a communication standard optimized for low-power devices, the WBAN can serve a variety of applications such as medical, consumer electronics, and personal entertainment~\cite{movassaghi2014wireless}. Because of the energy constrained power supplies of tiny sensor nodes, effective energy consumption is a key challenge in WBAN. SemCom prompts us to think about whether we can save energy and increase the lifetime of wearable devices by reducing the number of actual bits transmitted. In WBAN, it has been shown that the on-board extraction of features on modern low-power wearables is both feasible and beneficial for system lifetime improvement~\cite{elsts2018board}. Although a resource-constrained sensing system needs to strike the balance between the accuracy of the semantic features output and the cost of analyzing the data for extraction, the benefits from reducing the radio duty cycle which is used for transmission, vastly outweigh the cost of increasing the processor duty cycle which is used for semantic features extraction~\cite{zalewski2020bits}. As knowledge extraction from the raw data can significantly reduce the information that needs to be transmitted, the SemCom-based method increases the lifetime of the wearable device by one order of magnitude, at the cost of approximately 5\% degradation of classification accuracy~\cite{zalewski2020bits}. The development of SemCom and the deeper integration with WBAN will give rise to longer-lasting and more convenient wearable devices.

\subsubsection{Collaborative robots}
A group of cooperative robots can explore, interact with, and perceive the environments far more efficiently than a single robot working alone~\cite{bragancca2019brief}. 
In scenarios such as disaster management, warehouse automation, and surveillance, the application of collaborative robots is rapidly increasing. However, the limited computation capability of each robot limits the widespread deployment in computation intensive tasks~\cite{yue2020collaborative}. One promising solution is to apply the SemCom techniques to achieve
efficient data exchanging and processing. SEMIoTICS, a new
SemCom-based control system architecture is proposed in~\cite{milis2017semiotics}, which enables the utilization of logic-based reasoning over declarative language models to reduce the decision-making time. In \cite{milis2017semiotics}, SEMIoTICS is deployed in a building that consists of 15 IoT components for temperature regulation control. The results show that the overall control processing time can be maintained in $6$ minutes, which is only $24\%$ of that of the traditional fuzzy logic control-based method~\cite{kolokotsa2003comparison}. To further reduce the communication overhead among collaborative robots, a lite distributed SemCom system, named $L$-DeepSC~\cite{xie2020lite}, can be used. When the communication environment among collaborative robots experiences low SNR, $L$-DeepSC can enable efficient information exchange. In particular, with $L$-DeepSC, the amount of data needed to interact among robots can be compressed to $2.5\%$ of the information that is needed by the traditional method.
\subsubsection{Hyper-intelligent IoT}
Hyper-Intelligence (HI) refers to higher- and super-intelligent abilities to accomplish complex tasks. The combination of HI and IoT will lead to a smarter and data-driven society~\cite{da2014internet,gao2021mac}. A general and reasonably predictable trend in the next few years will be a rise in the native intelligence of networks, network nodes, and linked devices. Devices that are formerly utilized merely as sense-and-transmit entities will be endowed with various levels of embedded intelligence that operate directly on the data acquired. The necessity for progressively smarter communication parties opens the door to the creation of ``smarter" content to exchange and reason, as well as innovative methods for ensuring that it is done effectively and accurately~\cite{popovski2021internet}, which coincides with the booming development of SemCom. With the help of SemCom, only the most useful data is transmitted, and therefore the communication effort is optimized. To deploy the SemCom techniques in HT IoT to promote  efficiency, a potential solution would be to consider whatever enables the receiver to effectively execute a given task, while relying on SemCom that extracts only the necessary information from the data. By giving HI the ability to process and reason information at the semantic and even effective level, the HI IoT will be more connected, further advancing the construction of an interlinked and connected society.
\subsection{SemCom-empowered 6G architecture}\label{Sec:6G Internet}
\begin{figure*}[t]
 \centering
 \includegraphics[scale = 0.46]{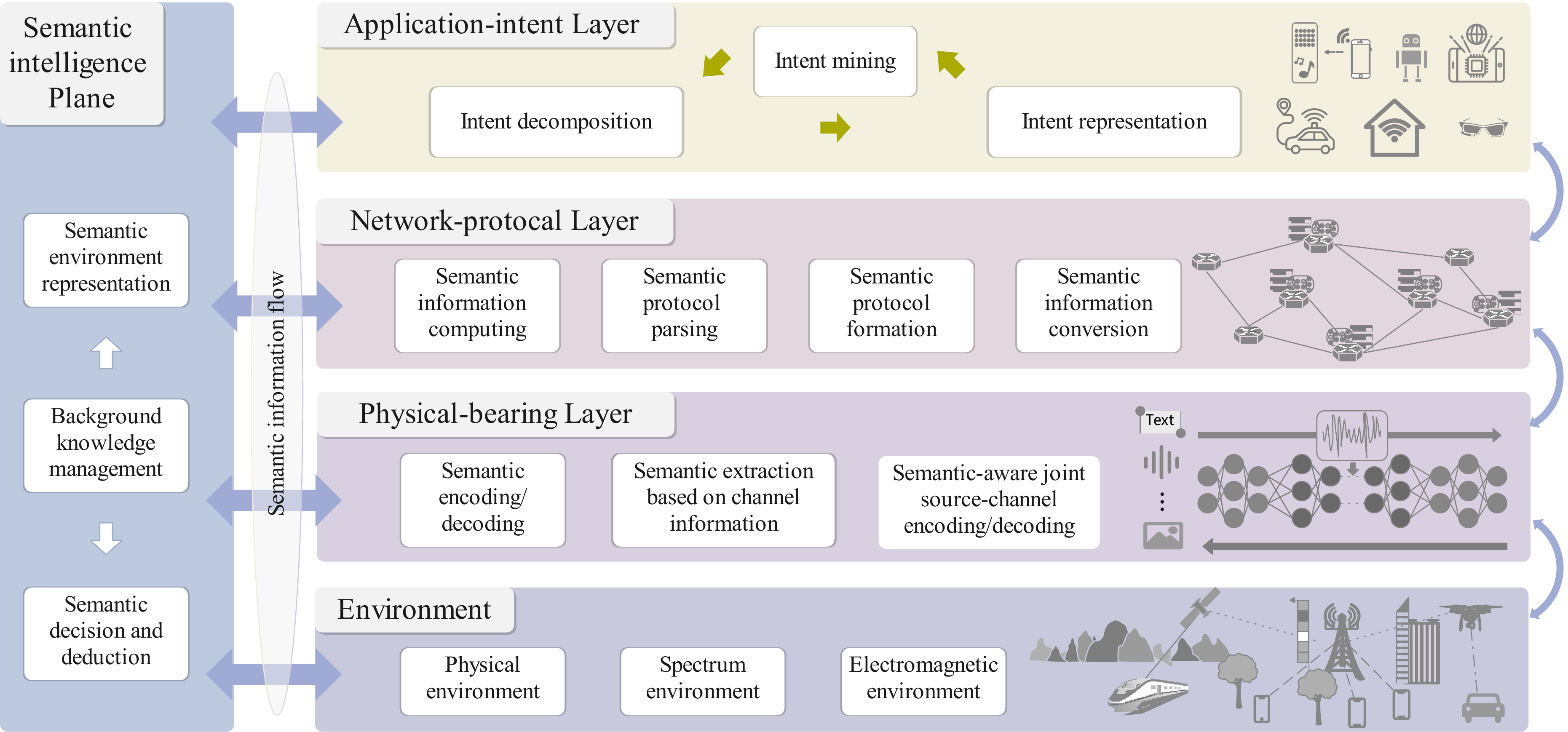}\\
 \caption{Semantic communication-empowered 6G architecture~\cite{zhang2021toward}.
 }
 \label{architecture}
\end{figure*}

In the traditional communication network, the network nodes are not concerned with what the data is trying to convey. The information exchanged between the inter-and intra-layer nodes in the network can be generally seen as homogeneous bit sequences. However, in the SemCom architecture with ubiquitous consciousness, the information representation can achieve a higher level. In line with the concept of ``Bit" in Shannon’s information theory, the authors in~\cite{zhang2021toward} introduce a new concept called ``Seb" for SI.  A SemCom system dimensioned by Seb is highly modulated compared to traditional communication systems. 

{In analogy to the construction process, a bit-flow-based network architecture is like a building constructed in a brick-by-brick manner, while a Seb-flow-based network architecture is similar to the building constructed by the laminboard and integrated window or door~\cite{zhang2021toward}. 
As the integration of raw materials simplifies the construction process, the SE of raw data also calls for a simplified network architecture to efficiently support SemCom.}

To this end, the authors in~\cite{zhang2021toward} propose a novel intelligent and efficient SemCom (IE-SC) architecture, as shown in~Fig.~\ref{architecture}. {In contrast to the well-known seven-layer open systems interconnection (OSI) model,} their proposed IE-SC architecture divides the network into three semantic-empowered layers: semantic application-intent (S-AI) layer, semantic network-protocol (S-NP) layer, and semantic physical-bearing (S-PB) layer. Meanwhile, the three layers, as well as the physical environment, are coordinated by a separated semantic intelligence plane via the semantic information flow (S-IF). In the following, we first introduce the three main functions of the semantic intelligence plane. Then, we present the other three layers with examples in detail.
\subsubsection{Semantic intelligence plane}
As a coordinator of the entire IE-SC network, the semantic intelligence plane contains three main functions:
\begin{itemize}
 \item \textbf{Semantic environment representation}: {This function deals with the extraction of SI from the raw data and communication context based on the semantic intent information provided by the S-AI layer, which is detailed in Section~\ref{S-AI}.}  Meanwhile, the semantic intelligence plane generates the semantic extraction components, {such as SemCom protocol design and semantic encoding and decoding,} and maps them to the functions of each layer. Meanwhile, SI is embedded into the S-IF to be transferred in the network through the semantic intelligence plane and the interfaces of different layers.
 \item \textbf{Background knowledge management}: {Just as human communication requires knowledge of each other's language and culture, SemCom requires that both communicating parties have the same background knowledge, which lays the foundation of the intent analysis of the S-AI layer.  The semantic intelligence plane serves as a coordinator, which performs the synchronization, integration, and storage of background knowledge.}
 \item \textbf{Semantic decision and deduction}: The semantic intelligence plane is also responsible for evaluating the achievable performance according to the results of the intent analysis fed by the S-AI layer and performing decision-making for all network layers.
\end{itemize}

\subsubsection{Semantic application-intent layer}
\label{S-AI}
{In the traditional OSI model, the application layer mainly allows users to download data from storage or dataset, or send data to a destination regardless of what and how the data is used. In contrast, the S-AI layer adds the function of intent analysis for a particular communication.} It can support  decomposing and translating the users' intent into the network’s deployment, configuration, or control policies.  The three main functions within the S-AI layer are highlighted below.
\begin{itemize}
 \item \textbf{Intent mining}: The S-AI layer is responsible for extracting, analyzing, aggregating, and synthesizing the original intents received from users or applications. {For example, for an image, some requesters may use it for target identification, some may be concerned with local details, and some may be interested only in image quality.  Different intents may correspond to different communication processes.}
 \item \textbf{Intent decomposition}: {In order to facilitate the subsequent SE from raw data,} the obtained intent after intent mining is decomposed into a set of sub-intents, {which may be the attention of the requester to different paragraphs, sentences, words in a text, or the importance of different features of an image. Moreover, recall an application in ITS, the intent of which is to provide vehicles that are about to pass through a walled subdivision with information about the traffic flow of the subdivision exit. During the intent decomposition, the sub-intents can be a series of continuous views of the subdivision exit.
 For guiding the implementation of the entire communication process, the sub-intents can be informed to the S-NP layer and S-PB layer and used to design protocols and SE, respectively.}
 \item \textbf{Semantic representation}: After obtaining the sub-intent set, the S-AI layer 
 gives the semantic representation of the sub-intents.  {The semantic representation is passed to the semantic intelligence plane, facilitating SE. }
\end{itemize}
\subsubsection{Semantic network-protocol layer}
The S-NP layer aims to efficiently serve the intents of upper-layer applications with intelligent network protocols. The design of this layer mainly concentrates on the strategies of semantic interaction, which includes experience accumulation for learning (e.g., multiple rounds of dialogues), real-time knowledge sharing, and simplified semantic interaction. In this sense, some key modules should be included in the S-NP layer:
\begin{itemize}
 \item \textbf{SI computation}: {In traditional communication, most protocols mainly focus on the destination address and port, and ignore the content of the data. However, in the SemCom-enabled network architecture, if the intermediate forwarding nodes still treat all data equally, it would compromise the analytical significance of the application.} To this end, this module is for identifying the intent information on the S-IF and obtaining knowledge from other modules. 
 \item \textbf{Semantic protocol parsing}: This module is used to analyze the functions available for the existing protocol.
 \item \textbf{Semantic protocol formation}: This module is responsible for optimizing the existing protocol or forming a new one to adapt to the intent of the application.
  \item \textbf{SI conversion}: {In the traditional network architecture, the protocols of the different network layers have individual frame encapsulation, packet encapsulation, segment encapsulation, etc. Similarly, in SemCom-enabled protocols, SI needs to be encapsulated separately.} This module is responsible for completing the SI encapsulation based on the generated protocol.
\end{itemize}
{Take the immersive AR application as an example, which is the application with the coexistence of video, audio, and haptic data and requires high bandwidth and ultra-reliable low latency. Suppose that by interacting with the semantic intelligent plane, the S-NP layer can map the sub-intents, such as the attention to different sound sources and different objects in the visual field, as well as the haptic data with highly critical latency requirements,  to the data within the S-IF. For the haptic packets, the traditional design of the medium access control frame usually reserves dedicated resources for ensuring that the delay QoS requirement is met. However, the request of the haptic packets is random and intermittent, which inevitably leads to low resource utilization or transmission failures. In order to address the above issues, a semantic protocol should be designed to coordinate the transmission of the data corresponding to different sub-intents. For example, due to the tightest time delay requirement, the haptic data are given the highest priority when being forwarded. Moreover, when the network is under heavy load, the redundant data with less attention from the user can be detected and removed adaptively, and resources can be freed for delay-sensitive traffic~\cite{popovski2020semantic}.}

\subsubsection{Semantic physical-bearing layer}
The S-PB layer is responsible for converting SI from the upper layers into physical signals. Different from conventional communication, SemCom aims to deliver SI instead of raw data, so the traditional source encoding/decoding and channel encoding/decoding are to be replaced. {Three possible types of semantic-related modules employed in the S-PB layer are listed below}:
\begin{itemize}
 \item \textbf{Semantic encoding/decoding}: Following the modular design method, the modules of semantic encoding and decoding are designed separately from other modules, such as channel coding. { For example, as presented in Section~\ref{sec:KB-assistant}, the authors in~\cite{yang2021semantic} propose a pair of semantic encoder and decoder based on a well-established KB. Meanwhile, the channel encoder and  decoder are still performed in the same way as in traditional communications.}
 \item \textbf{Semantic encoding/decoding based on channel information}: Integrate the channel information in terms of fading,  SNR, and interference into semantic encoding and decoding. {As discussed in Section~\ref{DL-based SE}, in~\cite{xu2021wireless}, the authors apply SNR to SE by introducing the attention module, which allows the SE method to be adaptive to the changes in channel gain.}
 \item \textbf{Semantic-aware joint source-channel encoding/decoding}: Following the integrated design method, channel encoding and decoding can be jointly integrated into the modules of semantic encoding/decoding. {In fact, most of the existing end-to-end SE methods~\cite{weng2021semantic,xie2021deep,lee2019deep} discussed in Section~\ref{DL-based SE} and Section~\ref{RL-based SE} belong to the type of semantic-aware joint source-channel encoding/decoding.}
\end{itemize}

\section{Future directions}\label{Section7}
In the previous sections, we review the potential SemCom applications in 6G and the state-of-the-art  techniques applied in SemCom.  In addition to the remaining issues discussed in Section~\ref{sec:4}--\ref{semantic metric}, several other SemCom-related directions can be explored further in terms of  system effectiveness, sustainability, and trustworthiness.

\subsection{Interpretability and explainability of SE}
    The communication environment is always experiencing a variety of uncertainties, such as unexpected changes in the network environment or completely new source information. The black box nature makes the SE model unpredictable for  the output corresponding to uncertain inputs in practice, which restricts the SE model's social acceptance and practicality, as well as leaves little basis to use as a guide for SE model optimization. Meanwhile, the available SE models are with little or no understanding of how and why the internal states in the hidden layers and the features contribute to a given example to produce a decision or outcome~\cite{montavon2018methods}, which fails to give  valuable insights into the design of SemCom systems and the SI transmission. Therefore, the issues related to interpretability and explainability  of SE have to be addressed.   
    
    As defined in~\cite{kim2016examples}, interpretability is used to measure the degree to which a human can consistently predict the model’s decisions. Gaining an insight into how and why the SE model arrives at a particular decision or outcome not only builds confidence in the model to deal with unknown situations, thereby reducing the risk of uncertainty, but also helps to understand the overall strengths and weaknesses of the models and guides improvements to the model~\cite{dong2017improving}. In contrast to interpretability, the study of explainable AI focuses on the hidden states in DNN and aims to open up the black box.  For example, the contribution of each input semantic feature to the accuracy of the semantic inference can be quantified by analyzing the gradient information of the semantic decoder.
    Based on this, the radio resource allocation at the sender can be achieved with a more flexible and fine-grained implementation, such as allocating the crucial semantic feature higher transmitting power to ensure its transmission reliability and the accuracy of the semantic inference.


\subsection{Tradeoff between SE accuracy and communication overhead}
Most of the existing works focus on how to perform accurate SE to save radio resources and enhance communication performance, while ignoring the extra communication overhead for SE. In fact, SE model training and updating require significant additional resources. For example,
the training of accurate semantic extraction models relies on a complete KB with both senders and receivers, which requires, first of all, adequate storage resources. In addition, as the communication context evolves, each user's local KB is constantly being updated individually. In this sense, ensuring that updates to the local database of all communicating participants can be shared  in real time is extremely challenging, especially for the case with a large number of participating users who are geographically distant, which can cause significant communication overhead.
Moreover, in an ideal case, retraining or fine-tuning of the SE model needs to be done promptly after the KB update. However, this is unrealistic for practical systems with limited computational resources. Therefore, making a favorable trade-off between SE accuracy and communication overhand is essential for the implementation of SemCom. 

For example, we can utilize  edge intelligence to train the SE model based on the shared KB of local senders and receivers stored in the MEC server of the local area first. Then, with the help of the  distributed learning paradigms, such as federated learning, a generalized SE model can be obtained by aggregating multiple well-trained SE models for different geographical areas. In this way, storage resources scattered around the edge can be efficiently utilized to reduce storage pressure on end devices or the central cloud, and the communication overhead caused by sharing data over long distances can be greatly reduced. However, the reasonable division of geographical areas and the strategic deployment of edge servers are still to be explored. Moreover, the aggregation period and the participants  selected in each round are also the essential issues that can be optimized in making a tradeoff between SE accuracy and communication overhead.

\subsection{Combination of SemCom and semantic caching}
In traditional communications, the implementation of data caching on the  router, MEC server, base station, etc., has already shown to be of great benefit in avoiding unnecessary delay and network overhead for  ~\cite{sheraz2020artificial}. By joint optimizing caching and communication, with a 1\% increase in cache hits, the perceived latency is reduced by 35\%~\cite{cidon2016cliffhanger}. However, traditional  caching in terms of raw data is no longer ideally suitable for SemCom systems, as the frequent and repetitive semantic extraction of the raw data results in redundancy and inefficiency of the system. Meanwhile, 
the data volume of SI extracted is much smaller compared to the raw data. In this sense, semantic caching strategies that fit with SemCom cannot only enhance system efficiency, but also save memory resources. 

However, there are  new issues raised for semantic caching. For example, different from the traditional data caching, such as that mainly focuses on the hit rate of the data content, semantic caching is more concerned with whether the SI in the cache can be accurately inferred by the requester. Since there may be multiple SI for the same data content, which ones to cache demands more prior knowledge, such as the popularity of the specific SI. Moreover, as the context of SemCom is constantly changing, the lifetime of SI is more difficult to determine. To this end, it also requires new estimate refreshing algorithms for semantic caching.

\subsection{Reasoning in implicit SemCom}
The majority of previous SemCom research focused on transferring explicit SI, such as the labels of things that can be directly identified from the source signals, e.g., images, voices, and texts. However, communication between users is not only limited to explicit information, but also contains rich implicit information that is difficult to express, recognize, or recover. For example in \cite{xiao2022reasoning}, a kid sends her father a voice message asking, ``What is a Tweety?" The major semantic part of this message, ``Tweety", might be interpreted in several ways, such as a smartphone app, a canary bird, or a character from a cartoon television program. Therefore, to deduce the message's exact meaning, the receiver must be able to infer the implicit information from the transmitter's context and background. Thus, it is unrealistic to assume that the destination user has a well-defined analytical expression, such as a reward function or utility function, which is directly optimized to maximize its understanding of the semantic meaning. 

A few works have considered this point and tried to propose solutions. A generative adversarial imitation learning-based reasoning mechanism learning (GAML) is designed in~\cite{xiao2022reasoning} for the destination user to learn and imitate the reasoning process of the source user to obtain the implicit semantic meaning. It is shown that GAML can achieve significant
error correction performance and offer 20\% of accuracy improvement over genetic algorithm (GA)-based reasoning solution. In another work~\cite{liang2022life}, the authors develop a novel inference function-based approach that can infer hidden information such as incomplete entities and relations that cannot be directly observed from the message, where the solution achieves 76\% and 48\% of accuracy in recovering missing information when using additive and linear inference functions, respectively. However, both solutions in~\cite{xiao2022reasoning,liang2022life} add additional inference overhead, and there is still room for further performance enhancement. Moreover, because explicit SI is typically dominant, the communication resources should be allocated proportionally between explicit and implicit SI, which inspires us to further design the joint optimization algorithms.

\subsection{Artificial intelligence in SemCom channel management}
In SemCom, AI is more often deployed on the transmitter and receiver for coding and decoding to serve upper-layer applications. However, in the 6G wireless communication that has a higher data rate and more frequent handover, channel modeling becomes more and more complex than the traditional stochastic or deterministic approaches \cite{wang20206g}. This prompted us to think about whether AI could be brought down to the SemCom channel layer to help model, estimate, and change channel conditions. Unlike simply applying AI to the end-to-end SemCom model training, the development of new intelligent materials gives AI more freedom in wireless channels~\cite{hu2022metasketch}. It is believed that the radio environment in the future generation of wireless communication networks will become controllable and intelligent by leveraging the emerging technologies of reconfigurable metasurface (RMS) and AI~\cite{gacanin2020wireless}. RMS can effectively control the wavefront, e.g., the phase, amplitude, frequency, and even polarization, of the impinging signals. Through the use of AI-enable programmable intelligent materials, SemCom networks can further surpass the limits predicted by the classical Shannon theory by jointly optimizing the transmitter, the receiver, and the environment. 


\subsection{Tradeoff between SemCom performance and security}
Data security and privacy issues are always significant topics in the field of wireless communications~\cite{letaief2021edge}. Due to the fact that SemCom requires only partial data to be transmitted and the decoding of SI relies on the receiver's background knowledge, it has also been regarded as a potential method for secure communications~\cite{basu2014preserving}. In addition, the security of the data can be further enhanced by encrypting the extracted SI. However, this also leads us to consider the tradeoff between computational resource overhead and data security. One possible solution is to use physical layer security technologies. Considering the success of covert communication~\cite{bloch2016covert}, we can make the data eavesdropper unsure whether the SemCom is ongoing by introduced interference to the physical layer for secure wireless transmission. However, although the computational resources for encrypting the data are reduced, we need to keep the transmitting power not too high to ensure the covertness of the communication. In addition, the interference signals have a negative impact on the transmission of SI, which brings a trade-off between covertness and signal quality.

\section{Conclusion}\label{Section8}
In this paper, we have provided a comprehensive survey of SemCom for 6G. First, we have highlighted the mutually reinforcing properties of 6G and SemCom. {Then, we introduced the development from SemCom-related theories and identified three types of SemCom.  Next, we organized the design of the communication system into three dimensions of
SI extraction,  SI transmission, and SI metrics, and discussed the state-of-the-art techniques and challenges, respectively.  Meanwhile, we have presented the potential applications of  SemCom in the 6G network as well as the promising SemCom-empowered network architecture. Moreover, we highlight some future directions with insights for further
in-depth investigations.}

\bibliographystyle{IEEEtran}
\bibliography{ref}{}

\begin{thebibliography}{100}
\providecommand{\url}[1]{#1}
\csname url@samestyle\endcsname
\providecommand{\newblock}{\relax}
\providecommand{\bibinfo}[2]{#2}
\providecommand{\BIBentrySTDinterwordspacing}{\spaceskip=0pt\relax}
\providecommand{\BIBentryALTinterwordstretchfactor}{4}
\providecommand{\BIBentryALTinterwordspacing}{\spaceskip=\fontdimen2\font plus
\BIBentryALTinterwordstretchfactor\fontdimen3\font minus
  \fontdimen4\font\relax}
\providecommand{\BIBforeignlanguage}[2]{{%
\expandafter\ifx\csname l@#1\endcsname\relax
\typeout{** WARNING: IEEEtran.bst: No hyphenation pattern has been}%
\typeout{** loaded for the language `#1'. Using the pattern for}%
\typeout{** the default language instead.}%
\else
\language=\csname l@#1\endcsname
\fi
#2}}
\providecommand{\BIBdecl}{\relax}
\BIBdecl

\bibitem{shannon1948mathematical}
C.~E. Shannon, ``A mathematical theory of communication,'' \emph{Bell Sys.
  Tech. J.}, vol.~27, no.~3, pp. 379--423, Oct. 1948.

\bibitem{li2017intelligent}
R.~Li, Z.~Zhao, X.~Zhou, G.~Ding, Y.~Chen, Z.~Wang, and H.~Zhang, ``Intelligent
  5{G}: {W}hen cellular networks meet artificial intelligence,'' \emph{IEEE
  Wireless Commun.}, vol.~24, no.~5, pp. 175--183, May 2017.

\bibitem{xu2022full}
M.~Xu, W.~C. Ng, W.~Y.~B. Lim, J.~Kang, Z.~Xiong, D.~Niyato, Q.~Yang, X.~S.
  Shen, and C.~Miao, ``A full dive into realizing the edge-enabled metaverse:
  {V}isions, enabling technologies, and challenges,'' \emph{arXiv preprint
  arXiv:2203.05471}, 2022.

\bibitem{lim2022realizing}
W.~Y.~B. Lim, Z.~Xiong, D.~Niyato, X.~Cao, C.~Miao, S.~Sun, and Q.~Yang,
  ``Realizing the metaverse with edge intelligence: {A} match made in heaven,''
  \emph{arXiv preprint arXiv:2201.01634}, 2022.

\bibitem{du2022attention}
H.~Du, J.~Liu, D.~Niyato, J.~Kang, Z.~Xiong, J.~Zhang, and D.~I. Kim,
  ``Attention-aware resource allocation and {QoE} analysis for metaverse
  x{URLLC} services,'' \emph{arXiv preprint arXiv:2208.05438}, 2022.

\bibitem{jeon2022blockchain}
H.-j. Jeon, H.-c. Youn, S.-m. Ko, and T.-h. Kim, ``Blockchain and ai meet in
  the metaverse,'' \emph{Advances in the Convergence of Blockchain and
  Artificial Intelligence}, p.~73, 2022.

\bibitem{du2022rethinking}
H.~Du, B.~Ma, D.~Niyato, and J.~Kang, ``Rethinking quality of experience for
  metaverse services: {A} consumer-based economics perspective,'' \emph{arXiv
  preprint arXiv:2208.01076}, 2022.

\bibitem{lan2021semantic}
Q.~Lan, D.~Wen, Z.~Zhang, Q.~Zeng, X.~Chen, P.~Popovski, and K.~Huang, ``What
  is semantic communication? a view on conveying meaning in the era of machine
  intelligence,'' \emph{Journal of Communications and Information Networks},
  vol.~6, no.~4, pp. 336--371, Apr. 2021.

\bibitem{shi2021semantic}
G.~Shi, Y.~Xiao, Y.~Li, and X.~Xie, ``From semantic communication to
  semantic-aware networking: {M}odel, architecture, and open problems,''
  \emph{IEEE Commun. Mag.}, vol.~59, no.~8, pp. 44--50, Aug. 2021.

\bibitem{kountouris2021semantics}
M.~Kountouris and N.~Pappas, ``Semantics-empowered communication for networked
  intelligent systems,'' \emph{IEEE Commun. Mag.}, vol.~59, no.~6, pp. 96--102,
  Jun. 2021.

\bibitem{strinati20216g}
E.~C. Strinati and S.~Barbarossa, ``6{G} networks: {B}eyond shannon towards
  semantic and goal-oriented communications,'' \emph{Comput. Netw.}, vol. 190,
  p. 107930, 2021.

\bibitem{qin2021semantic}
Z.~Qin, X.~Tao, J.~Lu, and G.~Y. Li, ``Semantic communications: {P}rinciples
  and challenges,'' \emph{arXiv preprint arXiv:2201.01389}, 2021.

\bibitem{luo2022semantic}
X.~Luo, H.-H. Chen, and Q.~Guo, ``Semantic communications: {O}verview, open
  issues, and future research directions,'' \emph{IEEE Wireless Commun.}, to
  appear, 2022.

\bibitem{zhang2022toward}
P.~Zhang, W.~Xu, H.~Gao, K.~Niu, X.~Xu, X.~Qin, C.~Yuan, Z.~Qin, H.~Zhao,
  J.~Wei \emph{et~al.}, ``Toward wisdom-evolutionary and primitive-concise
  {6G}: {A} new paradigm of semantic communication networks,''
  \emph{Engineering}, vol.~8, pp. 60--73, 2022.

\bibitem{wu2021toward}
J.~Wu, R.~Li, X.~An, C.~Peng, Z.~Liu, J.~Crowcroft, and H.~Zhang, ``Toward
  native artificial intelligence in 6{G} networks: {S}ystem design,
  architectures, and paradigms,'' \emph{arXiv preprint arXiv:2103.02823}, 2021.

\bibitem{zhong2017theory}
Y.~Zhong, ``A theory of semantic information,'' \emph{China Commun.}, vol.~14,
  no.~1, pp. 1--17, Jan. 2017.

\bibitem{morris1938foundations}
C.~W. Morris, ``Foundations of the theory of signs,'' in \emph{International
  encyclopedia of unified science}.\hskip 1em plus 0.5em minus 0.4em\relax
  Chicago University Press, 1938, pp. 1--59.

\bibitem{ch1971writings}
M.~Ch, ``Writings on the general theory of signs,'' \emph{Mouton, The Hague},
  1971.

\bibitem{weaver1953recent}
W.~Weaver, ``Recent contributions to the mathematical theory of
  communication,'' \emph{ETC: {A} review of general semantics}, pp. 261--281,
  1953.

\bibitem{zhang2021toward}
P.~Zhang, W.~Xu, H.~Gao, K.~Niu, X.~Xu, X.~Qin, C.~Yuan, Z.~Qin, H.~Zhao,
  J.~Wei \emph{et~al.}, ``Toward wisdom-evolutionary and primitive-concise
  6{G}: {A} new paradigm of semantic communication networks,''
  \emph{Engineering}, 2021.

\bibitem{bar1953semantic}
Y.~Bar-Hillel and R.~Carnap, ``Semantic information,'' \emph{Br. J. Philos.
  Sci.}, vol.~4, no.~14, pp. 147--157, 1953.

\bibitem{elias1954outline}
P.~Elias, ``An outline of a theory of semantic information,'' 1954.

\bibitem{floridi2004outline}
L.~Floridi, ``Outline of a theory of strongly semantic information,''
  \emph{Minds. Mach.}, vol.~14, no.~2, pp. 197--221, Feb. 2004.

\bibitem{d2011quantifying}
S.~D'Alfonso, ``On quantifying semantic information,'' \emph{Information},
  vol.~2, no.~1, pp. 61--101, 2011.

\bibitem{niiniluoto2012truthlikeness}
I.~Niiniluoto, \emph{Truthlikeness}.\hskip 1em plus 0.5em minus 0.4em\relax
  Springer Science \& Business Media, 2012, vol. 185.

\bibitem{barwise1997information}
J.~Barwise, J.~Seligman \emph{et~al.}, \emph{Information flow: {T}he logic of
  distributed systems}.\hskip 1em plus 0.5em minus 0.4em\relax Cambridge
  University Press, 1997.

\bibitem{oddie1986likeness}
G.~Oddie, ``Likeness to truth,'' 1986.

\bibitem{niiniluoto1987truthlikeness}
I.~Niiniluoto, ``Truthlikeness,'' 1987.

\bibitem{bao2011towards}
J.~Bao, P.~Basu, M.~Dean, C.~Partridge, A.~Swami, W.~Leland, and J.~A. Hendler,
  ``Towards a theory of semantic communication,'' in \emph{2011 IEEE Network
  Science Workshop}.\hskip 1em plus 0.5em minus 0.4em\relax IEEE, 2011, pp.
  110--117.

\bibitem{basu2014preserving}
P.~Basu, J.~Bao, M.~Dean, and J.~Hendler, ``Preserving quality of information
  by using semantic relationships,'' \emph{Pervasive and Mobile Computing},
  vol.~11, pp. 188--202, 2014.

\bibitem{juba2008universal1}
B.~Juba and M.~Sudan, ``Universal semantic communication i,'' in \emph{Proc.
  fortieth annual ACM symposium on Theory of computing}, 2008, pp. 123--132.

\bibitem{lund1992algebraic}
C.~Lund, L.~Fortnow, H.~Karloff, and N.~Nisan, ``Algebraic methods for
  interactive proof systems,'' \emph{J. ACM}, vol.~39, no.~4, pp. 859--868,
  Apr. 1992.

\bibitem{juba2008universal}
B.~Juba and M.~Sudan, ``Universal semantic communication {II}: {A} theory of
  goal-oriented communication,'' in \emph{Elec. Colloq. Comput. Complex.
  (ECCC)}, vol.~15, no. 095.\hskip 1em plus 0.5em minus 0.4em\relax Citeseer,
  2008.

\bibitem{goldreich2012theory}
O.~Goldreich, B.~Juba, and M.~Sudan, ``A theory of goal-oriented
  communication,'' \emph{J. ACM}, vol.~59, no.~2, pp. 1--65, Feb. 2012.

\bibitem{juba2011semantic}
B.~Juba and S.~Vempala, ``Semantic communication for simple goals is equivalent
  to on-line learning,'' in \emph{Proc. Int. Conf. Algo. Learn. Theory}.\hskip
  1em plus 0.5em minus 0.4em\relax Springer, 2011, pp. 277--291.

\bibitem{dretske1981knowledge}
F.~I. Dretske, ``Knowledge and the flow of information,'' 1981.

\bibitem{bista2018semantic}
H.~Bista, I.-L. Yen, F.~Bastani, M.~Mueller, and D.~Moore, ``Semantic-based
  information sharing in vehicular networks,'' in \emph{2018 IEEE International
  Conference on Web Services (ICWS)}.\hskip 1em plus 0.5em minus 0.4em\relax
  IEEE, 2018, pp. 282--289.

\bibitem{xie2021deep}
H.~Xie, Z.~Qin, G.~Y. Li, and B.-H. Juang, ``Deep learning enabled semantic
  communication systems,'' \emph{IEEE Trans. Signal Process.}, vol.~69, pp.
  2663--2675, 2021.

\bibitem{sana2021learning}
M.~Sana and E.~C. Strinati, ``Learning semantics: {A}n opportunity for
  effective 6{G} communications,'' \emph{arXiv preprint arXiv:2110.08049},
  2021.

\bibitem{merono2015semantic}
A.~Mero{\~n}o-Pe{\~n}uela, A.~Ashkpour, M.~Van~Erp, K.~Mandemakers, L.~Breure,
  A.~Scharnhorst, S.~Schlobach, and F.~Van~Harmelen, ``Semantic technologies
  for historical research: {A} survey,'' \emph{Semantic Web}, vol.~6, no.~6,
  pp. 539--564, 2015.

\bibitem{rachana2021literature}
L.~Rachana and S.~Shridevi, ``A literature survey: {S}emantic technology
  approach in machine learning,'' \emph{Advances in Smart Grid Technology}, pp.
  467--477, 2021.

\bibitem{thoma2016survey}
M.~Thoma, ``A survey of semantic segmentation,'' \emph{arXiv preprint
  arXiv:1602.06541}, 2016.

\bibitem{sheu2010semantic}
P.~C.-y. Sheu, ``Semantic computing,'' \emph{Semantic computing}, pp. 1--9,
  2010.

\bibitem{hitzler2021review}
P.~Hitzler, ``A review of the semantic web field,'' \emph{Communications of the
  ACM}, vol.~64, no.~2, pp. 76--83, 2021.

\bibitem{chen2021recommendation}
W.~Chen, ``Recommendation system based on semantic web,'' in \emph{Int. Conf.
  Machine Learning and Big Data Analytics for IoT Security and Privacy}.\hskip
  1em plus 0.5em minus 0.4em\relax Springer, 2021, pp. 511--517.

\bibitem{she2020deep}
C.~She, R.~Dong, Z.~Gu, Z.~Hou, Y.~Li, W.~Hardjawana, C.~Yang, L.~Song, and
  B.~Vucetic, ``Deep learning for ultra-reliable and low-latency communications
  in 6{G} networks,'' \emph{IEEE Netw.}, vol.~34, no.~5, pp. 219--225, 2020.

\bibitem{qin2019deep}
Z.~Qin, H.~Ye, G.~Y. Li, and B.-H.~F. Juang, ``Deep learning in physical layer
  communications,'' \emph{IEEE Wireless Commun.}, vol.~26, no.~2, pp. 93--99,
  2019.

\bibitem{o2017introduction}
T.~O’shea and J.~Hoydis, ``An introduction to deep learning for the physical
  layer,'' \emph{IEEE Trans. Cogn. Commun. Netw.}, vol.~3, no.~4, pp. 563--575,
  2017.

\bibitem{ye2017power}
H.~Ye, G.~Y. Li, and B.-H. Juang, ``Power of deep learning for channel
  estimation and signal detection in {OFDM} systems,'' \emph{IEEE Wireless
  Commun. Lett.}, vol.~7, no.~1, pp. 114--117, 2017.

\bibitem{chun2019deep}
C.-J. Chun, J.-M. Kang, and I.-M. Kim, ``Deep learning-based joint pilot design
  and channel estimation for multiuser {MIMO} channels,'' \emph{IEEE
  Communications Letters}, vol.~23, no.~11, pp. 1999--2003, 2019.

\bibitem{guo2021canet}
J.~Guo, C.-K. Wen, and S.~Jin, ``Canet: {U}plink-aided downlink channel
  acquisition in {FDD} massive {MIMO} using deep learning,'' \emph{arXiv
  preprint arXiv:2101.04377}, 2021.

\bibitem{park2020end}
S.~Park, O.~Simeone, and J.~Kang, ``End-to-end fast training of communication
  links without a channel model via online meta-learning,'' in \emph{2020 IEEE
  21st International Workshop on Signal Processing Advances in Wireless
  Communications (SPAWC)}.\hskip 1em plus 0.5em minus 0.4em\relax IEEE, 2020,
  pp. 1--5.

\bibitem{ye2021deep}
H.~Ye, G.~Y. Li, and B.-H.~F. Juang, ``Deep learning based end-to-end wireless
  communication systems without pilots,'' \emph{IEEE Trans. Cogn. Commun.
  Netw.}, 2021.

\bibitem{dorner2017deep}
S.~D{\"o}rner, S.~Cammerer, J.~Hoydis, and S.~Ten~Brink, ``Deep learning based
  communication over the air,'' \emph{IEEE J. Sel. Top. Signal Process.},
  vol.~12, no.~1, pp. 132--143, 2017.

\bibitem{mnih2014recurrent}
V.~Mnih, N.~Heess, A.~Graves \emph{et~al.}, ``Recurrent models of visual
  attention,'' in \emph{Advances in neural information processing systems},
  2014, pp. 2204--2212.

\bibitem{szegedy2015going}
C.~Szegedy, W.~Liu, Y.~Jia, P.~Sermanet, S.~Reed, D.~Anguelov, D.~Erhan,
  V.~Vanhoucke, and A.~Rabinovich, ``Going deeper with convolutions,'' in
  \emph{Proc. IEEE Conf. Comput. Vis. Pattern Recog.}, 2015, pp. 1--9.

\bibitem{wang2017residual}
F.~Wang, M.~Jiang, C.~Qian, S.~Yang, C.~Li, H.~Zhang, X.~Wang, and X.~Tang,
  ``Residual attention network for image classification,'' in \emph{Proc. IEEE
  Conf. Comput. Vis. Pattern Recog.}, 2017, pp. 3156--3164.

\bibitem{bahdanau2014neural}
D.~Bahdanau, K.~Cho, and Y.~Bengio, ``Neural machine translation by jointly
  learning to align and translate,'' \emph{arXiv preprint arXiv:1409.0473},
  2014.

\bibitem{luong2015effective}
M.-T. Luong, H.~Pham, and C.~D. Manning, ``Effective approaches to
  attention-based neural machine translation,'' \emph{arXiv preprint
  arXiv:1508.04025}, 2015.

\bibitem{vaswani2017attention}
A.~Vaswani, N.~Shazeer, N.~Parmar, J.~Uszkoreit, L.~Jones, A.~N. Gomez,
  {\L}.~Kaiser, and I.~Polosukhin, ``Attention is all you need,'' in
  \emph{Advances in neural information processing systems}, 2017, pp.
  5998--6008.

\bibitem{purwins2019deep}
H.~Purwins, B.~Li, T.~Virtanen, J.~Schl{\"u}ter, S.-Y. Chang, and T.~Sainath,
  ``Deep learning for audio signal processing,'' \emph{IEEE J. Sel. Top. Signal
  Process.}, vol.~13, no.~2, pp. 206--219, Feb. 2019.

\bibitem{ogunfunmi2019primer}
T.~Ogunfunmi, R.~P. Ramachandran, R.~Togneri, Y.~Zhao, and X.~Xia, ``A primer
  on deep learning architectures and applications in speech processing,''
  \emph{Circuits, Systems, and Signal Processing}, vol.~38, no.~8, pp.
  3406--3432, 2019.

\bibitem{haeb2019speech}
R.~Haeb-Umbach, S.~Watanabe, T.~Nakatani, M.~Bacchiani, B.~Hoffmeister, M.~L.
  Seltzer, H.~Zen, and M.~Souden, ``Speech processing for digital home
  assistants: {C}ombining signal processing with deep-learning techniques,''
  \emph{IEEE Signal Process. Mag.}, vol.~36, no.~6, pp. 111--124, 2019.

\bibitem{xie2020deep}
H.~Xie, Z.~Qin, G.~Y. Li, and B.-H. Juang, ``Deep learning based semantic
  communications: {A}n initial investigation,'' in \emph{GLOBECOM 2020-2020
  IEEE Global Communications Conference}.\hskip 1em plus 0.5em minus
  0.4em\relax IEEE, 2020, pp. 1--6.

\bibitem{lee2019deep}
C.-H. Lee, J.-W. Lin, P.-H. Chen, and Y.-C. Chang, ``Deep learning-constructed
  joint transmission-recognition for internet of things,'' \emph{IEEE Access},
  vol.~7, pp. 76\,547--76\,561, 2019.

\bibitem{he2016deep}
K.~He, X.~Zhang, S.~Ren, and J.~Sun, ``Deep residual learning for image
  recognition,'' in \emph{Proc. IEEE Conf. Comput. Vis. Pattern Recog.}, 2016,
  pp. 770--778.

\bibitem{xu2021wireless}
J.~Xu, B.~Ai, W.~Chen, A.~Yang, P.~Sun, and M.~Rodrigues, ``Wireless image
  transmission using deep source channel coding with attention modules,''
  \emph{IEEE Trans. Circuits Syst. Video Technol.}, 2021.

\bibitem{hu2022robust}
Q.~Hu, G.~Zhang, Z.~Qin, Y.~Cai, and G.~Yu, ``Robust semantic communications
  against semantic noise,'' \emph{arXiv preprint arXiv:2202.03338}, 2022.

\bibitem{he2021masked}
K.~He, X.~Chen, S.~Xie, Y.~Li, P.~Doll{\'a}r, and R.~Girshick, ``Masked
  autoencoders are scalable vision learners,'' \emph{arXiv preprint
  arXiv:2111.06377}, 2021.

\bibitem{farsad2018deep}
N.~Farsad, M.~Rao, and A.~Goldsmith, ``Deep learning for joint source-channel
  coding of text,'' in \emph{2018 IEEE Int. Conf. Acoustics, Speech and Signal
  Processing (ICASSP)}.\hskip 1em plus 0.5em minus 0.4em\relax IEEE, 2018, pp.
  2326--2330.

\bibitem{pennington2014glove}
J.~Pennington, R.~Socher, and C.~D. Manning, ``Glove: {G}lobal vectors for word
  representation,'' in \emph{Proc. 2014 conference on empirical methods in
  natural language processing (EMNLP)}, 2014, pp. 1532--1543.

\bibitem{wu2016google}
Y.~Wu, M.~Schuster, Z.~Chen, Q.~V. Le, M.~Norouzi, W.~Macherey, M.~Krikun,
  Y.~Cao, Q.~Gao, K.~Macherey \emph{et~al.}, ``Google's neural machine
  translation system: {B}ridging the gap between human and machine
  translation,'' \emph{arXiv preprint arXiv:1609.08144}, 2016.

\bibitem{graves2012sequence}
A.~Graves, ``Sequence transduction with recurrent neural networks,''
  \emph{arXiv preprint arXiv:1211.3711}, 2012.

\bibitem{mikolov2013efficient}
T.~Mikolov, K.~Chen, G.~Corrado, and J.~Dean, ``Efficient estimation of word
  representations in vector space,'' \emph{arXiv preprint arXiv:1301.3781},
  2013.

\bibitem{xie2020lite}
H.~Xie and Z.~Qin, ``A lite distributed semantic communication system for
  internet of things,'' \emph{IEEE J. Sel. Areas Commun.}, vol.~39, no.~1, pp.
  142--153, 2020.

\bibitem{zhou2021semantic}
Q.~Zhou, R.~Li, Z.~Zhao, C.~Peng, and H.~Zhang, ``Semantic communication with
  adaptive universal transformer,'' \emph{IEEE Wireless Commun. Lett.}, to
  apper, 2021.

\bibitem{dehghani2018universal}
M.~Dehghani, S.~Gouws, O.~Vinyals, J.~Uszkoreit, and {\L}.~Kaiser, ``Universal
  transformers,'' \emph{arXiv preprint arXiv:1807.03819}, 2018.

\bibitem{graves2016adaptive}
A.~Graves, ``Adaptive computation time for recurrent neural networks,''
  \emph{arXiv preprint arXiv:1603.08983}, 2016.

\bibitem{tong2021federated}
H.~Tong, Z.~Yang, S.~Wang, Y.~Hu, O.~Semiari, W.~Saad, and C.~Yin, ``Federated
  learning for audio semantic communication,'' \emph{Front. Commun. Netw.},
  vol.~2, 2021.

\bibitem{schneider2019wav2vec}
S.~Schneider, A.~Baevski, R.~Collobert, and M.~Auli, ``wav2vec: {U}nsupervised
  pre-training for speech recognition,'' \emph{arXiv preprint
  arXiv:1904.05862}, 2019.

\bibitem{weng2021semantic}
Z.~Weng, Z.~Qin, and G.~Y. Li, ``Semantic communications for speech signals,''
  in \emph{Proc. IEEE Intel. Conf. Commun.}, 2021, pp. 1--6.

\bibitem{weng2021semantic2}
Z.~Weng and Z.~Qin, ``Semantic communication systems for speech transmission,''
  \emph{IEEE J. Sel. Areas Commun.}, to appear, 2021.

\bibitem{weng2021Recognition}
Z.~Weng, Z.~Qin, and G.~Y. Li, ``Semantic communications for speech
  recognition,'' \emph{arXiv preprint arXiv:2107.11190}, 2021.

\bibitem{schuster1997bidirectional}
M.~Schuster and K.~K. Paliwal, ``Bidirectional recurrent neural networks,''
  \emph{IEEE Trans. Signal Process.}, vol.~45, no.~11, pp. 2673--2681, Nov.
  1997.

\bibitem{amodei2016deep}
D.~Amodei, S.~Ananthanarayanan, R.~Anubhai, J.~Bai, E.~Battenberg, C.~Case,
  J.~Casper, B.~Catanzaro, Q.~Cheng, G.~Chen \emph{et~al.}, ``Deep speech 2:
  End-to-end speech recognition in english and mandarin,'' in
  \emph{International conference on machine learning}.\hskip 1em plus 0.5em
  minus 0.4em\relax PMLR, 2016, pp. 173--182.

\bibitem{jiang2019deepturbo}
Y.~Jiang, S.~Kannan, H.~Kim, S.~Oh, H.~Asnani, and P.~Viswanath, ``Deepturbo:
  {D}eep turbo decoder,'' in \emph{2019 IEEE 20th International Workshop on
  Signal Processing Advances in Wireless Communications (SPAWC)}.\hskip 1em
  plus 0.5em minus 0.4em\relax IEEE, 2019, pp. 1--5.

\bibitem{xie2021task}
H.~Xie, Z.~Qin, and G.~Y. Li, ``Task-oriented semantic communications for
  multimodal data,'' \emph{arXiv preprint arXiv:2108.07357}, 2021.

\bibitem{xie2021task2}
H.~Xie, Z.~Qin, X.~Tao, and K.~B. Letaief, ``Task-oriented multi-user semantic
  communications,'' \emph{arXiv preprint arXiv:2112.10255}, 2021.

\bibitem{russakovsky2015imagenet}
O.~Russakovsky, J.~Deng, H.~Su, J.~Krause, S.~Satheesh, S.~Ma, Z.~Huang,
  A.~Karpathy, A.~Khosla, M.~Bernstein \emph{et~al.}, ``Imagenet large scale
  visual recognition challenge,'' \emph{International journal of computer
  vision}, vol. 115, no.~3, pp. 211--252, Mar. 2015.

\bibitem{hudson2018compositional}
D.~A. Hudson and C.~D. Manning, ``Compositional attention networks for machine
  reasoning,'' \emph{arXiv preprint arXiv:1803.03067}, 2018.

\bibitem{ranzato2015sequence}
M.~Ranzato, S.~Chopra, M.~Auli, and W.~Zaremba, ``Sequence level training with
  recurrent neural networks,'' \emph{arXiv preprint arXiv:1511.06732}, 2015.

\bibitem{rennie2017self}
S.~J. Rennie, E.~Marcheret, Y.~Mroueh, J.~Ross, and V.~Goel, ``Self-critical
  sequence training for image captioning,'' in \emph{Proc. IEEE Conf. Comput.
  Vis. Pattern Recog.}, 2017, pp. 7008--7024.

\bibitem{sutton2018reinforcement}
R.~S. Sutton and A.~G. Barto, \emph{Reinforcement learning: {A}n
  introduction}.\hskip 1em plus 0.5em minus 0.4em\relax MIT press, 2018.

\bibitem{bahdanau2016actor}
D.~Bahdanau, P.~Brakel, K.~Xu, A.~Goyal, R.~Lowe, J.~Pineau, A.~Courville, and
  Y.~Bengio, ``An actor-critic algorithm for sequence prediction,'' \emph{arXiv
  preprint arXiv:1607.07086}, 2016.

\bibitem{ren2017deep}
Z.~Ren, X.~Wang, N.~Zhang, X.~Lv, and L.-J. Li, ``Deep reinforcement
  learning-based image captioning with embedding reward,'' in \emph{Proc. IEEE
  Conf. Comput. Vis. Pattern Recog.}, 2017, pp. 290--298.

\bibitem{yu2017seqgan}
L.~Yu, W.~Zhang, J.~Wang, and Y.~Yu, ``Seqgan: {S}equence generative
  adversarial nets with policy gradient,'' in \emph{Proc. AAAI Conf. Artif.
  Intell.}, vol.~31, no.~1, 2017.

\bibitem{lu2021reinforcement}
K.~Lu, R.~Li, X.~Chen, Z.~Zhao, and H.~Zhang, ``Reinforcement learning-powered
  semantic communication via semantic similarity,'' \emph{arXiv preprint
  arXiv:2108.12121}, 2021.

\bibitem{lu2021rethinking}
K.~Lu, Q.~Zhou, R.~Li, Z.~Zhao, X.~Chen, J.~Wu, and H.~Zhang, ``Rethinking
  modern communication from semantic coding to semantic communication,''
  \emph{arXiv preprint arXiv:2110.08496}, 2021.

\bibitem{silver2016mastering}
D.~Silver, A.~Huang, C.~J. Maddison, A.~Guez, L.~Sifre, G.~Van Den~Driessche,
  J.~Schrittwieser, I.~Antonoglou, V.~Panneershelvam, M.~Lanctot \emph{et~al.},
  ``Mastering the game of {G}o with deep neural networks and tree search,''
  \emph{nature}, vol. 529, no. 7587, pp. 484--489, 2016.

\bibitem{konda2000actor}
V.~R. Konda and J.~N. Tsitsiklis, ``Actor-critic algorithms,'' in
  \emph{Advances in neural information processing systems}, 2000, pp.
  1008--1014.

\bibitem{carpi2019reinforcement}
F.~Carpi, C.~H{\"a}ger, M.~Martal{\`o}, R.~Raheli, and H.~D. Pfister,
  ``Reinforcement learning for channel coding: {L}earned bit-flipping
  decoding,'' in \emph{2019 57th Annual Allerton Conference on Communication,
  Control, and Computing (Allerton)}.\hskip 1em plus 0.5em minus 0.4em\relax
  IEEE, 2019, pp. 922--929.

\bibitem{luo2020better}
R.~Luo, ``A better variant of self-critical sequence training,'' \emph{arXiv
  preprint arXiv:2003.09971}, 2020.

\bibitem{farshbafan2022curriculum}
M.~K. Farshbafan, W.~Saad, and M.~Debbah, ``Curriculum learning for
  goal-oriented semantic communications with a common language,'' \emph{arXiv
  preprint arXiv:2204.10429}, 2022.

\bibitem{wang2021performance}
Y.~Wang, M.~Chen, W.~Saad, T.~Luo, S.~Cui, and V.~Poor, ``Performance
  optimization for semantic communications: {A}n attention-based learning
  approach,'' in \emph{Proc. IEEE Global Commun. Conf.}, 2021.

\bibitem{zhou2022cognitive}
F.~Zhou, Y.~Li, X.~Zhang, Q.~Wu, X.~Lei, and R.~Q. Hu, ``Cognitive semantic
  communication systems driven by knowledge graph,'' \emph{arXiv preprint
  arXiv:2202.11958}, 2022.

\bibitem{rosa2018knowledge}
R.~L. Rosa, G.~M. Schwartz, W.~V. Ruggiero, and D.~Z. Rodr{\'\i}guez, ``A
  knowledge-based recommendation system that includes sentiment analysis and
  deep learning,'' \emph{IEEE Trans. Industr. Inform.}, vol.~15, no.~4, pp.
  2124--2135, 2018.

\bibitem{lin2020kbpearl}
X.~Lin, H.~Li, H.~Xin, Z.~Li, and L.~Chen, ``Kbpearl: {A} knowledge base
  population system supported by joint entity and relation linking,''
  \emph{Proc. VLDB Endowment}, vol.~13, no.~7, pp. 1035--1049, Jul. 2020.

\bibitem{zheng2021knowledge}
W.~Zheng, L.~Yin, X.~Chen, Z.~Ma, S.~Liu, and B.~Yang, ``Knowledge base graph
  embedding module design for visual question answering model,'' \emph{Pattern
  Recognit.}, vol. 120, p. 108153, 2021.

\bibitem{yang2021semantic}
Y.~Yang, C.~Guo, F.~Liu, C.~Liu, L.~Sun, Q.~Sun, and J.~Chen, ``Semantic
  communications with {AI} tasks,'' \emph{arXiv preprint arXiv:2109.14170},
  2021.

\bibitem{selvaraju2017grad}
R.~R. Selvaraju, M.~Cogswell, A.~Das, R.~Vedantam, D.~Parikh, and D.~Batra,
  ``Grad-cam: {V}isual explanations from deep networks via gradient-based
  localization,'' in \emph{Proc. IEEE Int. Conf. computer vision}, 2017, pp.
  618--626.

\bibitem{fong2018net2vec}
R.~Fong and A.~Vedaldi, ``Net2vec: {Q}uantifying and explaining how concepts
  are encoded by filters in deep neural networks,'' in \emph{Proc. IEEE Conf.
  Comput. Vis. Pattern Recog.}, 2018, pp. 8730--8738.

\bibitem{johnson2016perceptual}
J.~Johnson, A.~Alahi, and L.~Fei-Fei, ``Perceptual losses for real-time style
  transfer and super-resolution,'' in \emph{European conference on computer
  vision}.\hskip 1em plus 0.5em minus 0.4em\relax Springer, 2016, pp. 694--711.

\bibitem{seo2021semantics}
H.~Seo, J.~Park, M.~Bennis, and M.~Debbah, ``Semantics-native communication
  with contextual reasoning,'' \emph{arXiv preprint arXiv:2108.05681}, 2021.

\bibitem{ogden1923meaning}
C.~K. Ogden and I.~A. Richards, ``The meaning of meaning: {A} study of the
  influence of thought and of the science of symbolism,'' 1923.

\bibitem{lazaridou2020emergent}
A.~Lazaridou and M.~Baroni, ``Emergent multi-agent communication in the deep
  learning era,'' \emph{arXiv preprint arXiv:2006.02419}, 2020.

\bibitem{lazaridou2016multi}
A.~Lazaridou, A.~Peysakhovich, and M.~Baroni, ``Multi-agent cooperation and the
  emergence of (natural) language,'' \emph{arXiv preprint arXiv:1612.07182},
  2016.

\bibitem{mu2021emergent}
J.~Mu and N.~Goodman, ``Emergent communication of generalizations,''
  \emph{arXiv preprint arXiv:2106.02668}, 2021.

\bibitem{bell1992pragmatic}
J.~Bell, ``Pragmatic reasoning--{A} model-based theory,'' 1992.

\bibitem{goodman2013knowledge}
N.~D. Goodman and A.~Stuhlm{\"u}ller, ``Knowledge and implicature: {M}odeling
  language understanding as social cognition,'' \emph{Topics in cognitive
  science}, vol.~5, no.~1, pp. 173--184, Jan. 2013.

\bibitem{kao2014formalizing}
J.~Kao, L.~Bergen, and N.~Goodman, ``Formalizing the pragmatics of metaphor
  understanding,'' in \emph{Proc. annual meeting of the Cognitive Science
  Society}, vol.~36, no.~36, 2014.

\bibitem{goodman2016pragmatic}
N.~D. Goodman and M.~C. Frank, ``Pragmatic language interpretation as
  probabilistic inference,'' \emph{Trends Cogn. Sci.}, vol.~20, no.~11, pp.
  818--829, Nov. 2016.

\bibitem{grice1975logic}
H.~P. Grice, ``Logic and conversation,'' in \emph{Speech acts}.\hskip 1em plus
  0.5em minus 0.4em\relax Brill, 1975, pp. 41--58.

\bibitem{lotfi2021semantic}
F.~Lotfi, O.~Semiari, and W.~Saad, ``Semantic-aware collaborative deep
  reinforcement learning over wireless cellular networks,'' \emph{arXiv
  preprint arXiv:2111.12064}, 2021.

\bibitem{yun2021attention}
W.~J. Yun, B.~Lim, S.~Jung, Y.-C. Ko, J.~Park, J.~Kim, and M.~Bennis,
  ``Attention-based reinforcement learning for real-time uav semantic
  communication,'' in \emph{2021 17th International Symposium on Wireless
  Communication Systems (ISWCS)}.\hskip 1em plus 0.5em minus 0.4em\relax IEEE,
  2021, pp. 1--6.

\bibitem{tao2021repaint}
Y.~Tao, S.~Genc, J.~Chung, T.~Sun, and S.~Mallya, ``Repaint: {K}nowledge
  transfer in deep reinforcement learning,'' in \emph{Int. Conf. Machine
  Learning}.\hskip 1em plus 0.5em minus 0.4em\relax PMLR, 2021, pp.
  10\,141--10\,152.

\bibitem{visus2021taxonomy}
{\'A}.~Vis{\'u}s, J.~Garc{\'\i}a, and F.~Fern{\'a}ndez, ``A taxonomy of
  similarity metrics for markov decision processes,'' \emph{arXiv preprint
  arXiv:2103.04706}, 2021.

\bibitem{goodfellow2016deep}
I.~Goodfellow, Y.~Bengio, and A.~Courville, \emph{Deep learning}.\hskip 1em
  plus 0.5em minus 0.4em\relax MIT press, 2016.

\bibitem{foerster2016learning}
J.~Foerster, I.~A. Assael, N.~De~Freitas, and S.~Whiteson, ``Learning to
  communicate with deep multi-agent reinforcement learning,'' \emph{Advances in
  neural information processing systems}, vol.~29, 2016.

\bibitem{kang2021task}
X.~Kang, B.~Song, J.~Guo, Z.~Qin, and F.~R. Yu, ``Task-oriented image
  transmission for scene classification in unmanned aerial systems,''
  \emph{arXiv preprint arXiv:2112.10948}, 2021.

\bibitem{calvanese20196g}
E.~Calvanese~Strinati, S.~Barbarossa, J.~L. Gonzalez-Jimenez, D.~Kt{\'e}nas,
  N.~Cassiau, and C.~Dehos, ``6{G}: {T}he next frontier,'' \emph{arXiv
  e-prints}, pp. arXiv--1901, 2019.

\bibitem{yacoub2007alpha}
M.~D. Yacoub, ``The $\alpha$-$\mu$ distribution: {A} physical fading model for
  the stacy distribution,'' \emph{IEEE Trans. Veh. Technol.}, vol.~56, no.~1,
  pp. 27--34, Jan. 2007.

\bibitem{du2020sum}
H.~Du, J.~Zhang, J.~Cheng, and B.~Ai, ``Sum of fisher-snedecor $\mathcal{F}$
  random variables and its applications,'' \emph{IEEE Open J. Commun. Soc.},
  vol.~1, pp. 342--356, 2020.

\bibitem{zhang2017new}
J.~Zhang, W.~Zeng, X.~Li, Q.~Sun, and K.~P. Peppas, ``New results on the
  fluctuating two-ray model with arbitrary fading parameters and its
  applications,'' \emph{IEEE Trans. Veh. Technol.}, vol.~67, no.~3, pp.
  2766--2770, Mar. 2017.

\bibitem{mirza2014conditional}
M.~Mirza and S.~Osindero, ``Conditional generative adversarial nets,''
  \emph{arXiv preprint arXiv:1411.1784}, 2014.

\bibitem{ye2020deep}
H.~Ye, L.~Liang, G.~Y. Li, and B.-H. Juang, ``Deep learning-based end-to-end
  wireless communication systems with conditional {GAN}s as unknown channels,''
  \emph{IEEE Trans. Wireless Commun.}, vol.~19, no.~5, pp. 3133--3143, May
  2020.

\bibitem{meinila2009winner}
J.~Meinil{\"a}, P.~Ky{\"o}sti, T.~J{\"a}ms{\"a}, and L.~Hentil{\"a}, ``Winner
  {II} channel models,'' in \emph{Radio Technologies and Concepts for
  IMT-Advanced}, 2009.

\bibitem{ding2021snr}
M.~Ding, J.~Li, M.~Ma, and X.~Fan, ``{SNR}-adaptive deep joint source-channel
  coding for wireless image transmission,'' in \emph{ICASSP 2021-2021 IEEE Int.
  Conf. Acoustics, Speech and Signal Processing (ICASSP)}.\hskip 1em plus 0.5em
  minus 0.4em\relax IEEE, 2021, pp. 1555--1559.

\bibitem{moon2020error}
T.~K. Moon, \emph{Error correction coding: {M}athematical methods and
  algorithms}.\hskip 1em plus 0.5em minus 0.4em\relax John Wiley \& Sons, 2020.

\bibitem{jiang2021deep}
P.~Jiang, C.-K. Wen, S.~Jin, and G.~Y. Li, ``Deep source-channel coding for
  sentence semantic transmission with {HARQ},'' \emph{arXiv preprint
  arXiv:2106.03009}, 2021.

\bibitem{shrivastava2013error}
P.~Shrivastava and U.~P. Singh, ``Error detection and correction using reed
  solomon codes,'' \emph{Int. J. Adv. Comput. Sci. Appl.}, vol.~3, no.~8, Aug.
  2013.

\bibitem{xuan2019multi}
Y.~Xuan, C.~Guo, C.~Feng, and Z.~Li, ``Multi-graph based spectrum sharing
  scheme in vehicular network with integration of heterogenous spectrum,'' in
  \emph{Proc. IEEE Intel. Conf. Commun. Workshops (ICC Workshops)}.\hskip 1em
  plus 0.5em minus 0.4em\relax IEEE, 2019, pp. 1--6.

\bibitem{liang2018graph}
L.~Liang, S.~Xie, G.~Y. Li, Z.~Ding, and X.~Yu, ``Graph-based resource sharing
  in vehicular communication,'' \emph{IEEE Trans. Wireless Commun.}, vol.~17,
  no.~7, pp. 4579--4592, Jul. 2018.

\bibitem{khan2011qoe}
A.~Khan, L.~Sun, and E.~Ifeachor, ``Qo{E} prediction model and its application
  in video quality adaptation over {UMTS} networks,'' \emph{IEEE Trans.
  Multimedia}, vol.~14, no.~2, pp. 431--442, Feb. 2011.

\bibitem{xu2012qoe}
C.~Xu, F.~Zhao, J.~Guan, H.~Zhang, and G.-M. Muntean, ``{QoE}-driven
  user-centric {VoD} services in urban multihomed {P2P}-based vehicular
  networks,'' \emph{IEEE Trans. Veh. Technol.}, vol.~62, no.~5, pp. 2273--2289,
  May 2012.

\bibitem{liew2021economics}
Z.~Q. Liew, Y.~Cheng, W.~Y.~B. Lim, D.~Niyato, C.~Miao, and S.~Sun, ``Economics
  of semantic communication system in wireless powered internet of things,''
  \emph{arXiv preprint arXiv:2110.01423}, 2021.

\bibitem{ramezani2017toward}
P.~Ramezani and A.~Jamalipour, ``Toward the evolution of wireless powered
  communication networks for the future internet of things,'' \emph{IEEE
  Netw.}, vol.~31, no.~6, pp. 62--69, Nov./Dec. 2017.

\bibitem{chae2018simultaneous}
S.~H. Chae, C.~Jeong, and S.~H. Lim, ``Simultaneous wireless information and
  power transfer for internet of things sensor networks,'' \emph{IEEE Internet
  Things J.}, vol.~5, no.~4, pp. 2829--2843, Aug. 2018.

\bibitem{devlin2018bert}
J.~Devlin, M.-W. Chang, K.~Lee, and K.~Toutanova, ``{BERT}: {P}re-training of
  deep bidirectional transformers for language understanding,'' \emph{arXiv
  preprint arXiv:1810.04805}, 2018.

\bibitem{liu2017learning}
Z.~Liu, J.~Li, Z.~Shen, G.~Huang, S.~Yan, and C.~Zhang, ``Learning efficient
  convolutional networks through network slimming,'' in \emph{Proc. IEEE Intel.
  Conf. Comput. Vis.}, 2017, pp. 2736--2744.

\bibitem{zhu2021video}
M.~Zhu, C.~Feng, J.~Chen, C.~Guo, and X.~Gao, ``Video semantics based resource
  allocation algorithm for spectrum multiplexing scenarios in vehicular
  networks,'' in \emph{2021 IEEE/CIC Int. Conf. Commun. China (ICCC
  Workshops)}.\hskip 1em plus 0.5em minus 0.4em\relax IEEE, 2021, pp. 31--36.

\bibitem{yang2022semantic}
W.~Yang, Z.~Q. Liew, W.~Y.~B. Lim, Z.~Xiong, D.~Niyato, X.~Chi, X.~Cao, and
  K.~B. Letaief, ``Semantic communication meets edge intelligence,''
  \emph{arXiv preprint arXiv:2202.06471}, 2022.

\bibitem{schwarz2007overview}
H.~Schwarz, D.~Marpe, and T.~Wiegand, ``Overview of the scalable video coding
  extension of the {H}.264/{AVC} standard,'' \emph{IEEE Trans. Circuits Syst.
  Video Technol.}, vol.~17, no.~9, pp. 1103--1120, Sep. 2007.

\bibitem{goyal2001multiple}
V.~K. Goyal, ``Multiple description coding: {C}ompression meets the network,''
  \emph{IEEE Signal Process Mag.}, vol.~18, no.~5, pp. 74--93, May 2001.

\bibitem{popovski2020semantic}
P.~Popovski, O.~Simeone, F.~Boccardi, D.~G{\"u}nd{\"u}z, and O.~Sahin,
  ``Semantic-effectiveness filtering and control for post-5{G} wireless
  connectivity,'' \emph{J. Indian Inst. Sci.}, vol. 100, no.~2, pp. 435--443,
  Feb. 2020.

\bibitem{papineni2002bleu}
K.~Papineni, S.~Roukos, T.~Ward, and W.-J. Zhu, ``B{L}{E}{U}: {A} method for
  automatic evaluation of machine translation,'' in \emph{Proc. 40th annual
  meeting of the Association for Computational Linguistics}, 2002, pp.
  311--318.

\bibitem{vedantam2015cider}
R.~Vedantam, C.~Lawrence~Zitnick, and D.~Parikh, ``{CIDE}r: {C}onsensus-based
  image description evaluation,'' in \emph{Proc. IEEE Conf. Comput. Vis.
  Pattern Recog.}, 2015, pp. 4566--4575.

\bibitem{vincent2006performance}
E.~Vincent, R.~Gribonval, and C.~F{\'e}votte, ``Performance measurement in
  blind audio source separation,'' \emph{IEEE Trans. Audio Speech Lang.
  Process.}, vol.~14, no.~4, pp. 1462--1469, Apr. 2006.

\bibitem{rix2001perceptual}
A.~W. Rix, J.~G. Beerends, M.~P. Hollier, and A.~P. Hekstra, ``Perceptual
  evaluation of speech quality ({PESQ})-a new method for speech quality
  assessment of telephone networks and codecs,'' in \emph{2001 IEEE Int. Conf.
  acoustics, speech, and signal processing. Proceedings (Cat. No. 01CH37221)},
  vol.~2.\hskip 1em plus 0.5em minus 0.4em\relax IEEE, 2001, pp. 749--752.

\bibitem{cox1997three}
R.~V. Cox, ``Three new speech coders from the {ITU} cover a range of
  applications,'' \emph{IEEE Commun. Mag.}, vol.~35, no.~9, pp. 40--47, 1997.

\bibitem{zhang2018unreasonable}
R.~Zhang, P.~Isola, A.~A. Efros, E.~Shechtman, and O.~Wang, ``The unreasonable
  effectiveness of deep features as a perceptual metric,'' in \emph{Proc. IEEE
  Conf. Comput. Vis. Pattern Recog.}, 2018, pp. 586--595.

\bibitem{yang2021deep}
M.~Yang, C.~Bian, and H.-S. Kim, ``Deep joint source channel coding for
  wirelessimage transmission with {OFDM},'' \emph{arXiv preprint
  arXiv:2101.03909}, 2021.

\bibitem{wang2004image}
Z.~Wang, A.~C. Bovik, H.~R. Sheikh, and E.~P. Simoncelli, ``Image quality
  assessment: {F}rom error visibility to structural similarity,'' \emph{IEEE
  Trans. Image Process.}, vol.~13, no.~4, pp. 600--612, Apr. 2004.

\bibitem{simonyan2014very}
K.~Simonyan and A.~Zisserman, ``Very deep convolutional networks for
  large-scale image recognition,'' \emph{arXiv preprint arXiv:1409.1556}, 2014.

\bibitem{gatys2016image}
L.~A. Gatys, A.~S. Ecker, and M.~Bethge, ``Image style transfer using
  convolutional neural networks,'' in \emph{Proc. IEEE Conf. Comput. Vis.
  Pattern Recog.}, 2016, pp. 2414--2423.

\bibitem{dosovitskiy2016generating}
A.~Dosovitskiy and T.~Brox, ``Generating images with perceptual similarity
  metrics based on deep networks,'' \emph{Advances in neural information
  processing systems}, vol.~29, 2016.

\bibitem{uysal2021semantic}
E.~Uysal, O.~Kaya, A.~Ephremides, J.~Gross, M.~Codreanu, P.~Popovski,
  M.~Assaad, G.~Liva, A.~Munari, T.~Soleymani \emph{et~al.}, ``Semantic
  communications in networked systems,'' \emph{arXiv preprint
  arXiv:2103.05391}, 2021.

\bibitem{maatouk2020age}
A.~Maatouk, M.~Assaad, and A.~Ephremides, ``The age of incorrect information:
  {A}n enabler of semantics-empowered communication,'' \emph{arXiv preprint
  arXiv:2012.13214}, 2020.

\bibitem{kaul2012real}
S.~Kaul, R.~Yates, and M.~Gruteser, ``Real-time status: {H}ow often should one
  update?'' in \emph{Proc. IEEE INFOCOM}.\hskip 1em plus 0.5em minus
  0.4em\relax IEEE, 2012, pp. 2731--2735.

\bibitem{yates2021age}
R.~D. Yates, Y.~Sun, D.~R. Brown, S.~K. Kaul, E.~Modiano, and S.~Ulukus, ``Age
  of information: {A}n introduction and survey,'' \emph{IEEE J. Sel. Areas
  Commun.}, vol.~39, no.~5, pp. 1183--1210, May 2021.

\bibitem{sun2019sampling}
Y.~Sun, Y.~Polyanskiy, and E.~Uysal, ``Sampling of the wiener process for
  remote estimation over a channel with random delay,'' \emph{IEEE Trans. Inf.
  Theory}, vol.~66, no.~2, pp. 1118--1135, Feb. 2019.

\bibitem{howard1966information}
R.~A. Howard, ``Information value theory,'' \emph{IEEE Trans. Syst. Man Cybern.
  Syst.}, vol.~2, no.~1, pp. 22--26, Mar. 1966.

\bibitem{ayan2019age}
O.~Ayan, M.~Vilgelm, M.~Kl{\"u}gel, S.~Hirche, and W.~Kellerer,
  ``Age-of-information vs. value-of-information scheduling for cellular
  networked control systems,'' in \emph{Proc. 10th ACM/IEEE Int. Conf.
  Cyber-Physical Systems}, 2019, pp. 109--117.

\bibitem{liang2008analysis}
Y.~J. Liang, J.~G. Apostolopoulos, and B.~Girod, ``Analysis of packet loss for
  compressed video: {E}ffect of burst losses and correlation between error
  frames,'' \emph{IEEE Trans. Circuits Syst. Video Technol.}, vol.~18, no.~7,
  pp. 861--874, Jul. 2008.

\bibitem{holm2021freshness}
J.~Holm, A.~E. Kal{\o}r, F.~Chiariotti, B.~Soret, S.~K. Jensen, T.~B. Pedersen,
  and P.~Popovski, ``Freshness on demand: {O}ptimizing age of information for
  the query process,'' in \emph{Proc. IEEE Int. Conf. Commun.}, 2021, pp. 1--6.

\bibitem{li2021adaptive}
M.~Li, J.~Gao, L.~Zhao, and X.~Shen, ``Adaptive computing scheduling for
  edge-assisted autonomous driving,'' \emph{{IEEE} Trans. Veh. Technol},
  vol.~70, no.~6, pp. 5318--5331, Jun. 2021.

\bibitem{ye2021joint}
Q.~Ye, W.~Shi, K.~Qu, H.~He, W.~Zhuang, and X.~Shen, ``Joint {RAN} slicing and
  computation offloading for autonomous vehicular networks: {A}
  learning-assisted hierarchical approach,'' \emph{{IEEE} Open J. Veh.
  Technol.}, vol.~2, pp. 272--288, Feb. 2021.

\bibitem{de2021survey}
C.~De~Alwis, A.~Kalla, Q.-V. Pham, P.~Kumar, K.~Dev, W.-J. Hwang, and
  M.~Liyanage, ``Survey on 6{G} frontiers: {T}rends, applications,
  requirements, technologies and future research,'' \emph{IEEE Open J. Commun.
  Soc.}, vol.~2, pp. 836--886, 2021.

\bibitem{nanda2019internet}
A.~Nanda, D.~Puthal, J.~J. Rodrigues, and S.~A. Kozlov, ``Internet of
  autonomous vehicles communications security: {O}verview, issues, and
  directions,'' \emph{IEEE Wireless Commun.}, vol.~26, no.~4, pp. 60--65, Apr.
  2019.

\bibitem{samarakoon2019distributed}
S.~Samarakoon, M.~Bennis, W.~Saad, and M.~Debbah, ``Distributed federated
  learning for ultra-reliable low-latency vehicular communications,''
  \emph{IEEE Trans. Commun.}, vol.~68, no.~2, pp. 1146--1159, Feb. 2019.

\bibitem{zhang2021optimizing}
W.~Zhang, D.~Yang, W.~Wu, H.~Peng, N.~Zhang, H.~Zhang, and X.~Shen,
  ``Optimizing federated learning in distributed industrial {I}o{T}: {A}
  multi-agent approach,'' \emph{{IEEE} J. Sel. Areas Commun.}, vol.~39, no.~12,
  pp. 3688--3703, Dec. 2021.

\bibitem{amiri2020federated}
M.~M. Amiri and D.~G{\"u}nd{\"u}z, ``Federated learning over wireless fading
  channels,'' \emph{IEEE Trans. Wireless Commun.}, vol.~19, no.~5, pp.
  3546--3557, 2020.

\bibitem{jiang2019model}
Y.~Jiang, S.~Wang, V.~Valls, B.~J. Ko, W.-H. Lee, K.~K. Leung, and
  L.~Tassiulas, ``Model pruning enables efficient federated learning on edge
  devices,'' \emph{arXiv preprint arXiv:1909.12326}, 2019.

\bibitem{hayat2016survey}
S.~Hayat, E.~Yanmaz, and R.~Muzaffar, ``Survey on unmanned aerial vehicle
  networks for civil applications: {A} communications viewpoint,'' \emph{IEEE
  Commun. Surv. Tutor.}, vol.~18, no.~4, pp. 2624--2661, Apr. 2016.

\bibitem{cheng2019space}
N.~Cheng, F.~Lyu, W.~Quan, C.~Zhou, H.~He, W.~Shi, and X.~Shen,
  ``Space/aerial-assisted computing offloading for {I}o{T} applications: {A}
  learning-based approach,'' \emph{{IEEE} J. Sel. Areas Commun.}, vol.~37,
  no.~5, pp. 1117--1129, May 2019.

\bibitem{zhou2020deep}
C.~Zhou, W.~Wu, H.~He, P.~Yang, F.~Lyu, N.~Cheng, and X.~Shen, ``Deep
  reinforcement learning for delay-oriented iot task scheduling in {SAGIN},''
  \emph{IEEE Trans. Wirel. Commun.}, vol.~20, no.~2, pp. 911--925, Feb. 2020.

\bibitem{laneman2004cooperative}
J.~N. Laneman, D.~N. Tse, and G.~W. Wornell, ``Cooperative diversity in
  wireless networks: {E}fficient protocols and outage behavior,'' \emph{IEEE
  Trans. Inf. Theory}, vol.~50, no.~12, pp. 3062--3080, Dec. 2004.

\bibitem{luo2021autoencoder}
X.~Luo, Z.~Chen, B.~Xia, and J.~Wang, ``Autoencoder-based semantic
  communication systems with relay channels,'' \emph{arXiv preprint
  arXiv:2111.10083}, 2021.

\bibitem{du2022exploring}
H.~Du, J.~Wang, D.~Niyato, J.~Kang, Z.~Xiong, D.~I. Kim \emph{et~al.},
  ``Exploring attention-aware network resource allocation for customized
  metaverse services,'' \emph{arXiv preprint arXiv:2208.00369}, 2022.

\bibitem{liu2022slicing4meta}
Y.-J. Liu, H.~Du, D.~Niyato, G.~Feng, J.~Kang, and Z.~Xiong, ``Slicing4{M}eta:
  {A}n intelligent integration framework with multi-dimensional network
  resources for metaverse-as-a-service in web 3.0,'' \emph{arXiv preprint
  arXiv:2208.06081}, 2022.

\bibitem{ismail2022semantic}
L.~Ismail, D.~Niyato, S.~Sun, D.~I. Kim, M.~Erol-Kantarci, and C.~Miao,
  ``Semantic information market for the metaverse: {A}n auction based
  approach,'' \emph{arXiv preprint arXiv:2204.04878}, 2022.

\bibitem{movassaghi2014wireless}
S.~Movassaghi, M.~Abolhasan, J.~Lipman, D.~Smith, and A.~Jamalipour, ``Wireless
  body area networks: {A} survey,'' \emph{IEEE Commun. Surv. Tutor.}, vol.~16,
  no.~3, pp. 1658--1686, Mar. 2014.

\bibitem{elsts2018board}
A.~Elsts, R.~McConville, X.~Fafoutis, N.~Twomey, R.~J. Piechocki,
  R.~Santos-Rodriguez, and I.~Craddock, ``On-board feature extraction from
  acceleration data for activity recognition.'' in \emph{EWSN}, 2018, pp.
  163--168.

\bibitem{zalewski2020bits}
P.~Zalewski, L.~Marchegiani, A.~Elsts, R.~Piechocki, I.~Craddock, and
  X.~Fafoutis, ``From bits of data to bits of knowledge—an on-board
  classification framework for wearable sensing systems,'' \emph{Sensors},
  vol.~20, no.~6, p. 1655, 2020.

\bibitem{bragancca2019brief}
S.~Bragan{\c{c}}a, E.~Costa, I.~Castellucci, and P.~M. Arezes, ``A brief
  overview of the use of collaborative robots in industry 4.0: {H}uman role and
  safety,'' \emph{Occup. Environ. Saf. Health}, pp. 641--650, 2019.

\bibitem{yue2020collaborative}
Y.~Yue, C.~Zhao, Z.~Wu, C.~Yang, Y.~Wang, and D.~Wang, ``Collaborative semantic
  understanding and mapping framework for autonomous systems,'' \emph{IEEE/ASME
  Trans. Mechatron.}, vol.~26, no.~2, pp. 978--989, Feb. 2020.

\bibitem{milis2017semiotics}
G.~M. Milis, C.~G. Panayiotou, and M.~M. Polycarpou, ``Semiotics:
  {S}emantically enhanced {I}o{T}-enabled intelligent control systems,''
  \emph{IEEE Internet Things J.}, vol.~6, no.~1, pp. 1257--1266, Jan. 2017.

\bibitem{kolokotsa2003comparison}
D.~Kolokotsa, ``Comparison of the performance of fuzzy controllers for the
  management of the indoor environment,'' \emph{Build. Environ.}, vol.~38,
  no.~12, pp. 1439--1450, Dec. 2003.

\bibitem{da2014internet}
L.~Da~Xu, W.~He, and S.~Li, ``Internet of things in industries: {A} survey,''
  \emph{IEEE Trans. Industr. Inform.}, vol.~10, no.~4, pp. 2233--2243, Apr.
  2014.

\bibitem{gao2021mac}
J.~Gao, W.~Zhuang, M.~Li, X.~Shen, and X.~Li, ``{MAC} for machine-type
  communications in industrial {I}o{T}—{P}art {I}: {P}rotocol design and
  analysis,'' \emph{{IEEE} Internet Things J.}, vol.~8, no.~12, pp. 9945--9957,
  Aug. 2021.

\bibitem{popovski2021internet}
P.~Popovski, F.~Chiariotti, V.~Croisfelt, A.~E. Kal{\o}r, I.~Leyva-Mayorga,
  L.~Marchegiani, S.~R. Pandey, and B.~Soret, ``Internet of things ({I}o{T})
  connectivity in 6{G}: {A}n interplay of time, space, intelligence, and
  value,'' \emph{arXiv preprint arXiv:2111.05811}, 2021.

\bibitem{montavon2018methods}
G.~Montavon, W.~Samek, and K.-R. M{\"u}ller, ``Methods for interpreting and
  understanding deep neural networks,'' \emph{Digit. Signal Process.}, vol.~73,
  pp. 1--15, 2018.

\bibitem{kim2016examples}
B.~Kim, R.~Khanna, and O.~O. Koyejo, ``Examples are not enough, learn to
  criticize! criticism for interpretability,'' \emph{Adv. Neural Inf. Process.
  Syst. (NIPS)}, vol.~29, 2016.

\bibitem{dong2017improving}
Y.~Dong, H.~Su, J.~Zhu, and B.~Zhang, ``Improving interpretability of deep
  neural networks with semantic information,'' in \emph{Proc. IEEE Comput. Soc.
  Conf. (CVPR)}, 2017, pp. 4306--4314.

\bibitem{sheraz2020artificial}
M.~Sheraz, M.~Ahmed, X.~Hou, Y.~Li, D.~Jin, Z.~Han, and T.~Jiang, ``Artificial
  intelligence for wireless caching: {S}chemes, performance, and challenges,''
  \emph{{IEEE} Commun. Surv. Tutor.}, vol.~23, no.~1, pp. 631--661, Jan. 2020.

\bibitem{cidon2016cliffhanger}
A.~Cidon, A.~Eisenman, M.~Alizadeh, and S.~Katti, ``Cliffhanger: {S}caling
  performance cliffs in web memory caches,'' in \emph{13th USENIX Symposium on
  Networked Systems Design and Implementation (NSDI 16)}, 2016, pp. 379--392.

\bibitem{xiao2022reasoning}
Y.~Xiao, Y.~Li, G.~Shi, and H.~V. Poor, ``Reasoning on the air: {A}n implicit
  semantic communication architecture,'' \emph{arXiv preprint
  arXiv:2202.01950}, 2022.

\bibitem{liang2022life}
J.~Liang, Y.~Xiao, Y.~Li, G.~Shi, and M.~Bennis, ``Life-long learning for
  reasoning-based semantic communication,'' \emph{arXiv preprint
  arXiv:2202.01952}, 2022.

\bibitem{wang20206g}
C.-X. Wang, J.~Huang, H.~Wang, X.~Gao, X.~You, and Y.~Hao, ``6{G} wireless
  channel measurements and models: {T}rends and challenges,'' \emph{IEEE Veh.
  Technol. Mag.}, vol.~15, no.~4, pp. 22--32, Apr. 2020.

\bibitem{hu2022metasketch}
J.~Hu, H.~Zhang, K.~Bian, Z.~Han, H.~V. Poor, and L.~Song, ``Metasketch:
  {W}ireless semantic segmentation by reconfigurable intelligent surfaces,''
  \emph{IEEE Trans. Wireless Commun.}, to apper, 2022.

\bibitem{gacanin2020wireless}
H.~Gacanin and M.~Di~Renzo, ``Wireless 2.0: {T}oward an intelligent radio
  environment empowered by reconfigurable meta-surfaces and artificial
  intelligence,'' \emph{IEEE Veh. Technol. Mag.}, vol.~15, no.~4, pp. 74--82,
  Apr. 2020.

\bibitem{letaief2021edge}
K.~B. Letaief, Y.~Shi, J.~Lu, and J.~Lu, ``Edge artificial intelligence for
  6{G}: {V}ision, enabling technologies, and applications,'' \emph{IEEE Journal
  on Selected Areas in Communications}, vol.~40, no.~1, pp. 5--36, 2021.

\bibitem{bloch2016covert}
M.~R. Bloch, ``Covert communication over noisy channels: {A} resolvability
  perspective,'' \emph{IEEE Trans. Inf. Theory}, vol.~62, no.~5, pp.
  2334--2354, May 2016.

\end{thebibliography}

\end{document}